\documentclass[fleqn,a4paper,usenatbib]{mnras}

\usepackage{newtxtext,newtxmath}

\usepackage[T1]{fontenc}
\usepackage{ae,aecompl}

\usepackage{graphicx}	
\usepackage{amsmath}	
\usepackage{amssymb}	

\usepackage{tikz} 
\tikzset{
  load/.style   = {ultra thick,-latex},
  stress/.style = {-latex},
  dim/.style    = {latex-latex},
  axis/.style   = {-latex},
}
\definecolor{oxfordblue}{rgb}{0.0, 0.13, 0.28}
\definecolor{fluorescentorange}{rgb}{1.0, 0.85, 0.0}
\definecolor{seagreen}{RGB}{46,139,87} 
\usepackage{subfig}

\newcommand{\vb}[1]{\boldsymbol{#1}}
\newcommand{\nablab}{\boldsymbol{\nabla}}
\newcommand{\xib}{\boldsymbol{\xi}}
\newcommand{\dpartial}[2]{\frac{\partial #1}{\partial #2}}
\newcommand{\ddpartial}[2]{\frac{\partial^2 #1}{\partial #2^2}}
\newcommand{\dtotal}[2]{\frac{\mathrm{d} #1}{\mathrm{d} #2}}
\newcommand{\ddtotal}[2]{\frac{\mathrm{d}^2 #1}{\mathrm{d} #2^2}}

\graphicspath{{./figures/}}

\title[Wave propagation in density staircases]{Wave propagation in semi-convective regions of giant planets}

\author[C. M. Pontin et al.]{
C. M. Pontin,$^{1}$\thanks{E-mail: mmcmp@leeds.ac.uk}
A. J. Barker,$^{1}$
R. Hollerbach,$^{1}$
Q. Andr\'e$^{2}$
and S. Mathis$^{2}$
\\
$^{1}$ Department of Applied Mathematics, School of Mathematics, University of Leeds, Leeds, LS2 9JT, UK \\
$^{2}$ AIM, CEA, CNRS, Universit\'e Paris-Saclay, Universit\'{e} Paris Diderot, Sorbonne Paris Cit\'e, F-91191 Gif-sur-Yvette, France \\
}

\date{Accepted XXX. Received YYY; in original form ZZZ}

\pubyear{2020}

\begin{document}
\label{firstpage}
\pagerange{\pageref{firstpage}--\pageref{lastpage}}
\maketitle

\begin{abstract}
Recent observations of Jupiter and Saturn suggest that heavy elements may be diluted in the gaseous envelope, providing a compositional gradient that could stabilise ordinary convection and produce a stably-stratified layer near the core of these planets. This region could consist of semi-convective layers with a staircase-like density profile, which have multiple convective zones separated by thin stably-stratified interfaces, as a result of double-diffusive convection. These layers could have important effects on wave propagation and tidal dissipation that have not been fully explored. We analyse the effects of these layers on the propagation and transmission of internal waves within giant planets, extending prior work in a local Cartesian model.
We adopt a simplified global Boussinesq planetary model in which we explore the internal waves in a non-rotating spherical body. We begin by studying the free modes of a region containing semi-convective layers. We then analyse the transmission of internal waves through such a region. The free modes depend strongly on the staircase properties, and consist of modes with both internal and interfacial gravity wave-like behaviour. We determine the frequency shifts of these waves as a function of the number of steps to explore their potential to probe planetary internal structures. We also find that wave transmission is strongly affected by the presence of a staircase. Very large-wavelength waves are transmitted efficiently, but small-scale waves are only transmitted if they are resonant with one of the free modes. The effective size of the core is therefore larger for non-resonant modes. 
\end{abstract}

\begin{keywords}
planets and satellites: gaseous planets -- hydrodynamics -- waves -- planets and satellites: physical evolution -- asteroseismology -- methods: analytical
\end{keywords}



\section{Introduction}
\label{sec:intro}

Understanding the internal structures of giant planets and stars is an important topic in astrophysics and planetary sciences. The interior structures of stars are generally well understood, helped in a large part by progress in helio- and asteroseismology (e.g.~\citealt{JCD2002,Chaplin2013}). High-precision photometric data from space missions such as CoRoT and \textit{Kepler} has enabled much recent progress in asteroseismology, which has extended our knowledge of the interior structure of stars to those outside the solar system \citep{Baglin2002,Kepler2010}. These methods rely on analysing the frequencies of stellar photometric variability, which allow us to probe the interior structure of a star if the internal free oscillation modes are well understood. It is however much more difficult to explore giant planets in a similar way, even those in our solar system (but see \citealt{Gaulme2011} who have detected a signal compatible with global acoustic modes using ground-based instrumentation). On the other hand, space missions such as Juno and Cassini have allowed high-precision measurements of the gravity fields of the giant planets in our solar system, Jupiter and Saturn \citep{Miguel2016,BoltonJUNO2017}. This information has allowed us to constrain planetary interior models, as well as their internal differential rotation (e.g.~\citealt{Wahl17,Guillot2018,Iess2019}). 

The interiors of giant planets are traditionally modelled with a three-layer structure, consisting of a rocky/icy core underneath a convective envelope of metallic hydrogen and helium, which is surrounded by a molecular envelope (e.g.~\citealt{Stevenson1982,Guillot2005,Fortney2010}). Each layer is usually assumed to be chemically homogeneous, with the heavy elements concentrated in the core. However, the sizes of each region, and the exact nature of the transitions between them are uncertain. Furthermore, recent observational evidence from Juno gravity field measurements indicates that heavy elements are probably distributed throughout the inner regions of the planet \citep{Wahl17,Helled2017,Debras2019}. As a result, there has been much ongoing research in recent years to explore planetary models incorporating compositional gradients or non-adiabatic structures \citep{ChabrierBaraffe2007,Leconte2012,Lozovsky2017,Vazan2016,Berardo2017,Vazan2018,Debras2019}.

Standard models with chemically homogeneous layers also assume convection to be efficient so that the entropy profile is adiabatic. Compositional gradients can however interact with ordinary convection, and inhibit it if the concentration of heavy elements decreases sufficiently rapidly with radius. 
In fluids with a stabilising compositional gradient and a destabilising entropy gradient, double-diffusive convection (also known as ``semi-convection" in astrophysics) may occur instead (e.g.~\citealt{Garaud2018}). This is possible if temperature diffuses more rapidly than composition, as expected in giant planet interiors. Double-diffusive convection is an oscillatory linear instability (or overstability) that excites internal gravity waves. It exhibits fascinating nonlinear dynamics, and often leads to the formation of layers in the density field (e.g.~\citealt{Wood2013,Garaud2018}). This layered state consists of thin convective layers (probably much smaller than a pressure scale height) that are sandwiched by much thinner diffusive (stably-stratified) interfaces, so that the density profile resembles a staircase. This leads to a non-adiabatic, stably-stratified, entropy profile in these parts of the planetary interior. Density staircases have also been observed in the Artic oceans on Earth, where there is a stabilising salinity gradient and a destabilising thermal gradient (e.g.~\citealt{Arctic2016,Shibley2017}). They may also be present outside the cores of massive stars, where heavy elements generated through nuclear reactions can diffuse into the neighbouring convective region \citep{Maeder2009,Kippenhahn2012}. 

The transport of heat by double-diffusive convection is much less efficient than that by ordinary convection (in the absence of compositional gradients), so its occurrence has important consequences for the evolution of giant planets \citep{ChabrierBaraffe2007,Leconte2012,Lozovsky2017,Vazan2016,Berardo2017,Vazan2018} and ice giant planets \citep{PodolakUrNep2019,HelledUrNep2019}. In particular, this could contribute to the inflated radii of some hot Jupiters (e.g.~\citealt{ChabrierBaraffe2007}). Saturn's observed luminosity is also larger than predicted using standard models at its present age, and the delayed cooling caused by compositional gradients is one possible explanation \citep{Leconte2013}. 

In addition to the gravity field measurements of Jupiter and Saturn, there is further indirect evidence for the possible existence of a stably-stratified layer in the interiors of these planets. 
Certain density waves in Saturn's rings are believed to be excited by gravitational forcing due to global oscillation modes inside Saturn \citep{MarleyPorco1993,Hedman2013,Hedman2019}. \cite{Fuller2014} showed that their frequencies could only be reproduced in his models if there is a (sufficiently large and strongly) stably-stratified region outside the core, which modifies the frequencies of the f-modes (surface gravity modes -- strictly speaking these are mixed modes). While these models may not contain all of the relevant physics at present, they provide independent evidence for the existence of a stably-stratified region near the core of Saturn.

There are two regions in the deep interiors of giant planets which could be stably stratified and potentially contain density staircases. The first is the region outside the core, where a compositional gradient could be produced by the erosion or dissolution of the core  \citep{Guillot2004,Wilson2012,Moll2017}, or perhaps exist as a remnant of the formation of these planets \citep{Stevenson1982,Liu2019}. The second region is located further from the centre, occurring near the transition between the metallic and molecular hydrogen and helium layers, where the conditions may be suitable for helium rain to occur \citep{Stevenson1977,Nettelmann2015}. In particular, this is thought to reduce the helium content of Saturn's outer envelope, and may create a stabilising compositional gradient. A different type of giant planet interior model with an extended stably-stratified layer near the surface has also been explored by \cite{IoannouLindzen1993p1,IoannouLindzen1993p2}.

The long-term tidal evolution of star, planet and moon systems can also be used to constrain the internal structures of these bodies (e.g.~\citealt{OgilvieLin2004,Mathis2013,Ogilvie2014}). This is because the rates of tidal dissipation are believed to depend strongly on their interior structure. Astrometric observations of the moons of both Jupiter and Saturn indicate that their moons are migrating outwards at rates that require efficient tidal dissipation inside these planets \citep{Lainey2009,Lainey2012,Lainey2017}. It is uncertain how such efficient tidal dissipation can be explained theoretically. One possibility, which motivates the present paper, is that stably-stratified layers in giant planets could play a key role, by enabling the excitation (and subsequent dissipation) of gravity waves. However, the presence of a density staircase could modify the properties of these waves, and this has not yet been fully explored. Alternative possibilities to explain the observations include the dissipation of inertial waves in convective regions \citep{OgilvieLin2004,Favier2014}, visco-elastic dissipation in a rocky/icy core \citep{Remus2012}, and the resonant locking of tidal gravito-inertial modes \citep{Fuller2016}. The latter mechanism may  require a stable layer to operate effectively. Note that the effective viscosity of turbulent convection acting on the non-wavelike tidal flows is unlikely to be important (e.g.~\citealt{GN1977,Duguid2019}).

Motivated by the potential importance of stably-stratified layers in giant planets, and of their possible density staircases, we set out to analyse the effects of these layers on wave propagation, and ultimately also on tidal dissipation. The effect of a density staircase on the free oscillation modes of a stratified region was studied by \cite{Belyaev2015} using a local Cartesian model. The free modes were found to differ from those of a continuously stratified medium, with those waves with wavelengths that are comparable with a step-size being affected the most. The transmission of internal waves through a density staircase in a similar Cartesian model was studied by \cite{Sutherland2016}, who adopted the ``traditional approximation" to incorporate rotation (this assumes that the buoyancy force dominates the Coriolis acceleration in the direction of stratification, thereby disallowing inertial waves), and subsequently \cite{Andre2017} studied the free modes and transmission of internal and inertial waves in a local model that included the full Coriolis acceleration at any latitude in a planet. The density staircase was found to strongly affect the transmission of waves through such a region in a frequency and wavelength-dependent manner. In particular, incident gravito-inertial waves are preferentially transmitted if they have large wavelengths relative to a step size, or if they are resonant with one of its free modes. Inertial waves are also strongly affected by a staircase, and are primarily reflected unless they have a large wavelength relative to the size of the entire stratified region, except for those modes that are resonant with a free mode of the staircase, or if their frequencies match the local inertial frequency \citep{Andre2017}.

In this paper, we set out to analyse the effects of a density staircase on the propagation and transmission of internal waves within giant planets. We build upon these prior works by adopting a simplified global (spherical) Boussinesq model. This allows us to study the propagation of waves with wavelengths comparable with the radius of the stratified layer, which may be important for the inner regions of these planets, and also those with small harmonic degrees (therefore large horizontal wavelengths) such as those that may be the easiest to observe. We neglect rotation in this study partly because we focus on internal waves, and partly for simplicity, because including the full effects of rotation makes the problem inherently two-dimensional (e.g.~\citealt{Dintrans1999}). Incorporating the full effects of rotation in a global model is an important topic for future work (e.g. \citealt{OgilvieLin2004,Ogilvie2007}).

The structure of this paper is as follows: in \S~\ref{sec:set_up}, we outline our model and the governing equations. In \S~\ref{sec:dispersion_realation}, we derive the dispersion relation governing the free modes of a density staircase, and discuss its properties. We compare the modes of a staircase to those of a continuously stratified medium in \S~\ref{sec:mode_diff}, and we explore the transmission of internal waves in \S~\ref{sec:transmission}.  Finally, we present our conclusions in \S~\ref{Conclusions}.

\section{Model}
\label{sec:set_up}

We consider the propagation of internal waves through a region consisting of well-mixed convective layers separated by infinitesimally thin interfaces, i.e.~a density staircase. This work extends \cite{Andre2017} to spherical geometry.

We adopt the Boussinesq approximation \citep{Spiegel1960} for simplicity, and to facilitate understanding before we progress to a more complicated physical model. This is appropriate for studying waves with shorter wavelengths than a pressure or density scale height, and with phase speeds that are slow relative to the sound speed. This is likely to be a reasonable approximation for studying the free modes of a density staircase, though it is strictly not valid for studying the largest wavelength waves in a planet. We also adopt the Cowling approximation \citep{Cowling1941}, thereby neglecting perturbations to the gravitational potential, which is a reasonable approximation for studying internal waves, particularly those with (horizontal and radial) wavelengths that are shorter than the planetary radius.

\subsection{Governing equations}\label{sec:eqnofmotion}
We briefly outline the derivation of the linear adiabatic equations of motion describing the non-radial oscillations of a non-rotating spherical planet \citep{Gough1993,JCDLectures,Thompson2006}. We use spherical polar coordinates $(r,\theta,\phi)$, where $r=0$ corresponds to the centre of the planet, and adopt a basic state that is a spherically-symmetric planetary model in hydrostatic equilibrium, with density $\rho_0(r)$, pressure $p_0(r)$ and gravitational potential $\Phi_0(r)$.
We consider linear perturbations to this basic state of the form  
\begin{equation*}
p(\vb{r},t)=p_0(r)+p'(\vb{r},t),
\end{equation*}
and similarly for other variables, where a prime denotes the Eulerian perturbation, $\vb{\xi}$ is the Eulerian displacement and $\vb{u}=\partial \vb{\xi}/\partial t$ is the fluid velocity. The resulting linearised adiabatic (thus far fully compressible) equations of motions are,
\begin{align}
\label{eq:linrho}
 & \rho' + \rho_0 \nablab \cdot \vb{\xi}=0, \\
 & \rho_0 \ddpartial{\vb{\xi}}{t} = -\nablab p' +\rho'\vb{g}_0, \\ 
\label{eq:linen}
 & p'+\vb{\xi} \cdot \nablab p_0 = \frac{\Gamma_{1} p_0}{\rho_0}\left(\rho'+\vb{\xi} \cdot \nablab \rho_0\right),
\end{align}
where $\Gamma_1=\left(\frac{\partial \ln p_0}{\partial \ln \rho_0}\right)_{\mathrm{ad}}$ is the first adiabatic exponent and $\vb{g}_0=-\nabla\Phi_0$. The displacement is split into radial and horizontal components, 
\begin{equation*}
\vb{\xi} = \xi_r \vb{\hat{r}} + \xib_h,
\end{equation*}
where $\vb{\hat{r}}\cdot \xib_h=0$, and $\vb{\hat{r}}$ is the radial unit vector. Since the basic state is static and spherically-symmetric, we may expand perturbations using spherical harmonics with harmonic time-dependence, i.e.~
\begin{equation*}
 \xi_r(r,\theta,\phi,t)=\tilde{\xi}_r(r) Y_l^m(\theta,\phi)\mathrm{e}^{-i\omega t},
\end{equation*}
and similarly for other variables, where the physical quantity is the real part of this expression, and we use orthonormalised spherical harmonics $Y_l^m$. Substituting this into Eqs.~(\ref{eq:linrho})~to~(\ref{eq:linen}), and using these to eliminate $\tilde{\xib}_h$ and $\tilde{\rho}$, we obtain:
\begin{align}
& \dtotal{\tilde{\xi}_r}{r}=-\bigg(\frac{2}{r}+ \frac{1}{\Gamma_{1} p_0}\dtotal{p_0}{r}\bigg)\tilde{\xi}_r + \frac{1}{\rho_0 \omega^2c^2}\bigg(S_l^2-\omega^2\bigg)\tilde{p}', \\
& \dtotal{\tilde{p}'}{r}=\rho_0\bigg(\omega^2 - N^2\bigg)\tilde{\xi}_r + \frac{1}{\Gamma_{1} p_0}\dtotal{p_0}{r}\tilde{p}', 
\end{align}
where the squared adiabatic sound speed is
\begin{equation}
c^2 = \Gamma_{1} \frac{p_0}{\rho_0},
\end{equation}
the squared Lamb frequency is
\begin{equation}
S_l^2=\frac{l(l+1)c^2}{r^2},
\end{equation}
and the squared buoyancy frequency, or Brunt-V\"ais\"al\"a frequency, is
\begin{equation}
N^2=g\left(\frac{1}{\Gamma_{1}}\frac{\ln\mathrm{d}p_0}{\mathrm{d}r}-\frac{\mathrm{d}\ln\rho_0}{\mathrm{d}r}\right).
\end{equation}
We have also defined $\vb{g}_0=-g(r)\hat{\vb{r}}$. The radial dependence of $g(r)$ involves the density structure of the entire region within that radius, not just the staircase. 

To simplify our analysis we assume that the background variations in density and pressure are much smaller than their maximum values, and that the wave speed is much smaller than the adiabatic sound speed, or equivalently, that $\omega^2 \ll S_l^2$.
The above system then reduces to
\begin{align}\label{eq:dxi}
& \dtotal{\tilde{\xi}_r}{r} =
-\frac{2 \tilde{\xi}_r }{r}+ \frac{1}{\rho_0 \omega^2}\frac{l(l+1)}{r^2}\tilde{p}', \\
\label{eq:dp}
& \dtotal{\tilde{p}'}{r}=\rho_0 \omega^2 \bigg(1 - \frac{N^2}{\omega^2}\bigg)\tilde{\xi}_r,
\end{align}
which can be combined to give
\begin{equation}
\label{EQ1}
\ddtotal{\tilde{\xi}_r}{r} + \frac{4}{r}\dtotal{\tilde{\xi}_r}{r}+ \Bigg[ \bigg(\frac{N^2}{\omega^2}-1\bigg)l(l+1)+ 2\Bigg]\frac{\tilde{\xi}_r}{r^2}=0.
\end{equation}
We confirm in Appendix \ref{app:B_approx} that this equation can also be obtained by adopting the Boussinesq approximation from the outset. 

We note that  Eqn.~(\ref{EQ1}) can be simplified using the substitution $\chi=r^2 \tilde{\xi}_r$, reducing it to the form, 
\begin{equation}
\frac{\mathrm{d}^2 \chi}{\mathrm{d}r^2} + \bigg(\frac{N^2}{\omega^2}-1\bigg)l(l+1)\frac{\chi}{r^2}=0,
\end{equation}
where the effective radial wavenumber can be identified as, 
\begin{equation}
\label{krcalc}
    k_r^2=\frac{l(l+1)}{r^2}\bigg(\frac{N^2}{\omega^2}-1\bigg). 
\end{equation}

Note that in general
\begin{equation}
    N^2=-\frac{T_0\alpha_T}{c_p}\vb{g}_0\cdot \nabla s_0,
\end{equation}
where $\alpha_T$ is the coefficient of thermal expansion, $c_p$ is the specific heat capacity at constant pressure, and $T_0$ and $s_0$ are the temperature and specific entropy profiles for the basic state. This means that $N^2\propto \partial_r s_0$. In the next section, we will specify a background profile of $s_0(r)$ that represents a layered profile. For clarity of presentation and comparison with prior work, we will refer to this as a ``density staircase",  which will be represented by a particular choice of $\rho_0(r)$, that is related to the buoyancy frequency in the incompressible limit by 
\begin{equation}
N^2\approx-\frac{g}{\rho_0}\frac{\mathrm{d}\rho_0}{\mathrm{d}r}.
\end{equation}
However, it should be remembered that we are strictly considering an ``entropy staircase", once we correctly account for the difference between the density and entropy of the gas.

\pgfdeclareradialshading{pizza}{\pgfpointorigin}{
  color(0cm)=(red!80);
  color(1.1cm)=(red!80);
  color(1.2cm)=(red!70); 
  color(1.3cm)=(red!60); 
  color(1.5cm)=(orange!75);  
  color(2.3cm)=(fluorescentorange!60);
  color(4.1cm)=(fluorescentorange!45);
  color(4.5cm)=(seagreen!50);
  color(4.9cm)=(oxfordblue!50);
  color(6.0cm)=(oxfordblue!80)
}

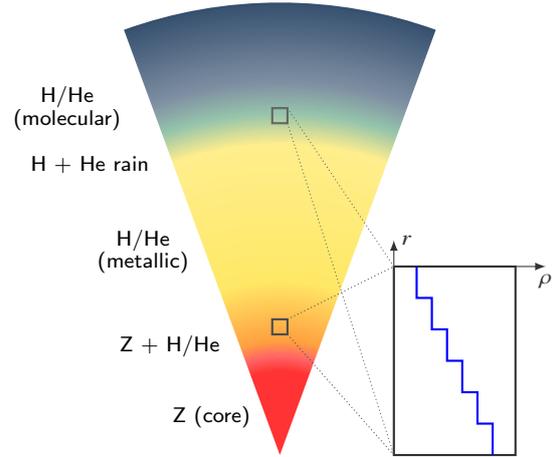
\begin{figure}
\subfloat{
\begin{tikzpicture}[scale=1]
  \begin{scope}
    \clip (0,0) -- +(70:6) arc (70:110:6) --cycle;
    \pgfuseshading{pizza}
  \end{scope}
  \draw[thick, black!70] (-0.1,1.6) -- (0.1,1.6) -- (0.1,1.8) -- (-0.1,1.8) -- cycle;
  \draw[thick, black!60] (-0.1,4.4) -- (0.1,4.4) -- (0.1,4.6) -- (-0.1,4.6) -- cycle;
  \draw[thick, black!80] (1.5,0) -- (3.1,0) -- (3.1,2.5) -- (1.5,2.5) -- cycle;
  \draw[axis, black!80] (1.5,0) -- (1.5,2.85) node[right] {$r$};
  \draw[axis, black!80] (1.5,2.5) -- (3.5,2.5) node[below] {$\rho$};
  \draw[densely dotted, black!70] (0.1,1.6) -- (1.5,0);
  \draw[densely dotted, black!70] (0.1,1.8) -- (1.5,2.5);
  \draw[densely dotted, black!60] (0.1,4.4) -- (1.5,0);
  \draw[densely dotted, black!60] (0.1,4.6) -- (1.5,2.5);
  \pgfmathsetmacro{\d}{2.5/6}
  \draw[thick, color=blue] (1.8,2.5) -- ++ (0,-\d) -- ++ (0.2,0) -- ++ (0,-\d) -- ++ (0.2,0) -- ++ (0,-\d) -- ++ (0.2,0) -- ++ (0,-\d) -- ++ (0.2,0) -- ++ (0,-\d) -- ++ (0.2,0) -- ++ (0,-\d);
  \node at (-0.9,0.5) {\small \textsf{Z (core)}};
  \node at (-1.45,1.45) {\small \textsf{Z + H/He}};
  \node at (-1.8,2.85) {\small \textsf{H/He}};
    \node at (-1.8,2.55) {\small \textsf{(metallic)}};
  \node at (-2.5,3.85) {\small \textsf{H + He rain}};
  \node at (-2.8,4.75) {\small \textsf{H/He}};
  	\node at (-2.8,4.45) {\small \textsf{(molecular)}};
\end{tikzpicture}}
    \caption{Diagram showing the expected radial locations of stably-stratified layers in giant planet interiors. Figure adapted from \citet{Andre2017}.}
    \label{fig:planet}
\end{figure}

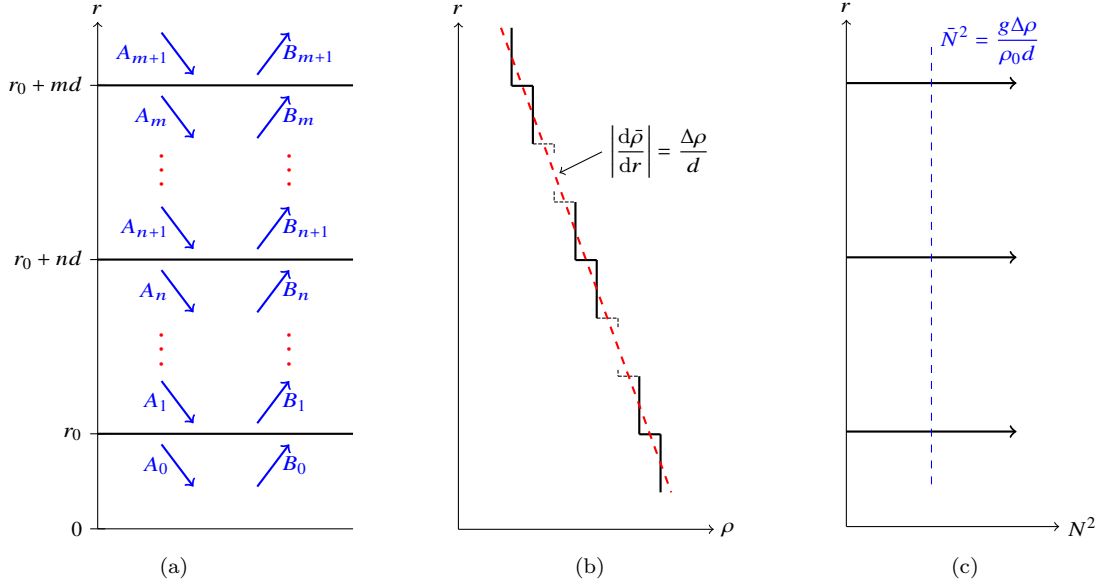
\begin{figure*}
\subfloat[]{
	\begin{tikzpicture}[scale=2.8]
		\draw[-] (0,0) -- (1.2,0) coordinate (x axis);
		\draw[->] (0,0) -- (0,2.4) coordinate (y axis) node[above]{$r$};
		\foreach \y/\ytext in {0/0, 0.45/r_0, 1.275/r_0+nd, 2.1/r_0+md} 
		\draw (1pt,\y cm) -- (-1pt,\y cm) node[anchor=east,fill=white] {$\ytext$};
		\draw[thick] (0,0.45)-- (1.2,0.45);
		\draw[thick] (0,1.275)-- (1.2,1.275);
		\draw[thick] (0,2.1)-- (1.2,2.1);
		\draw[->,blue,thick] (0.3,0.4) -- ++(0.15,-0.2) node[left,pos=0.5]{$A_0$};
		\draw[<-,blue,thick] (0.9,0.4) -- ++(-0.15,-0.2) node[right,pos=0.5]{$B_0$};
		\draw[->,blue,thick] (0.3,0.7) -- ++(0.15,-0.2) node[left,pos=0.5]{$A_1$};
		\draw[<-,blue,thick] (0.9,0.7) -- ++(-0.15,-0.2) node[right,pos=0.5]{$B_1$};
		\draw[->,blue,thick] (0.3,1.225) -- ++(0.15,-0.2) node[left,pos=0.5]{$A_n$};
		\draw[<-,blue,thick] (0.9,1.225) -- ++(-0.15,-0.2) node[right,pos=0.5]{$B_n$};
		\draw[->,blue,thick] (0.3,1.525) -- ++(0.15,-0.2) node[left,pos=0.5]{$A_{n+1}$};
		\draw[<-,blue,thick] (0.9,1.525) -- ++(-0.15,-0.2) node[right,pos=0.5]{$B_{n+1}$};
		\draw[->,blue,thick] (0.3,2.05) -- ++(0.15,-0.2) node[left,pos=0.5]{$A_m$};
		\draw[<-,blue,thick] (0.9,2.05) -- ++(-0.15,-0.2) node[right,pos=0.5]{$B_m$};
		\draw[->,blue,thick] (0.3,2.35) -- ++(0.15,-0.2) node[left,pos=0.5]{$A_{m+1}$};
		\draw[<-,blue,thick] (0.9,2.35) -- ++(-0.15,-0.2) node[right,pos=0.5]{$B_{m+1}$};
		\path (0.3,0.45) -- (0.3,1.275) node [red, font=\Large, midway, sloped] {$\dots$};
		\path (0.9,0.45) -- (0.9,1.275) node [red, font=\Large, midway, sloped] {$\dots$};
		\path (0.3,2.1) -- (0.3,1.275) node [red, font=\Large, midway, sloped] {$\dots$};
		\path (0.9,2.1) -- (0.9,1.275) node [red, font=\Large, midway, sloped] {$\dots$};
	\end{tikzpicture}\label{fig:steps_diagram}}
	\hspace{1cm}
\subfloat[]{
	\begin{tikzpicture}[scale=2.8]
		\draw[->] (0,0) -- (1.2,0) coordinate (x axis) node[right]{$\rho$};
		\draw[->] (0,0) -- (0,2.4) coordinate (y axis) node[above]{$r$};
		\draw[thick] (0.25,2.375) -- (0.25,2.1);
		\draw[thick] (0.25,2.1) -- (0.35,2.1);
		\draw[thick] (0.35,2.1) -- (0.35,1.825);
		\draw[dashed,thin,dash pattern=on 1.4pt off 0.6pt] (0.35,1.825) -- (0.45,1.825);
		\draw[dashed,thin,dash pattern=on 1.4pt off 0.6pt] (0.45,1.825) -- (0.45,1.775);
		\draw[dashed,thin,dash pattern=on 1.4pt off 0.6pt] (0.45,1.6) -- (0.45,1.55);		
		\draw[dashed,thin,dash pattern=on 1.4pt off 0.6pt] (0.45,1.55) -- (0.55,1.55);
		\draw[thick] (0.55,1.55) -- (0.55,1.275);
		\draw[thick] (0.55,1.275) -- (0.65,1.275);
		\draw[thick] (0.65,1.275) -- (0.65,1);
		\draw[dashed,thin,dash pattern=on 1.4pt off 0.6pt] (0.65,1) -- (0.75,1);
		\draw[dashed,thin,dash pattern=on 1.4pt off 0.6pt] (0.75,1) -- (0.75,0.95);
		\draw[dashed,thin,dash pattern=on 1.4pt off 0.6pt] (0.75,0.755) -- (0.75,0.725);
		\draw[dashed,thin,dash pattern=on 1.4pt off 0.6pt] (0.75,0.725) -- (0.85,0.725);
		\draw[thick] (0.85,0.725) -- (0.85,0.45);
		\draw[thick] (0.85,0.45) -- (0.95,0.45);
		\draw[thick] (0.95,0.45) -- (0.95,0.175);
		\draw[dashed,thick,red] (0.2,2.375) -- (1.0,0.175);
		\draw[<-] (0.475,1.6875) -- ++ (0.2,0.1) node[right] {\footnotesize 
		$\displaystyle \left|\frac{\text{d}\bar{\rho}}{\text{d}r}\right| = \frac{\Delta\rho}{d}$};
	\end{tikzpicture}\label{fig:denisty_diagram}}
	\hspace{1cm}
\subfloat[]{
	\begin{tikzpicture}[scale=2.8]
		\draw[->] (0,0) -- (1.0,0) coordinate (x axis) node[right]{$N^2$};
		\draw[->] (0,0) -- (0,2.4) coordinate (y axis) node[above]{$r$};
		\draw[->,thick] (0,0.45)-- (0.8,0.45);
		\draw[->,thick] (0,1.275)-- (0.8,1.275);
		\draw[->,thick] (0,2.1)-- (0.8,2.1);
		\draw[dashed,blue] (0.4,0.2) -- (0.4,2.3) node[right]
	{\footnotesize $\displaystyle \bar{N}^2=\frac{g\Delta\rho}{\rho_0 d}$};
		\end{tikzpicture}\label{fig:N_diagram}}
    \caption{Illustrations of our model, which consists of $m$ steps of size $d$, separated by $m+1$ interfaces with density jumps of $\Delta \rho$, with initial radius $r_0$. \protect\subref{fig:steps_diagram} Amplitudes of the downward ($A_n$) and upward ($B_n$) propagating waves in each layer. \protect\subref{fig:denisty_diagram} Density profile with a mean gradient $\frac{\Delta \rho}{d}$, shown by the red dashed line. \protect\subref{fig:N_diagram} Corresponding buoyancy frequency squared, consisting of $\delta$-functions with mean value $\bar{N}^2=\frac{g \Delta \rho}{\rho_0 d}$. Figures adapted from \citet{Andre2017}.}
    \label{fig:density_profile}
\end{figure*}

\subsection{Density profile}

We illustrate the regions in a giant planet where stable layers could be present in the top panel of Figure~\ref{fig:planet}. We are mainly interested in studying wave propagation in either the stable layer near the core of the planet or one near the H/He molecular to metallic transition radius where helium rain may occur. We define our (semi-convective) density staircase to have a typical radius $r_0$ (i.e.~1 in dimensionless radii) from the centre of the planet, which represents its inner radius.

We consider a staircase like that shown in the bottom panels of Figure~\ref{fig:density_profile}, consisting of $m$ steps of well mixed convective fluid layers with uniform depth $d$, in which $N=0$. These layers are separated by $m+1$ equal-sized density jumps, $\Delta \rho$. In reality, we might expect a staircase to possess a range of layer depths and density jumps, but we will primarily adopt equal sized layers with equal density jumps to simplify the analysis. Extending our model to explore a range of layer depths and density jumps  is straightforward, and is partly explored later in \S~\ref{nonuniform} (see also \cite{Sutherland2016} and \cite{Andre2017} in Cartesian geometry). We define a parameter 
\begin{equation}
\epsilon = \frac{d}{r_0},
\end{equation}
which represents the fractional depth of each convective layer relative to the typical inner radius of the staircase. We usually expect $\epsilon \ll 1$ \citep[e.g.][]{Leconte2012}, though this need not be the case if the layer is close to the centre of the planet.

We will vary the properties of the end regions that connect to the inner and outer radii of the staircase. First, we will consider an isolated staircase in which the end regions are well-mixed convective layers with $N=0$, so that gravity waves are evanescent in these layers. In the absence of a solid core, if we include $r=0$, a regularity condition must be imposed there. We will generally adopt a core of radius $r_c \ll r_0$, which we will treat as perfectly absorbing for the purposes of calculating the transmission of waves through the staircase.

The density profile is modelled as a series of $\delta$-functions at each interface between adjacent steps, such that the mean buoyancy frequency is $\bar{N}$, i.e.,
\begin{equation}\label{eq:Nsq}
N^2=\displaystyle\sum_{n=0}^m d \bar{N}^2 \delta(r_0+n d-r),
\end{equation}
where we define
\begin{equation}
\bar{N}^2 \equiv \frac{g \Delta \rho}{\rho_0 d},
\end{equation}
$\rho_0$ is the (constant) reference density, and $\Delta \rho$ is the density jump at each interface. \footnote{The factor of $d$ in Eqn.~(\ref{eq:Nsq}) arises from combining the density gradient, $\dtotal{\rho}{r}= -\Delta\rho \delta (r-r_0-nd)$, and the given definition of $\bar{N}$. This preserves the overall dimensions of the quantity as it balances the inverse length units of the delta function when its argument has units of length.} As previously discussed, we are strictly considering entropy jumps and would not necessarily expect to have equal-sized density jumps, but we consider them here for clarity.

In what follows we non-dimensionalise quantities, using a mean buoyancy frequency $\bar{N}^{-1}$ as our unit of time, and a typical radius $r_0$ as our unit of length. However, we choose to retain (but set to 1 in calculations) $\bar{N}$ and $r_0$ in some formulae and figures, even if these strictly should not appear, so that they can be more easily tracked in the derivations.

\subsection{Solutions for the radial displacement in the staircase}
When we substitute Eqn.~(\ref{eq:Nsq}) into Eqn.~(\ref{EQ1}) we obtain a discontinuous differential equation, so we may obtain the solution in each region separately as long as we apply suitable matching conditions at the interfaces. Within the $n$-th convective step $N^2=0$, so that 
\begin{equation}
\ddtotal{\xi_n}{r} + \frac{4}{r}\dtotal{\xi_n}{r} = \frac{l(l+1)-2}{r^2}\xi_n,
\end{equation}
which has solutions for the radial displacement
\begin{equation}
\xi_n = A_n r^{l-1} + B_n r^{-l-2}.
\end{equation}
We have omitted the subscript $r$ from $\xi_r$, and replaced it with a new subscript $n$ to identify the appropriate step number to which the solution applies. The radial displacement across the entire region is therefore described by
\begin{equation}\label{eq:xidispersion}
\xi =
  \begin{cases}
     A_0 r^{l-1} + B_0 r^{-l-2}     & \quad \frac{r_c}{r_0} < r < 1,\\
     A_n r^{l-1} + B_n r^{-l-2} & \quad r_{n-1} < r < r_n,\\
     A_{m+1} r^{l-1} + B_{m+1} r^{-l-2}		& \quad r > 1+m\epsilon,
  \end{cases}
\end{equation}
where $r_n=1+n \epsilon$, and $n=1,\ldots,m$.

If we were to instead consider an extended region with a spatially uniform buoyancy frequency $N=\bar{N}$, then Eqn.~(\ref{EQ1}) would have the solution
\begin{equation}\label{eq:xir2}
\xi =A r^{\lambda_{+}} + B r^{\lambda_{-}},
\end{equation}
where $A$ and $B$ denote the amplitude of the downward/upward propagating wave, and
\begin{equation}
\label{lambdaplusminus}
\lambda_{\pm} = -\frac{3}{2} \pm \frac{1}{2}\sqrt{1+4\bigg(1-\frac{\bar{N}^2}{\omega^2}\bigg)l(l+1)}.
\end{equation}
We will later use this solution when we consider the transmission of waves through a staircase sandwiched by two stably-stratified layers, and also when we compare the frequencies of the free modes of a staircase with those of a uniformly stably-stratified layer.

\subsection{Interface conditions and transfer matrices}
\label{sec:BCs}

Since Eqn.~(\ref{EQ1}) is a second order differential equation in $r$, we must apply two boundary conditions at each interface. Here, we generalise those in \citet{Andre2017} to spherical geometry. Firstly, we must ensure that there is no separation of the fluid on either side of each interface, therefore $\xi$ must be continuous there. This requires
\begin{equation}
\label{xicont}
\xi_{n+1}(1+n\epsilon) = \xi_{n}(1+n\epsilon),
\end{equation}
and using Eqn.~(\ref{eq:xidispersion}) we find
\begin{equation}
\label{interface1}
A_{n+1}-A_{n} + \big(B_{n+1}-B_{n}\big) \big(1 + n\epsilon \big)^{-2l-1} =0.
\end{equation}
Our second condition follows from the requirement that the momentum flux, and therefore the pressure perturbation, is continuous across each interface. We may obtain this condition by integrating Eqn.~(\ref{EQ1}) over a small volume of radial extent $2\Delta$ around an interface. We then take the limit of vanishing volume, such that $\Delta$ tends to $0$. For the $n$-th interface, we obtain
\begin{multline}
\int_{1+n\epsilon-\Delta}^{1+n\epsilon+\Delta} r^2 \ddtotal{\xi}{r} \mathrm{d} r + \int_{1+n\epsilon-\Delta}^{1+n\epsilon+\Delta} 4 r \dtotal{\xi}{r} \mathrm{d} r = \\ \int_{1+n\epsilon-\Delta}^{1+n\epsilon+\Delta}(l(l+1)-2) \xi \mathrm{d}r - \int_{1+n\epsilon-\Delta}^{1+n\epsilon+\Delta}\frac{N^2}{\omega^2}l(l+1) \xi \mathrm{d}r.
\end{multline}
We use integration by parts on the left hand side (LHS) and apply the continuity of $\xi$ (Eqn.~(\ref{xicont})), so that the limit $\Delta \to 0$ leads to
\begin{equation}
\mathrm{LHS} = \big(1 + n\epsilon\big)^2 \Bigg[\dtotal{\xi_{n+1}}{r} - \dtotal{\xi_{n}}{r}\Bigg]_{r=1+n\epsilon}.
\end{equation}
On the right hand side (RHS), we also apply the continuity of $\xi$, so that on taking $\Delta \to 0$, the first term drops out and substitute Eqn.~(\ref{eq:Nsq}) to give, 
\begin{equation}
\mathrm{RHS}=-\frac{\bar{N}^2 \epsilon}{\omega^2}l(l+1)\int_{1+n\epsilon-\Delta}^{1+n\epsilon+\Delta} \delta(1+n\epsilon-r) \xi \mathrm{d}r.
\end{equation}
After integration we obtain our second interface condition:
\begin{equation}\label{eq:bc2}
\left[\dtotal{\xi_{n+1}}{r} - \dtotal{\xi_{n}}{r}\right]_{r=1+n\epsilon} = -\frac{\bar{N}^2 l(l+1)\epsilon}{\omega^2 (1+n\epsilon)^2}\xi_{n}\bigg\rvert_{r= 1+n\epsilon}.
\end{equation}
Using Eqn.~(\ref{eq:xidispersion}) we then find
\begin{multline}
\label{interface2}
(l-1)(A_{n+1}-A_{n}) -(l+2)\Big(B_{n+1}-B_{n}(1 + n\epsilon)^{-2l-1}\Big) \\ =\frac{-\bar{N}^2 l(l+1)\epsilon}{\omega^2 (1 + n\epsilon)^2}\Big[A_{n}(1 + n\epsilon) + B_{n}(1 + n\epsilon)^{-2l}\Big].
\end{multline}

The two interface conditions (Eqns.~\ref{interface1} and \ref{interface2}) allow the solution in each step to be written in terms of the solution in an adjacent step. Therefore, with some algebra, the coefficients in adjacent layers are related by
\begin{equation}
\begin{bmatrix}
A_{n+1} \\ B_{n+1}
\end{bmatrix}
= T_n 
\begin{bmatrix}
A_{n} \\ B_{n}
\end{bmatrix},
\end{equation}
where the transfer matrix $T_n$ is defined as,
\begin{equation}
T_n = \begin{bmatrix}
1 - \frac{\epsilon l (l+1) \bar{N}^2}{(2l+1)(1+n\epsilon)\omega^2} & 
\frac{- \epsilon l (l+1) \bar{N}^2}{(2l+1)\left(1+n\epsilon\right)^{2(l+1)}\omega^2} \\ 
\frac{\epsilon l (l+1)(1+n\epsilon)^{2l} \bar{N}^2}{(2l+1)\omega^2} & 
1 + \frac{\epsilon l (l+1) \bar{N}^2}{\left(2l+1)(1+n\epsilon\right)\omega^2}
\end{bmatrix}.
\end{equation}
This transfer matrix correctly recovers the Cartesian geometry results in \cite{Belyaev2015}, \cite{Sutherland2016} and \cite{Andre2017}, once we take the double limits $l \gg 1$ and $1 \gg n \epsilon$, and we identify 
\begin{equation}
    k_{\perp}^2 = \frac{l(l+1)}{r_0^2}.
\end{equation}
Note also that $T_n$ reduces to the identity matrix in the limit $\epsilon\rightarrow 0$.

This formalism allows us to determine the solution in the $(m+1)$-th layer in terms of the solution in the $0$-th layer by repeatedly applying the transfer matrix. Note that $T_n$ depends on the radius of the $n$-th interface, which complicates the following analysis compared with the Cartesian case (even with constant $d$ and $\Delta\rho$) in \cite{Andre2017}. But we may still define a $2 \times 2$ matrix such that,
\begin{equation}
\begin{bmatrix}
A_{m+1} \\ B_{m+1}
\end{bmatrix}
= X
\begin{bmatrix}
A_{0} \\ B_{0}
\end{bmatrix},
\end{equation}
where
\begin{equation}
X=T_m T_{m-1} \dots T_1 T_0
\end{equation}
relates the solution in the end regions. With appropriate choices of the end regions, this formalism allows us to analyse the free modes of a density staircase (\S~\ref{sec:dispersion_realation}), as well as the transmission of waves through a staircase (\S~\ref{sec:transmission}).

\section{Free modes of a density staircase} 
\label{sec:dispersion_realation}

We begin by deriving a dispersion relation that describes the free internal modes of a density staircase. We consider the case of a finite staircase confined between two well-mixed convective regions followed by a finite staircase with solid walls at either end. Finally, we consider the case of a finite staircase with a solid wall at the inner boundary and a well-mixed convective region at the outer boundary, which could represent a solid core and convective envelope. In each case we analyse the properties of the free modes and how they depend on the parameters describing the staircase. 

\begin{figure}
    \subfloat[$d=0.01$, $r_0=1$, $\bar{N}=1$]{
    \begin{tikzpicture}
		\node at (0,0) {\includegraphics[width=0.45\textwidth,trim={0 0 6.8cm 0},clip]{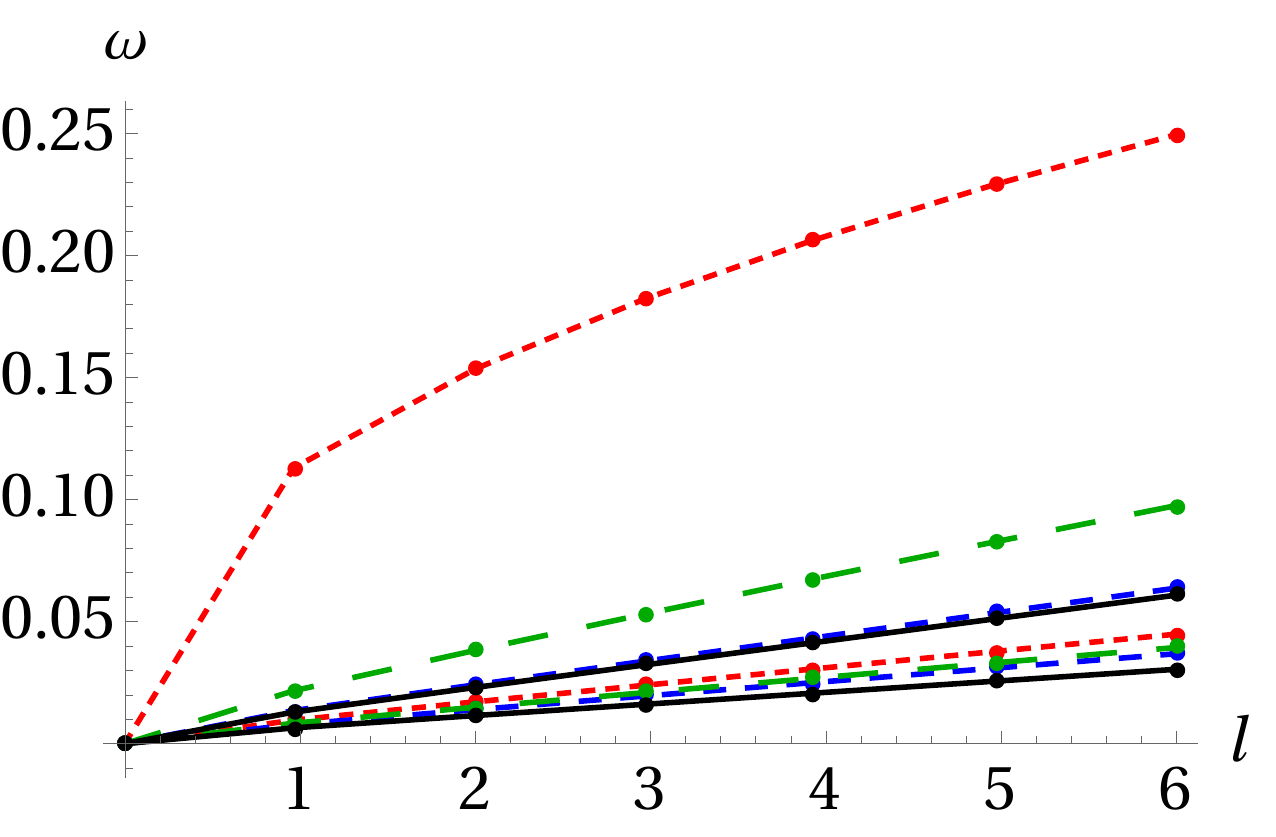}};
		\node at (-2.8,2.35) {\small \textsf{$/ \bar{N}$}};
	\end{tikzpicture}}\\
    \subfloat[$r_0=1$, $l=2$, $\bar{N}=1$]{
        \begin{tikzpicture}
            \node at (0,0)
            {\includegraphics[width=0.45\textwidth,trim={0 0 6.8cm 0},clip]{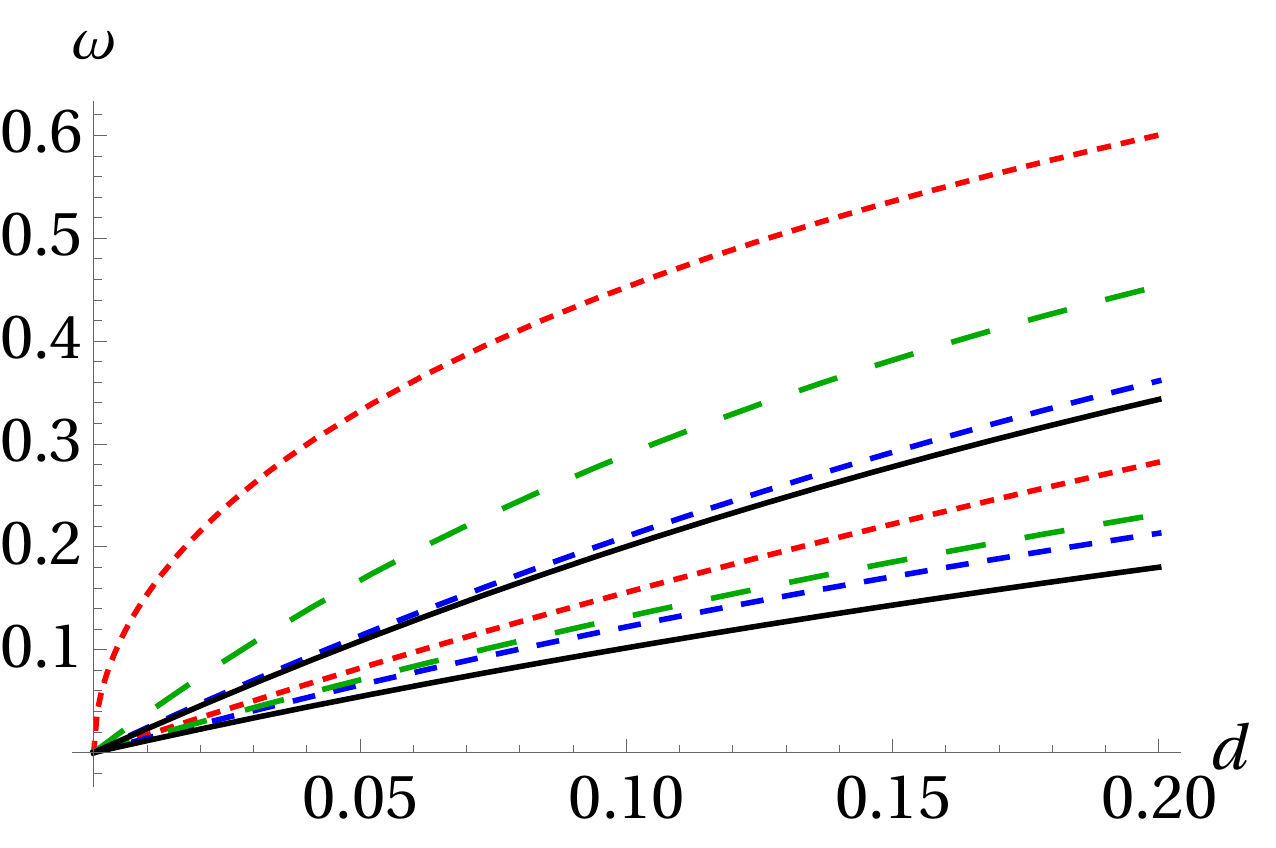}};
            \node at (-2.97,2.37) {\small \textsf{$/ \bar{N}$}};
        \end{tikzpicture}}
    \caption{Dependence of the mode frequency $\omega$ for each mode on the parameters of the staircase and the boundary conditions on the end regions, shown for the single step, $m=1$, case. The red (thin-dashed) lines show $\omega$ for each mode of a finite staircase embedded in a convective medium, blue (dashed) lines show a finite staircase with solid wall boundary conditions. The green (thick-dashed) lines show a finite staircase with a solid wall at the centre and a convective medium above. Finally, the black (solid) line shows the solution with a constant stratification between two solid walls. Top: dependence of $\omega$ for each mode on the angular wave number $l$. Bottom: dependence on step size $d$.}
    \label{fig:dispersion_parameters}
\end{figure}

\begin{figure*}
    \subfloat[$m=1, l=2$, $r_0=1$, $\bar{N}=1$, $d=0.01$]{
        \includegraphics[width=0.48\textwidth]{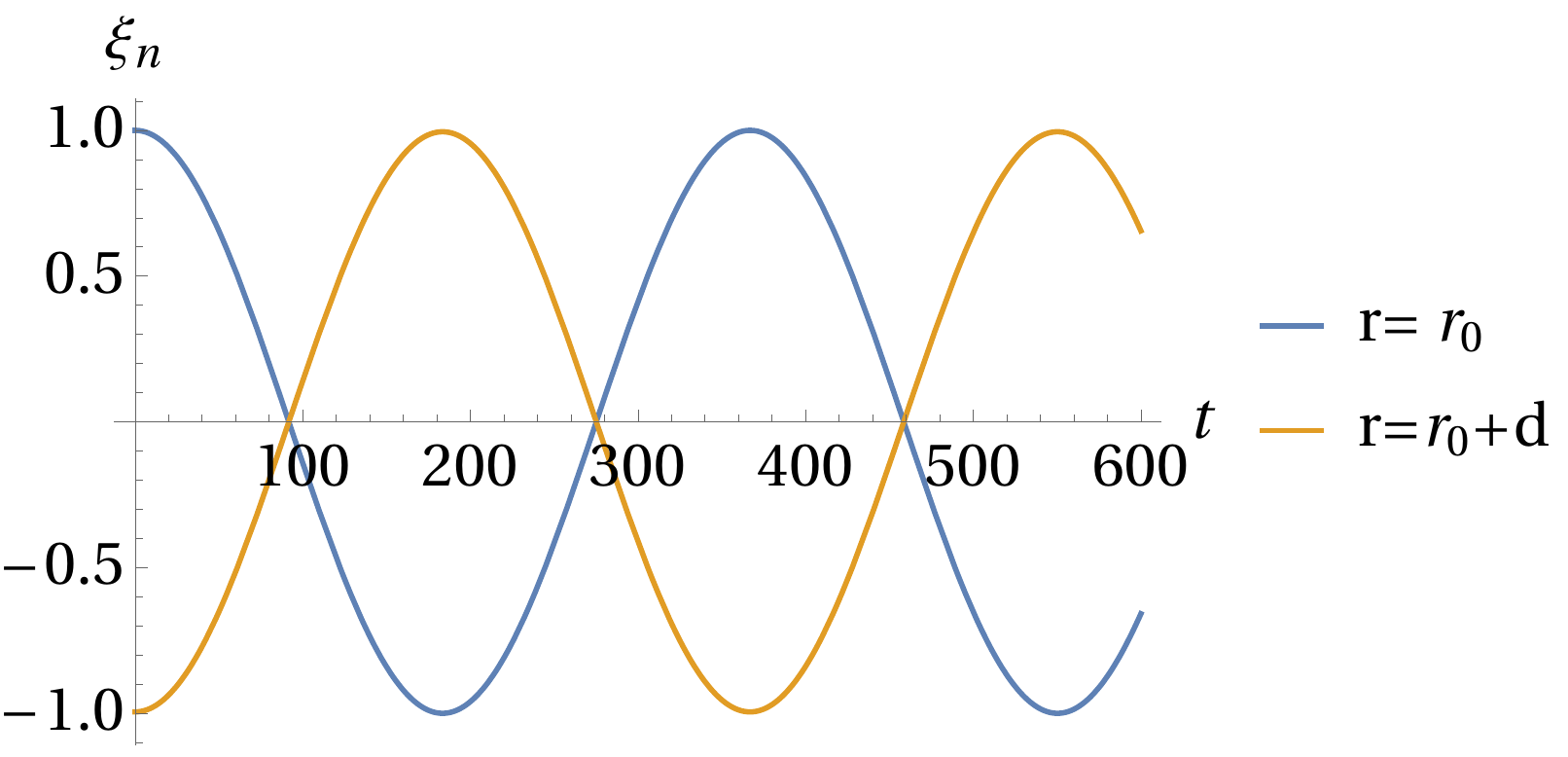}\label{m1outphase}}
	\subfloat[$m=6, l=2$, $r_0=1$, $\bar{N}=1$, $d=0.01$]{
        \includegraphics[width=0.48\textwidth]{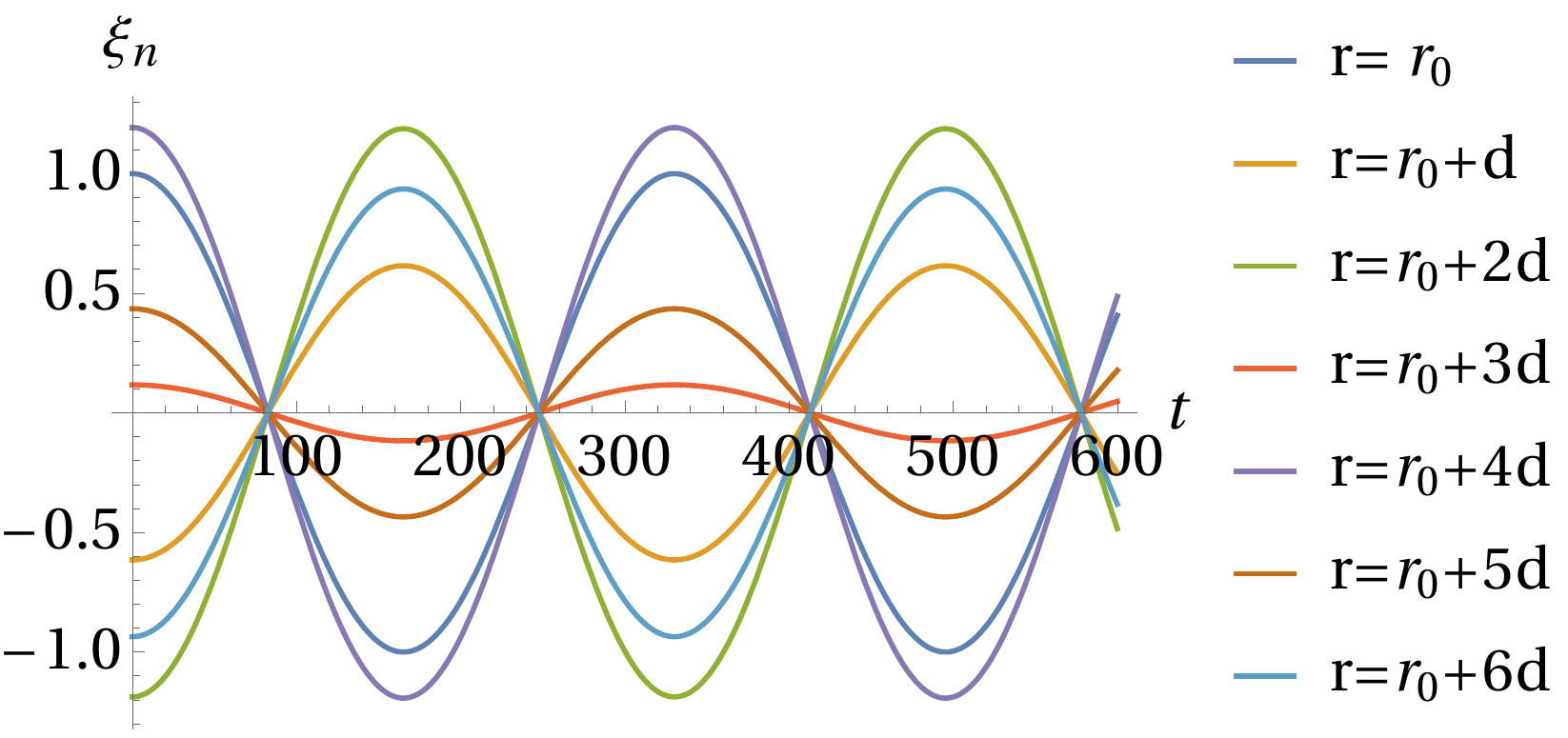}\label{m6outphase}} \\
    \subfloat[$m=1, l=2$, $r_0=1$, $\bar{N}=1$, $d=0.01$]{
        \includegraphics[width=0.48\textwidth]{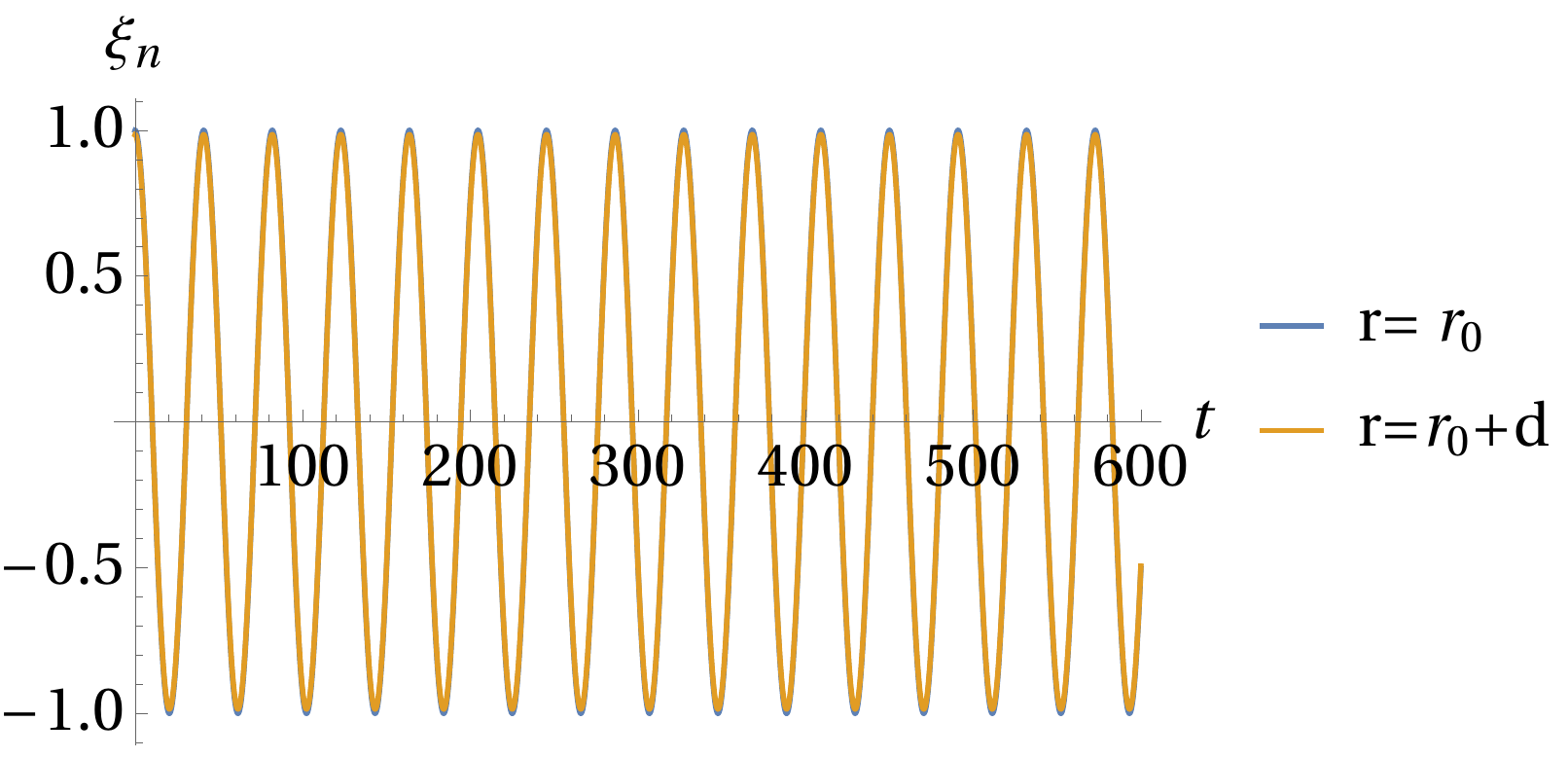}\label{m1inphase}}
	\subfloat[$m=6, l=2$, $r_0=1$, $\bar{N}=1$, $d=0.01$]{
        \includegraphics[width=0.48\textwidth]{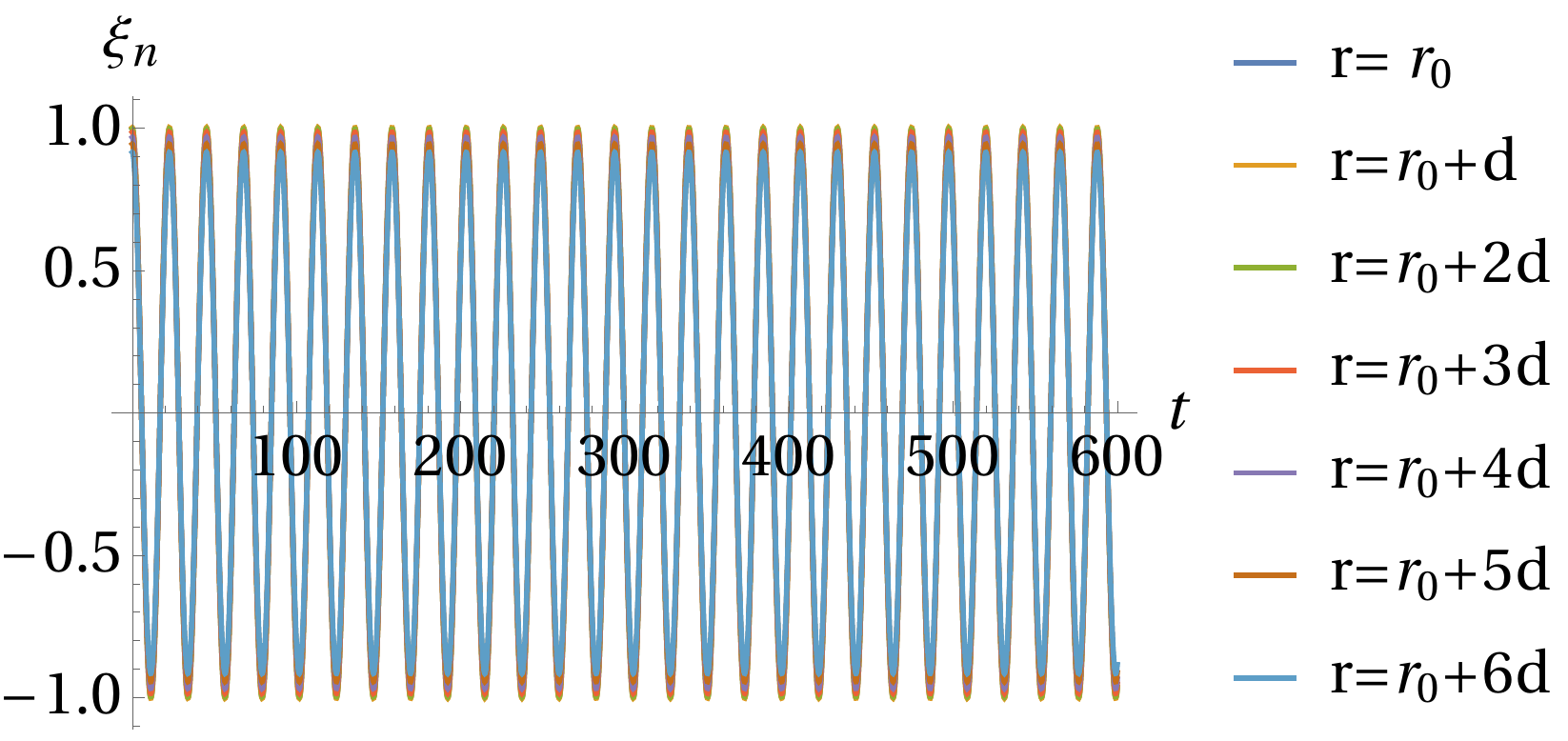}\label{m6inphase}}
	\caption{Radial displacement at each interface for the case of a staircase embedded within a convective medium. \protect\subref{m1outphase} and \protect\subref{m1inphase} show the one step ($m=1$) case where \protect\subref{m1outphase} shows the internal wave-like solution with the two interfaces oscillating out of phase, and \protect\subref{m1inphase} shows the interfacial wave solution with both interfaces in phase. \protect\subref{m6outphase} and \protect\subref{m6inphase} similarly show the interfaces for the six step  ($m=6$) case.}
    \label{fig:ifw_igw}
\end{figure*}

\subsection{Finite staircase embedded in a convective medium}\label{sec:free_modes_decay}

Our first example considers a finite staircase embedded in a convective medium, which could represent a staircase in the helium rain region, for example. We enforce boundary conditions such that the solution decays away from the first and last interface, corresponding with setting $B_0=0$ and $A_{m+1}=0$. The top left entry of $X$ is then required to be $0$, i.e.,
\begin{equation}\label{eq:T11}
X_{1,1}=0.
\end{equation}
This represents a polynomial in $\omega^2$, which is the dispersion relation describing the free modes of the staircase. The polynomial has degree $(m+1)$, implying that there are an equal number of (oppositely-signed pairs of) free modes in the system as there are interfaces in the staircase (see also \citealt{Belyaev2015,Andre2017}). 

\subsubsection{Single step ($m=1$)} \label{sec:one_step}

Solving Eqn.~(\ref{eq:T11}) for a staircase consisting of a single convective step and two interfaces ($m=1$), gives the dispersion relation 
\begin{multline}
\label{decayingsinglemodedisprel}
\omega^2 = \frac{\epsilon l(l+1)\bar{N}^2(1+2\epsilon)^{-1 - l}}{2(2l+1)(1+\epsilon)} \\ \bigg((1+2\epsilon)^l(2+3\epsilon) \pm \sqrt{4(1+\epsilon)^{2l+2} + \epsilon^2(1+2\epsilon)^{2l}}\bigg).
\end{multline}
This describes the frequencies of two (pairs of oppositely-signed) free modes. We can further analyse the two solutions by expanding in the small parameter $\epsilon$. The first solution is
\begin{equation} \label{eq:IGWExpan}
\omega^2 = \frac{1}{2}l(l+1)\bar{N}^2 \epsilon^2 + O(\epsilon^3),
\end{equation}
and therefore $\omega^2 \propto l^2$ for large $l$. This is similar to the behaviour of an internal gravity wave. The second solution is
\begin{equation} \label{eq:IfGWExpan}
\omega^2 = \frac{2l(l+1)\bar{N}^2 \epsilon}{2l+1}+O(\epsilon^2),
\end{equation}
so that $\omega^2 \propto l $ for large $l$. This can be compared with the properties of an interfacial gravity wave.
Figure \ref{fig:dispersion_parameters} shows the dependence of the mode frequencies on $l$ and $d$ (red line).
To justify our assertions, we consider that internal gravity waves in spherical geometry are described by the following dispersion relation \citep{JCDLectures}:
\begin{equation}
\omega^2 \approx \bar{N}^2 \frac{k_\perp^2 d^2}{k_r^2 d^2+k_\perp^2 d^2} 
\approx \frac{l(l+1)}{k_r^2 d^2} \bar{N}^2 \epsilon^2,
\label{eq:IGW}
\end{equation} 
in the ``plane-wave limit" in which $k_r\gg k_\perp$, and we identify $k_\perp^2=l(l+1)/r_0^2$. For large $l$, we find $\omega^2 \propto l^2$, just like in Eqn.~(\ref{eq:IGWExpan}). By comparing Eqn.~(\ref{eq:IGW}) to Eqn.~(\ref{eq:IGWExpan}), we observe that these are equivalent if $k_r^2 \approx \frac{2}{d^2}$. Indeed, we have confirmed numerically that the free modes in the single step case are well described by Eqn.~(\ref{eq:IGW}) if $k_r\approx 145$, which is just slightly higher than $\sqrt{2}/d\approx 141$. The corresponding wavelength $\lambda_r=\frac{2 \pi}{k_r}>d$, as we would expect for a mode with the character of an internal gravity wave. 

On the other hand, the dispersion relation describing an interfacial gravity mode, which is the solution we obtain in the case of a single interface ($m=0$) is
\begin{equation}
\omega^2 = \frac{l(l+1) \bar{N}^2 \epsilon}{(2l+1)}=\frac{l(l+1)}{(2l+1)} \frac{g \Delta \rho}{r_0 \rho_0},
\end{equation}
For large $l$, we find $\omega^2 \propto l$, which behaves similarly to Eqn.~(\ref{eq:IfGWExpan}). This appears to differ from Eqn.~(\ref{eq:IfGWExpan}) by a factor of $2$, but this only arises because it is the total density jump (across both steps) that is relevant, and this is twice as large in Eqn.~(\ref{eq:IfGWExpan}).

We show the radial displacement as a function of time at both interfaces in Figures~\ref{m1outphase}~and~\ref{m1inphase} for both types of solution. Note that the overall amplitude is arbitrary but the relative amplitudes are meaningful. Figure~\ref{m1outphase} shows the solution corresponding with Eqn.~(\ref{eq:IGWExpan}), in which both interfaces oscillate out of phase with each other, as we would expect if they are located either side of a node in a corresponding internal gravity mode. Figure~\ref{m1inphase} shows the solution corresponding with Eqn.~(\ref{eq:IfGWExpan}). This solution clearly has interfacial wave character because both interfaces oscillate in phase with one other, behaving as an ``extended interface". 

\subsubsection{Multiple steps ($m>1$)} \label{multiplesteps}

We can also explore the free modes of an $m$-step staircase in a similar way when $m>1$, except that we now obtain a polynomial of degree $(m+1)$. The solutions are too complicated to gain any insight from writing them down, but we can use a computer algebra package (e.g.~Mathematica) to analyse their properties. The solutions for multiple steps exhibit similar behaviour to the case of a single step. We again find that the highest frequency mode is an interfacial gravity-like mode, in which all of the interfaces oscillate in phase, so that the whole staircase behaves like a single extended interface. The other modes behave more like internal gravity modes, in which the interfaces do not all oscillate in phase, and the number of interfaces that are in phase can be related to the number of nodes in the corresponding gravity mode.

For the case with $m=6$ steps, we show the radial displacement at each interface (again, with an arbitrary overall amplitude) in Figure~\ref{m6inphase} for the one interfacial mode in which all interfaces oscillate in phase, and one example (chosen from 6) of an internal gravity-like mode in Figure~\ref{m6outphase}. In the latter, the interfaces do not all oscillate in phase, indicating that this is like an internal gravity mode (with a continuous uniform stratification) with 3 nodes. For multiple steps, the dependence on $l$, $d$ and $r_0$ is qualitatively similar to that of a single step. Series expansions to explore the dependence of the frequencies of the waves on the parameters were not carried out in this case because the behaviour can be obtained qualitatively.

\begin{figure}
	\subfloat[$x=(m+1)\epsilon=0.1$]{
        \includegraphics[width=0.45\textwidth]{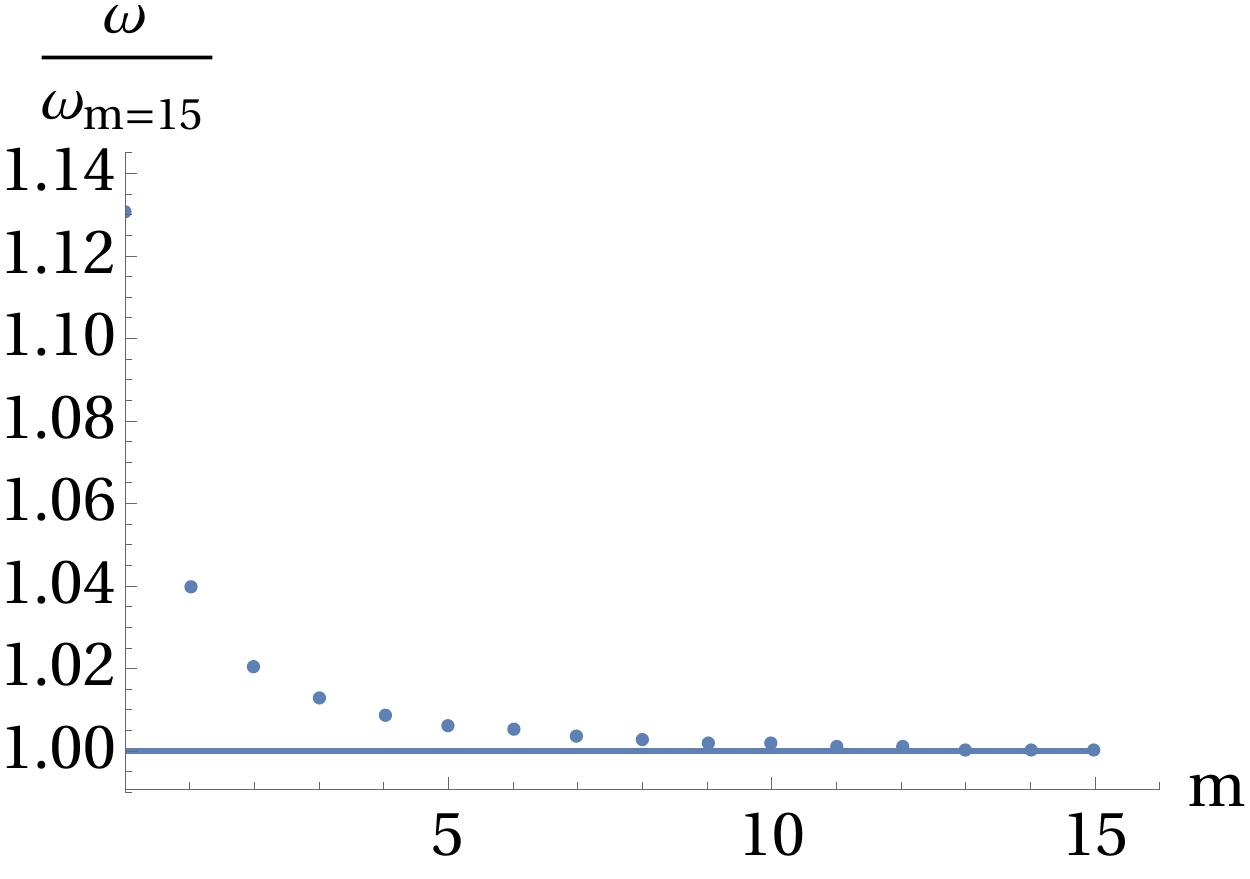}}\\
    \subfloat[$\epsilon=0.01$]{
        \includegraphics[width=0.45\textwidth]{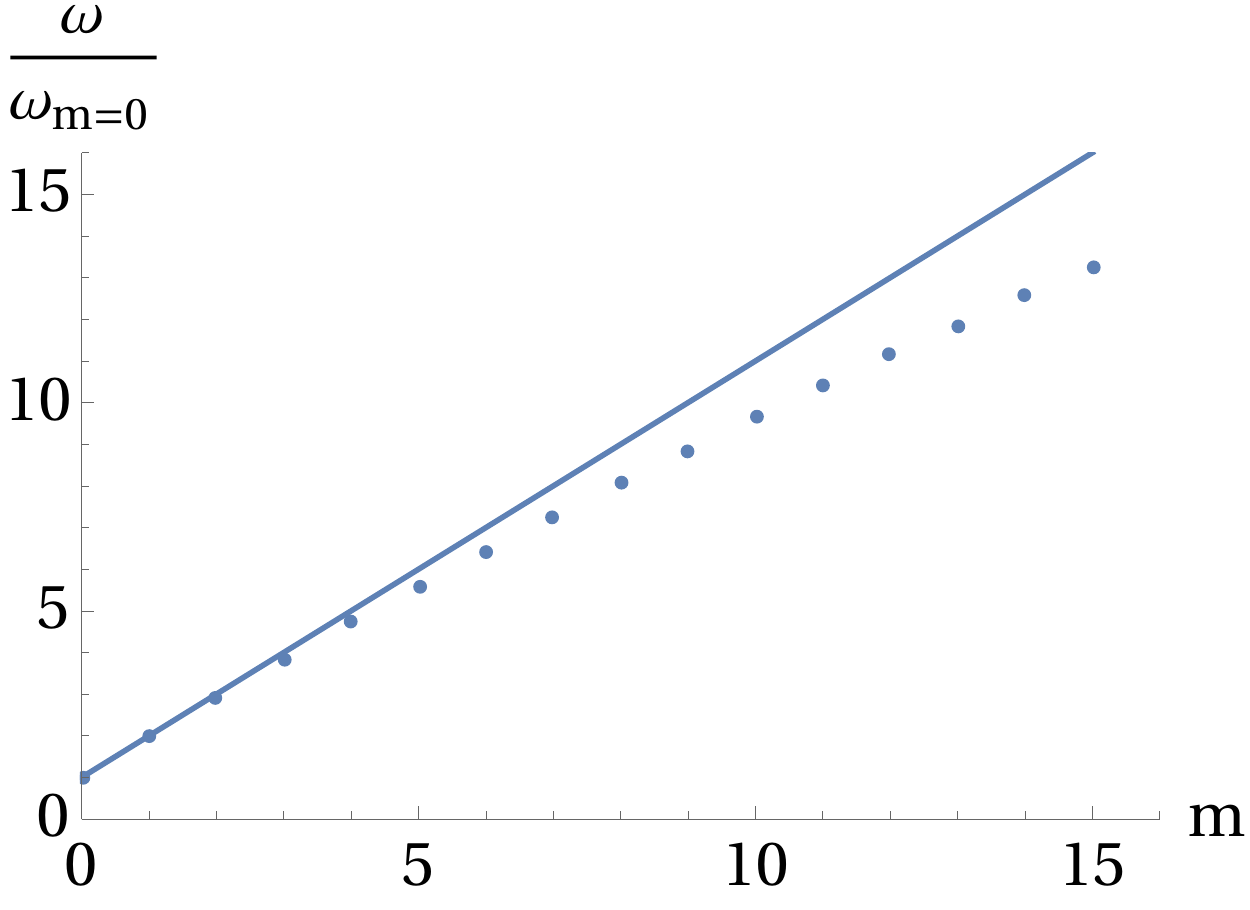}}
    \caption{Dependence of the frequency of the interfacial mode for each step number $m$, for $l=2$, $\bar{N}=1$. Top: interfacial-like mode for a fixed staircase size $x=(m+1)\epsilon=0.1$, such that the step size $\epsilon$ and density jump $\Delta \rho$ decrease as the step number increases. Solid line showing $y=1$. Bottom: interfacial-like mode for a fixed step size $\epsilon=0.01$, such that the total length of the staircase and total density jump increase as the step number increases. Solid line showing $y=m+1$.}
    \label{fig:dispersion_m}
\end{figure}

There are two ways to explore how the dispersion relation depends on the number of steps. If we fix the mean stratification, the total density jump and total length of the staircase, $x$, but we increase the number of steps, then $\epsilon$ and $\Delta \rho$ will decrease as steps are added such that, $\Delta \rho= \frac{1}{(m+1)}\Delta \rho_{\text{total}}$ and $\epsilon = \frac{1}{(m+1)} x$. The top panel of Figure \ref{fig:dispersion_m} shows the interfacial wave solutions dependence on $m$ for the case of a staircase with fixed size and total density jump. All solutions have been normalised by the $m=15$ solution and tend to 1 as $m$ is increased. This suggests that the total density jump $\Delta \rho_{\text{total}}$ is an important quantity for the dispersion relation.

If the step size and mean stratification are maintained, this will lead to a longer staircase and increased total density jump; the frequency therefore increases. The bottom panel of Figure~\ref{fig:dispersion_m} shows the solution for $\omega$ for different numbers of steps, which corresponds to the interfacial wave solution, normalised by the $m=0$ solution. We would expect to see a roughly linear dependence on $m$. We can see the trend falls away from the $y=m+1$ line for large $m$. As the total staircase gets larger we would expect the approximation to one thin interface to be less accurate and therefore expect the solution to depart from this prediction.

\subsection{Finite staircase with solid walls}\label{sec:free_modes_wall}

We now consider the case of a finite staircase confined between solid walls at both ends, which might be relevant for the case of a stably-stratified terrestrial planetary core, for example. In particular, we consider solid walls at $r_0$ and $r_0+(m+2)d$, on which we enforce $\xi_r=0$. The first interface is at $r_0+d$, and the buoyancy frequency is defined as,
\begin{equation}
N^2=\displaystyle\sum_{n=1}^{m+1} \bar{N}^2 \epsilon \delta(1+n\epsilon-r).
\end{equation}

The interface conditions remain unchanged and, as before, we construct a transfer matrix to relate our coefficients in the first and last layer,
\begin{equation}
\begin{bmatrix}
A_{m+1} \\ B_{m+1}
\end{bmatrix}
= X'
\begin{bmatrix}
A_{0} \\ B_{0}
\end{bmatrix},
\end{equation}
where
\begin{equation}
X'=T_{m+1} T_{m} \dots T_1.
\end{equation}
Instead of considering decaying solutions we now consider solid wall boundary conditions such that the radial displacement at either end of the staircase is zero, i.e.
\begin{equation}
\xi_0(r=1)=\xi_{m+1}(r=1+(m+2)\epsilon)=0.
\end{equation}
These combine to give four simultaneous equations,
\begin{eqnarray}
&& \hspace{-2cm} A_0+B_0=0, \\
&& \hspace{-2cm} A_{m+1}(1+(m+2)\epsilon)^{l-1}+B_{m+1}(1+(m+2)\epsilon)^{-l-2} =0, \\
&& \hspace{-2cm} A_{m+1}=A_0 X'_{1,1}+B_0 X'_{1,2}, \\
&& \hspace{-2cm} B_{m+1}=A_0 X'_{2,1}+B_0 X'_{2,2}.
\end{eqnarray}
We seek non-trivial solutions, which requires
\begin{multline}
     X'_{1,2}+(1+(m+2)\epsilon)^{-2l-1}X'_{2,2}=\\(X'_{1,1}+(1+(m+2)\epsilon)^{-2l-1}X'_{2,1}).
\end{multline}
This allows us to determine the dispersion relation describing the free modes of the staircase. We again obtain a polynomial of degree $(m+1)$, and so we obtain $(m+1)$ (pairs of) free modes.

The solution can be found for the single step case, and we also expand each solution assuming $\epsilon\ll 1$ to obtain,
\begin{equation}
\omega^2=l(l+1)\bar{N}^2\epsilon^2 + O(\epsilon^3),
\end{equation}
and 
\begin{equation}
\omega^2=\frac{1}{3}l(l+1)\bar{N}^2\epsilon^2 + O(\epsilon^3).
\end{equation}
Just like in \S~\ref{sec:one_step}, we observe that there are modes for which $\omega^2 \propto l^2$ for large $l$, which is the expected behaviour for an internal gravity wave. However, the highest frequency mode no longer corresponds with an interfacial wave, and in fact none of the waves have the dependence $\omega^2 \propto l$ for large $l$ expected of such waves. This is due to the boundary conditions that we have adopted. The highest frequency mode still has all of its interfaces oscillating in phase, but it no longer behaves as an interfacial wave. Instead, it behaves more like a gravity mode with no internal nodes. We show the roots of the dispersion relation in Figure~\ref{fig:dispersion_parameters} (blue dashed line).

\subsection{Finite staircase with mixed boundary conditions}\label{sec:free_modes_mixed}
Finally, we consider the case where the staircase has a solid wall at the lower boundary and lies below a convective region. This case might be a better representation of a stratified layer at the edge of a solid inner core, which connects onto a convective envelope at its outer radius.

The method used is a combination of the previous two methods, with a solid wall at $r_0$ and modes that decay above the staircase. The buoyancy profile $N^2$ and matrix $X'$ are unchanged from \S~\ref{sec:free_modes_wall}. 

Considering zero radial displacement at the bottom of the staircase to give,
\begin{equation}
\xi_0(r_0=1)=0.
\end{equation}
And forcing purely decaying solutions at the top of the staircase requires, 
\begin{equation}
A_{m+1}=0. 
\end{equation}
These combine to give three simultaneous equations,
\begin{eqnarray}
&& A_0+B_0=0, \\
&& A_0 X'_{1,1}+B_0 X'_{1,2}=0, \\
&& B_{m+1}=A_0 X'_{2,1}+B_0 X'_{2,2}.
\end{eqnarray}
Non-trivial solutions require
\begin{equation}
     X'_{1,2}=X'_{1,1}.
\end{equation}
Similarly to the previous cases this allows us to determine the dispersion relation describing the free modes of the staircase. We again obtain a polynomial of degree $(m+1)$, and so we obtain $(m+1)$ (pairs of) free modes.

Expanding each solution in the single step case, assuming $\epsilon \ll 1$, we obtain the two solutions,
\begin{equation}
    \omega^2=\frac{1}{2}(3+\sqrt{5})l(l+1)\bar{N}^2 \epsilon^2+ O(\epsilon^3),
\end{equation}
and
\begin{equation}
    \omega^2=\frac{1}{2}(3-\sqrt{5})l(l+1)\bar{N}^2 \epsilon^2+ O(\epsilon^3).
\end{equation}
As in \S~\ref{sec:free_modes_wall} we observe only modes where $\omega^2 \propto l^2$ for large $l$, corresponding to internal gravity wave behaviour. The highest frequency modes with all interfaces oscillating in phase also act as an internal mode with no nodes instead of an interfacial mode. The roots of these two solutions are shown in Figure \ref{fig:dispersion_parameters} (green thick-dashed line), which shows that they lie between the two previous cases.

\subsection{Comparison with a continuously-stratified medium}
\label{sec:mode_diff}

\begin{figure}
    \subfloat{	
    \begin{tikzpicture}
		\node at (0,0) {\includegraphics[width=0.45\textwidth]{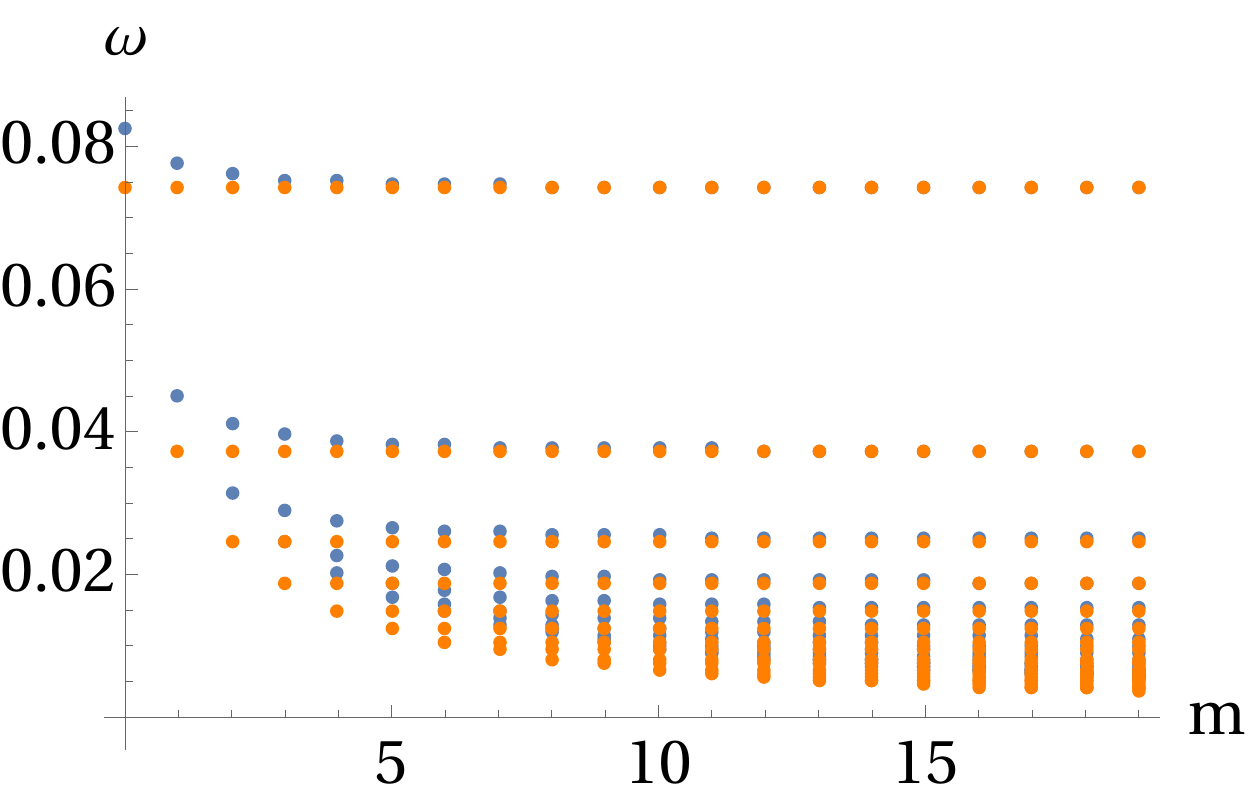}};
		\node at (-2.8,2.35) {\small \textsf{$/ \bar{N}$}};
	\end{tikzpicture}} \\
	\subfloat{
        \includegraphics[width=0.45\textwidth]{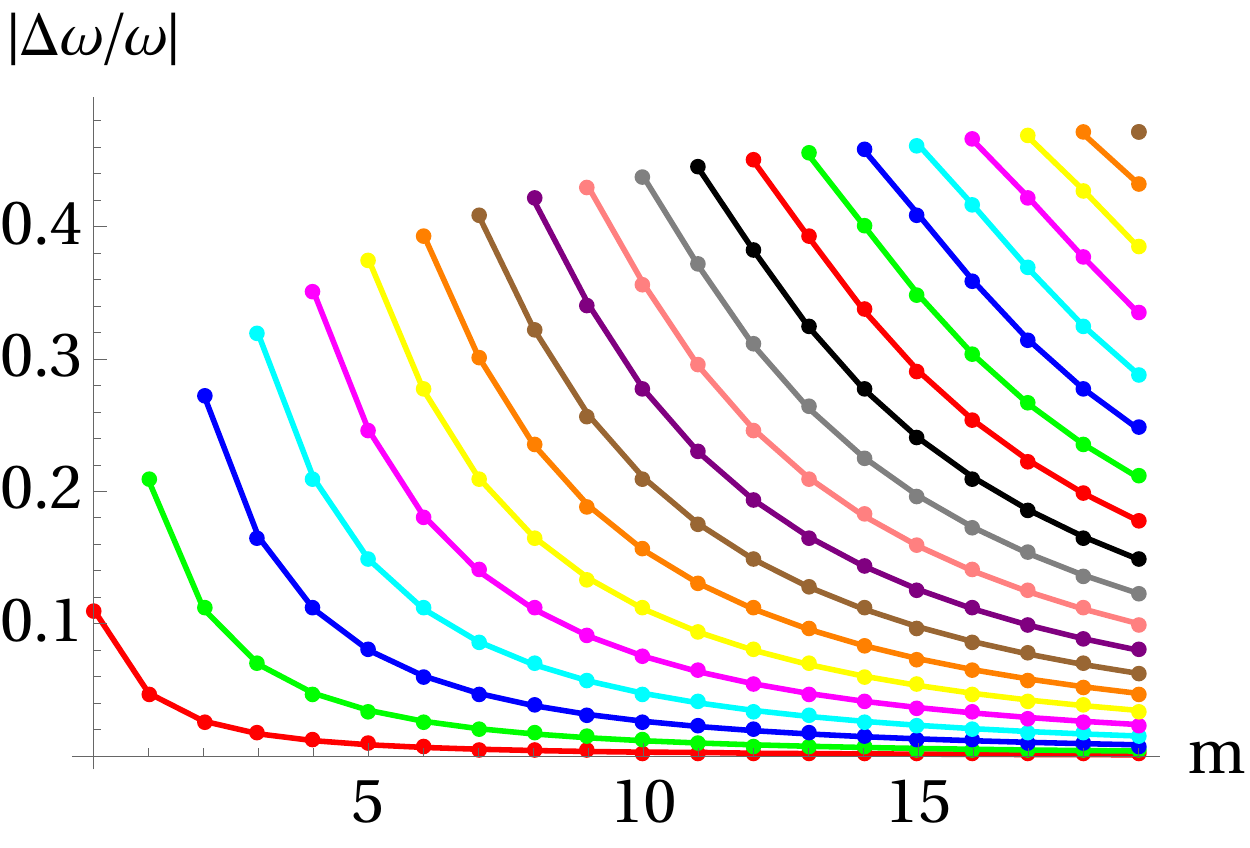}}\\
	\subfloat{
	    \includegraphics[width=0.49\textwidth]{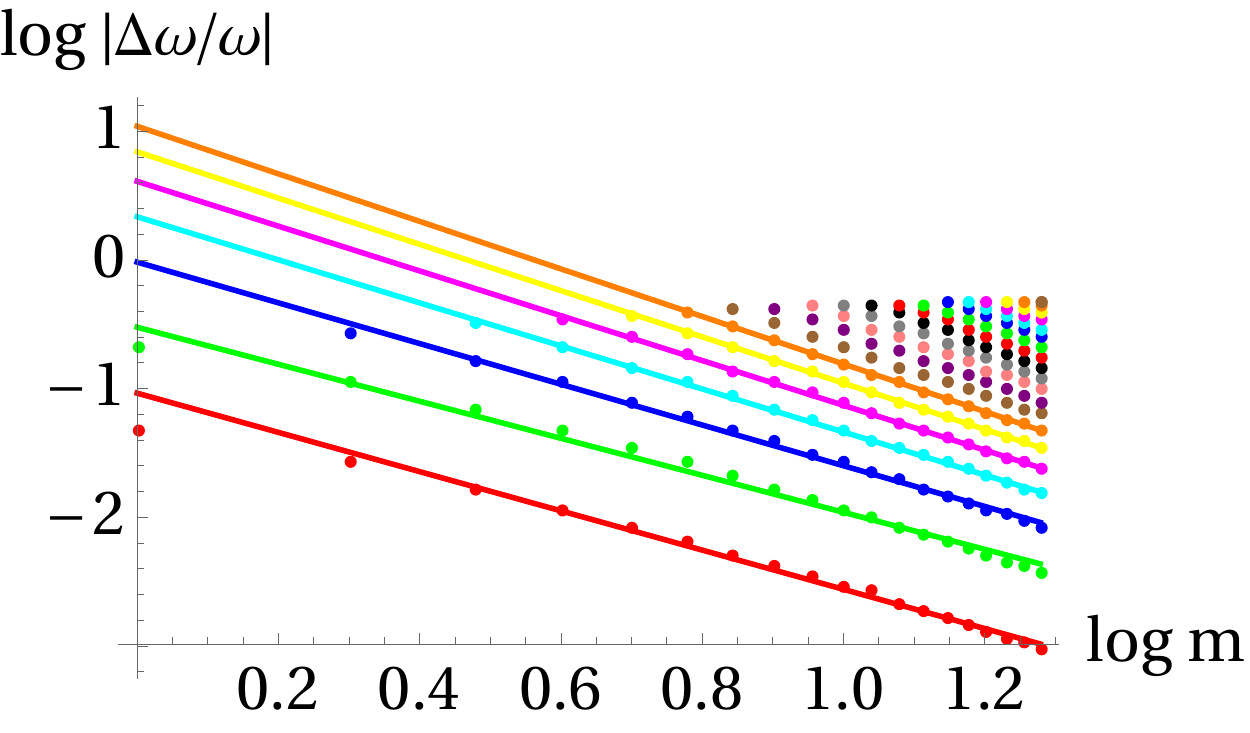}}
    \caption{Comparison of the frequencies of the free modes of a uniformly-stratified layer with a density staircase with the same $\bar{N}$. Top: frequencies vs $m$ for a uniformly-stratified layer (orange) and a staircase  (blue). Middle: fractional frequency difference (Eqn.~\ref{eq:diff}) between the modes of a uniformly-stratified layer and a staircase vs $m$. Bottom: same, but on a log-log (base 10) plot. This shows that the frequencies of the modes of a staircase approach those of a uniformly-stratified layer (behaving like $m^{-2}$) as the number of steps is increased, but that there is consistently positive frequency shift.}
    \label{fig:staricase_const_diff}
\end{figure}

Here we compare the frequencies of the free modes of a staircase with those of a continuously-stratified medium with the same mean (constant) buoyancy frequency. We choose to compare the case with solid wall boundary conditions at either end (i.e. $\xi_r(1)=\xi_r(1+(m+2)\epsilon)=0$), which we have already computed for a staircase in \S~\ref{sec:free_modes_wall}. 
We apply these boundary conditions to the solution given by Eqns.~(\ref{eq:xir2}) and~(\ref{lambdaplusminus}) to obtain an infinite set of modes in the continuous case. We index these by a positive integer $n$ which refers to the number of radial nodes in the solution. The resulting frequencies are
\begin{equation}\label{eq:eigs}
\omega = \pm \sqrt{\frac{4 l (l+1) \bar{N}^2 (\log (1+(m+2)\epsilon))^2}{(2 l+1)^2 (\log (1+(m+2)\epsilon))^2+4 \pi ^2 n^2}}. 
\end{equation}
For these calculations we fix the total size of the region and the total density jump across the staircase, and vary the number of steps $m$.

To compare the infinite set of free modes found in the stratified case to the free modes of the staircase, we take the first $m+1$ modes of the uniformly-stratified layer and compare these to the free modes of the staircase. The top panel of Figure~\ref{fig:staricase_const_diff} shows the wave frequencies for all modes as a function of the number of steps $m$. It is clear that as $m$ increases, the difference between the uniformly-stratified case and the staircase decreases.

To more clearly and quantitatively analyse the differences in frequency between a staircase and a uniformly-stratified medium, we define the fractional difference as
\begin{equation}\label{eq:diff}
\frac{\Delta\omega}{\omega} = \frac{\omega_{c}-\omega_{s}}{\omega_{c}},
\end{equation}
where $\omega_s$ is the frequency of the staircase mode and $\omega_c$ is the frequency of the constant stratification mode. The magnitude of this quantity is plotted in the middle panel of Figure~\ref{fig:staricase_const_diff}, and is re-plotted using a log-log scale (base 10) in the bottom panel to determine its scaling behaviour. We find the mode frequencies in the case of a uniformly-stratified layer are always smaller than those in the staircase. Similar results are also expected with mixed boundary conditions, which might be considered the most astrophysically-relevant case (e.g.~\S~\ref{sec:free_modes_mixed}).

\begin{figure}
    \subfloat{
        \includegraphics[width=0.45\textwidth]{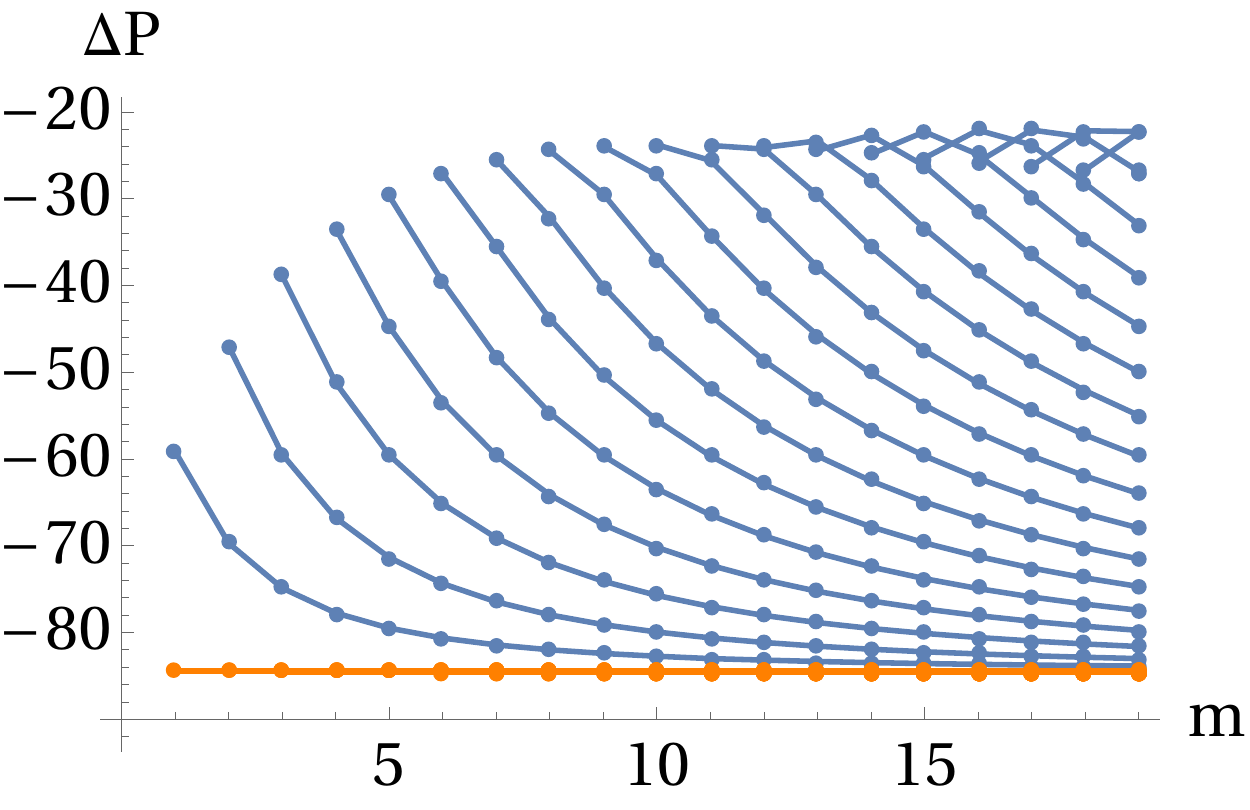}} \\
	\subfloat{
        \includegraphics[width=0.45\textwidth]{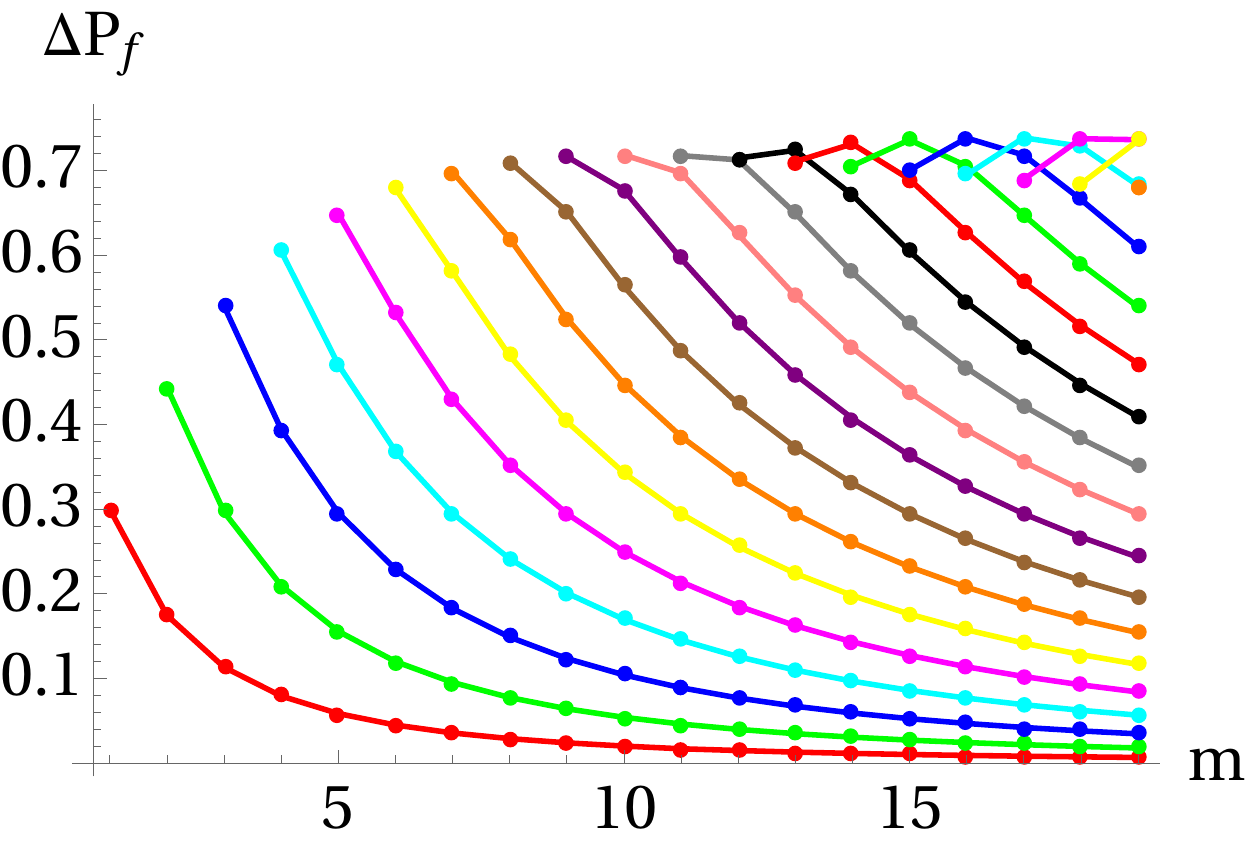}}\\
	\subfloat{
        \includegraphics[width=0.49 \textwidth]{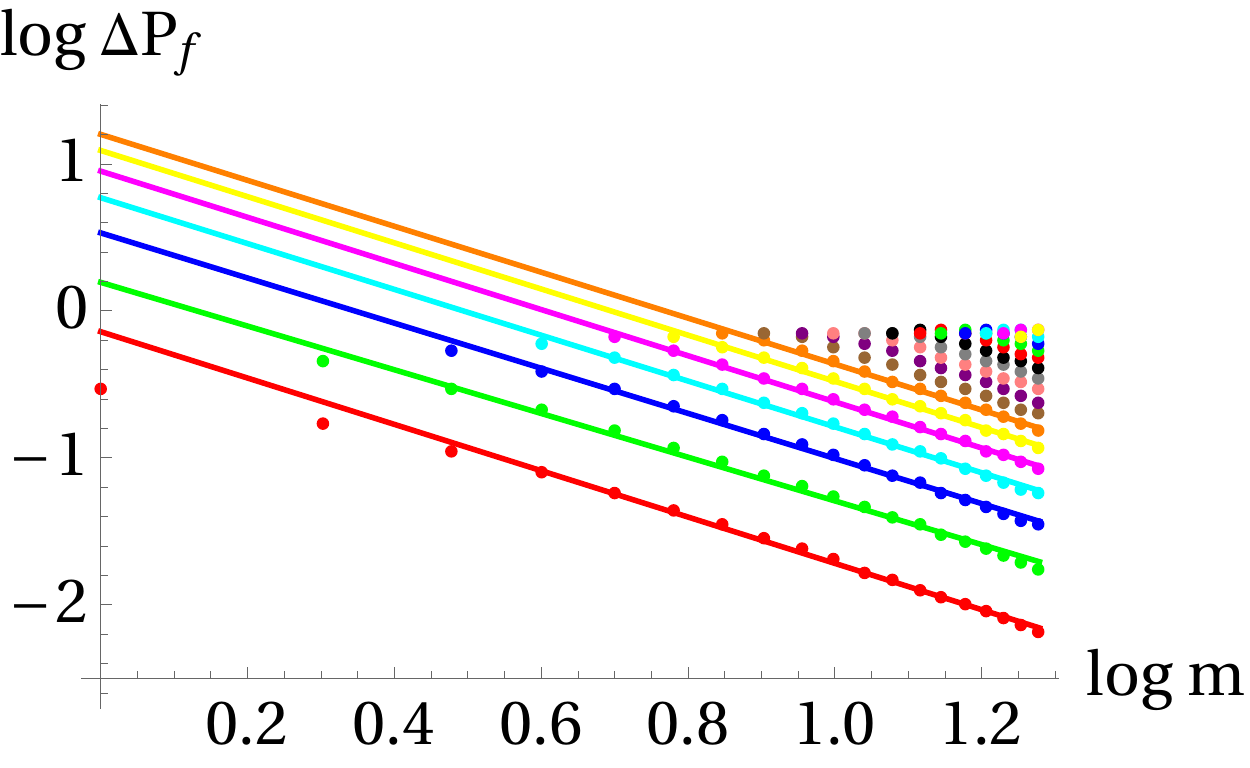}}\\
    \caption{Comparison of the period spacing of the adjacent free modes for a uniformly-stratified layer with a density staircase with the same $\bar{N}$. Top: period spacing vs $m$ for a uniformly-stratified layer (orange) and a staircase  (blue). Middle: fractional period difference (Eqn.~\ref{eq:diff_2}) between the modes of a uniformly-stratified layer and a staircase vs $m$. Bottom: same, but on a log-log (base 10) plot.}
    \label{fig:staricase_const_diff_period}
\end{figure}

Figure~\ref{fig:staricase_const_diff} shows that as the number of steps is increased, the fractional difference decreases, indicating that the free modes of a staircase converge to those of a uniformly-stratified medium with the same mean buoyancy frequency. This agrees with the results in Cartesian geometry found by \cite{Belyaev2015}. As steps are added, the number of modes in the staircase increases. The fractional difference for each mode with a given number of radial nodes decreases as we increase the number of steps. However, the lowest frequency mode with the shortest corresponding radial wavelength (largest number of radial nodes) is always the most affected by the staircase, and has the largest fractional frequency difference.
This is expected as when the wavelength is sufficiently large it ``sees the staircase" as a continuous medium with constant buoyancy frequency $\bar{N}$.

The dependence of the fractional frequency difference can be fitted with a power law for the purposes of extrapolation to a staircase with a large number of steps. We find 
\begin{equation}
    \frac{\Delta\omega}{\omega} \propto (m+1)^{-\alpha} \sim \epsilon^\alpha,
\end{equation}
with a range in exponent $\alpha\approx 1.7-2.3$ found for the highest frequency modes. This is consistent with \cite{Belyaev2015}, who found in their Cartesian model that $\alpha=2$. This power law is useful as it allows us to extrapolate the frequency shifts to a large number of steps. This is important since the number of steps in a stably-stratified layer of a giant planet is uncertain (e.g.~\citealt{Leconte2012}).

The staircase also alters the period spacing between two adjacent modes \citep{Belyaev2015}. This is interesting to analyse because the period spacing between adjacent internal gravity modes in a continuously-stratified medium is independent of the number of nodes (i.e~the mode frequency) in the short-wavelength limit. However, the presence of a staircase may modify this relation and lead to potentially observable shifts in the period spacings. To analyse the period spacing between adjacent modes, we define
\begin{equation}
\Delta P_x = 2 \pi \bigg(\frac{1}{\omega_{x,n}}-\frac{1}{\omega_{x,n+1}}\bigg).
\end{equation}
Therefore, the dependence on a staircase can be analysed by considering the fractional difference, 
\begin{equation}\label{eq:diff_2}
\Delta P_f=\frac{\Delta P_c - \Delta P_s}{\Delta P_c},
\end{equation}
where a subscript $s$ refers to a staircase mode, and a subscript $c$ refers to a continuous stratification mode.
The top panel in Figure~\ref{fig:staricase_const_diff_period} shows the staircase decreases the period spacing between adjacent modes (blue symbols and lines), and the constant stratification result is independent of node number (orange). As found in the analysis of the frequency shifts above, the fractional difference between a stably-stratified medium and a staircase structure decreases as the number of steps increases, and is largest for the lowest frequency modes with the shortest wavelengths in each case.
The fitted dependence is also found, for the purpose of extrapolation, 
\begin{equation}
    \Delta P_f \propto (m+1)^{-\beta} \sim \epsilon^\beta,
\end{equation}
where $\beta \approx 1.8-2$ for the highest frequency modes. This is also consistent with \citet{Belyaev2015}, who found the staircase decreases the spacing with a squared dependence in $\epsilon$.

\section{Wave transmission through a staircase}
\label{sec:transmission}

We now turn to explore the transmission of an internal gravity wave through a density staircase in spherical geometry, which extends prior work in Cartesian geometry \citep{Sutherland2016,Andre2017}. One motivation for these calculations is that if only part of a stratified layer has a layered density structure, then an internal gravity wave (that may be excited by tidal forcing or by interaction with neighbouring convection zones) can propagate in the continuously-stratified parts. It is important to analyse how the density staircase affects the transmission of these waves from/to the envelope to/from the interior of the planet to determine where these waves can propagate, and where they may dissipate.

\subsection{Model}\label{transmission_set_up}

We now consider a staircase-like structure embedded within a stably-stratified layer which permits the propagation of internal gravity waves. To do so, we must alter the density profile used in \S~\ref{sec:set_up} to have non-zero buoyancy frequency in each end region. We now define the buoyancy frequency as,
\begin{equation}\label{eq:DesnsityProfile2}
N^2 =
  \begin{cases}
     N_a^2      & \quad \frac{r_c}{r_0} \ll r < 1,\\
     \displaystyle\sum_{n=0}^m \bar{N}^2 \epsilon \delta (1+n \epsilon-r ) & \quad 1 < r< 1+m \epsilon, \\
     N_b^2		& \quad r > 1+m \epsilon,
  \end{cases}
\end{equation}
where $N_a$ and $N_b$ are assumed to be constants. We continue to consider a perfectly absorbing core to exist at a small radius $r_c\ll r_0$, which removes the requirement to impose a regularity condition at $r=0$. If we were to include $r=0$, then we would simply be modelling the transmission of a wave from a radius $r_0$, to the centre, and back again. Since we neglect dissipative processes, this would not be an informative calculation.

By combining Eqns.~(\ref{EQ1}) and~(\ref{eq:DesnsityProfile2}) and solving as before, the entire solution for the radial displacement is
\begin{equation}\label{eq:xi}
\xi_n =
  \begin{cases}
     A_0 r^{\lambda_{a_+}} + B_0 r^{\lambda_{a_-}}      & \quad \frac{r_c}{r_0} \ll r < 1,\\
     A_n r^{l-1} + B_n r^{-l-2} & \quad r_{n-1} < r< r_n,\\
     A_{m+1} r^{\lambda_{b_+}} + B_{m+1} r^{\lambda_{b_-}}		& \quad r > 1+m \epsilon,
  \end{cases}
\end{equation}
where $r_n=1+n \epsilon$, $n=1,\ldots,m$, and 
\begin{equation}\label{eq:lambda}
\lambda_{a/b_{\pm}} = -\frac{3}{2} \pm \frac{1}{2}\sqrt{1+4\Bigg(1-\frac{N_{a/b}^2}{\omega^2}\Bigg)l(l+1)}.
\end{equation}

For the wave to propagate in the end regions, we require $\operatorname{Im}[\lambda_{a/b_{\pm}}] \neq 0$. Therefore, from Eqn.~(\ref{eq:lambda}), the following condition must be satisfied for waves to exist in an end region:
\begin{equation}\label{eq:limit}
\frac{\omega^2}{N_x^2} < \frac{4 l (l+1)}{4 l (l+1) +1},
\end{equation}
where $N_x$ takes the appropriate value for the region considered. This restricts the allowable values of $k_\perp$ and $\omega$ that permit wave-like solutions in the end regions. We will later mark these limits on our plots showing the transmission of waves.

\begin{figure*}
    \subfloat{
	\begin{tikzpicture}
		\node at (0,0) {\includegraphics[width=0.41\textwidth]{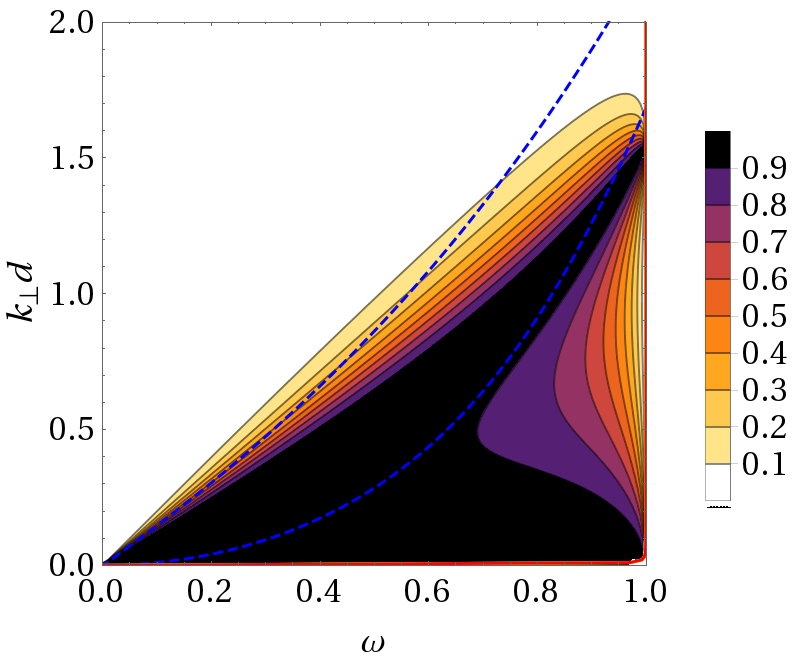}};
		\node at (-0.2,3.1) {\small \textsf{$m=1$}};
		\node at (3.1,2.3) {\small \textsf{$T_{\rm down}$}};
		\node at (0.15,-2.8) {\small \textsf{$/ \bar{N}$}};
	\end{tikzpicture}}
    \subfloat{
	\begin{tikzpicture}
		\node at (0,0) {\includegraphics[width=0.41\textwidth]{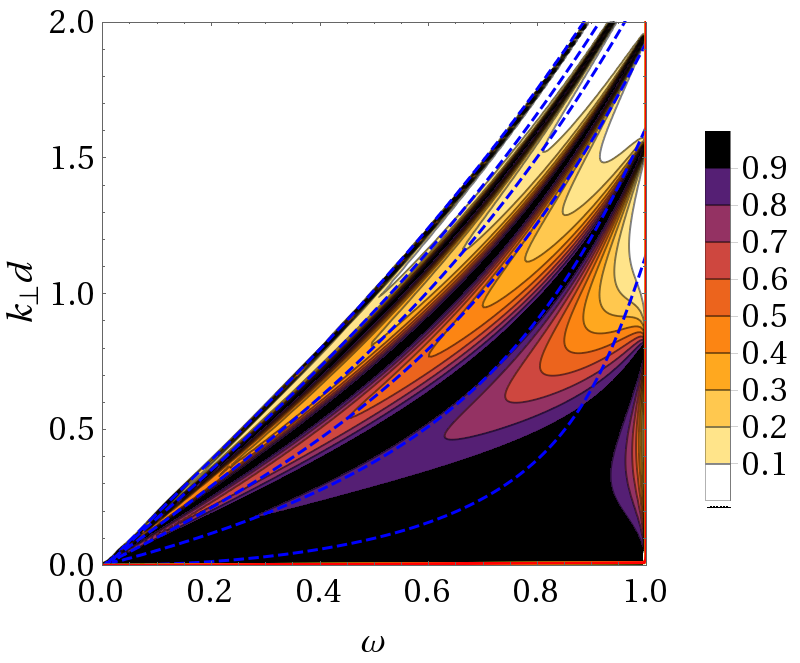}};
		\node at (-0.2,3.1) {\small \textsf{$m=5$}};
		\node at (3.1,2.3) {\small \textsf{$T_{\rm down}$}};
		\node at (0.15,-2.8) {\small \textsf{$/ \bar{N}$}};
	\end{tikzpicture}}\\
	\subfloat{
	\begin{tikzpicture}
		\node at (0,0) {\includegraphics[width=0.41\textwidth]{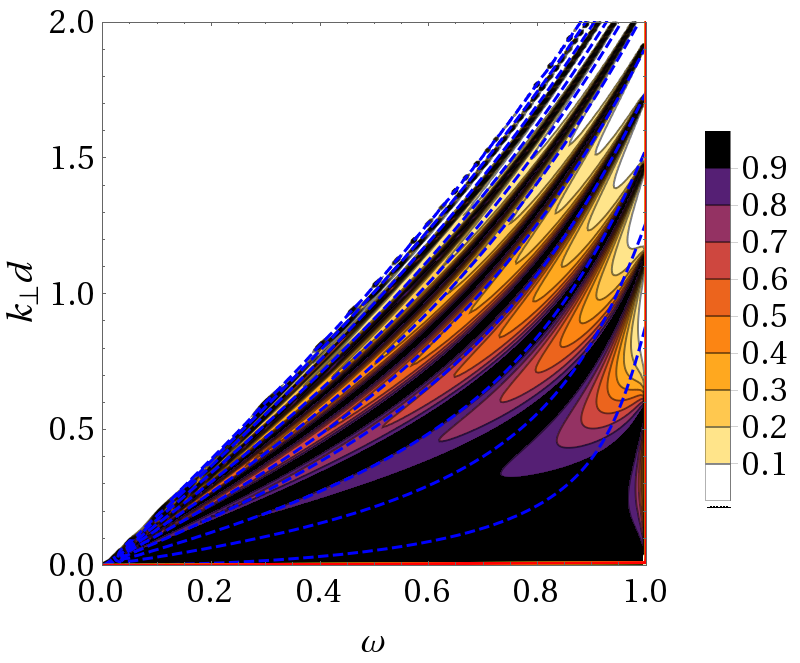}};
		\node at (-0.2,3.1) {\small \textsf{$m=10$}};
		\node at (3.1,2.3) {\small \textsf{$T_{\rm down}$}};
		\node at (0.15,-2.8) {\small \textsf{$/ \bar{N}$}};
	\end{tikzpicture}}
	\subfloat{
	\begin{tikzpicture}
		\node at (0,0) {\includegraphics[width=0.33\textwidth]{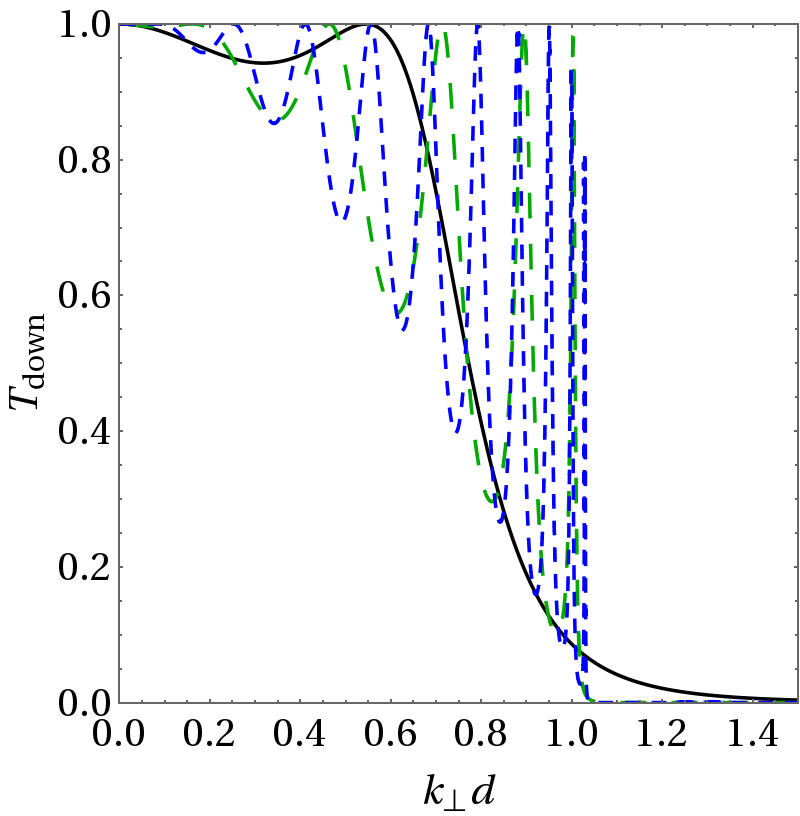}};
		\node at (2.25,2.3) {\includegraphics[width=0.07\textwidth]{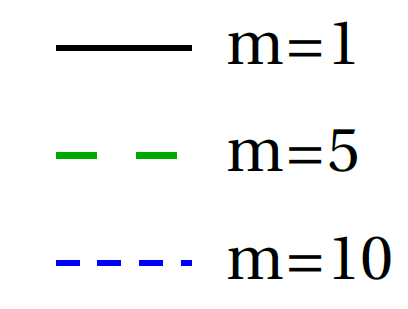}};
	\end{tikzpicture}}
    \caption{Transmission coefficient for a downward propagating wave $T_\mathrm{down}$ as a function of incident wave frequency ($\omega/\bar{N}$) and scaled horizontal wavenumber $k_\perp d=\sqrt{l(l+1)}\epsilon$, for a range of step numbers and a fixed small staircase size $(m+1)\epsilon$. Top left panel shows $(m+1)\epsilon=0.01$, with $m=1$, the top right and bottom left panels show the same case with $m=5$ and $m=10$, respectively. Each panel has $N_a=N_b=\bar{N}=1$. Over-plotted are the free modes of the same staircase (blue dashed lines), the frequency limits for wave propagation in the end regions and for the staircase if this was instead uniformly-stratified (red, close to axis). The bottom right panel shows a 1D profile at $\omega=0.5$, for $m=1$ (black), $m=5$ (green) and $m=10$ (blue).}
    \label{fig:transmission_fixed_region_small}
\end{figure*}
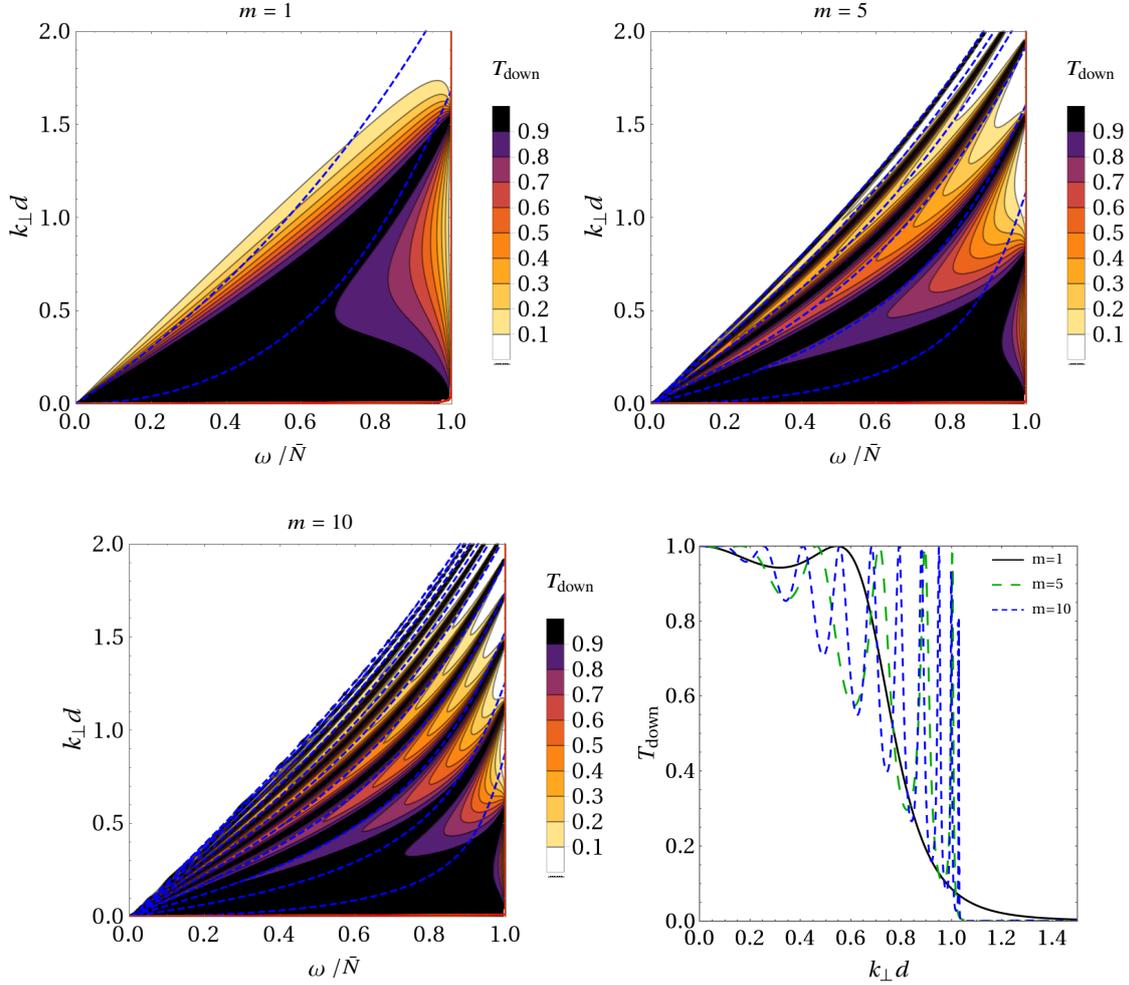

\begin{figure*}
    \subfloat{
	\begin{tikzpicture}
		\node at (0,0) {\includegraphics[width=0.41\textwidth]{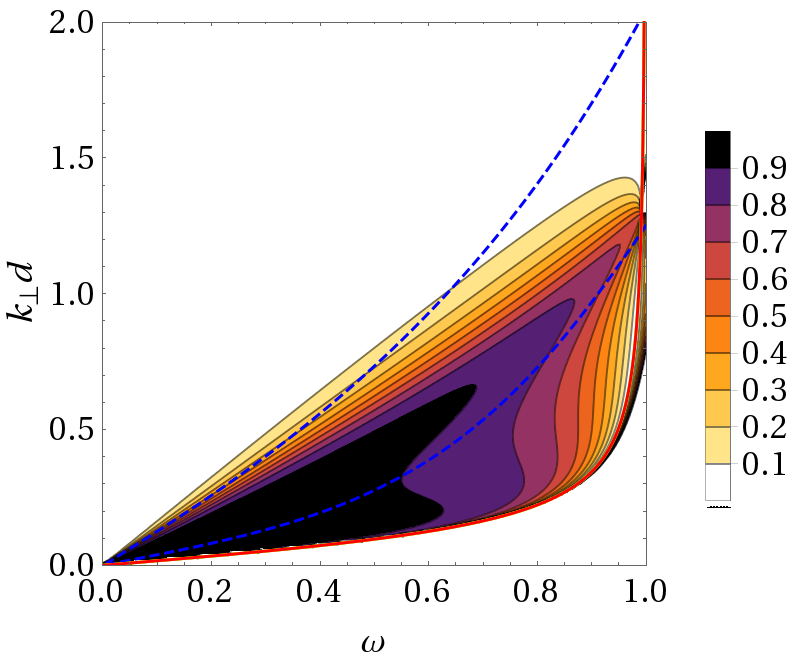}};
		\node at (-0.2,3.1) {\small \textsf{$m=1$}};
		\node at (3.1,2.3) {\small \textsf{$T_{\rm down}$}};
		\node at (0.15,-2.8) {\small \textsf{$/ \bar{N}$}};
	\end{tikzpicture}}
    \subfloat{
	\begin{tikzpicture}
		\node at (0,0) {\includegraphics[width=0.41\textwidth]{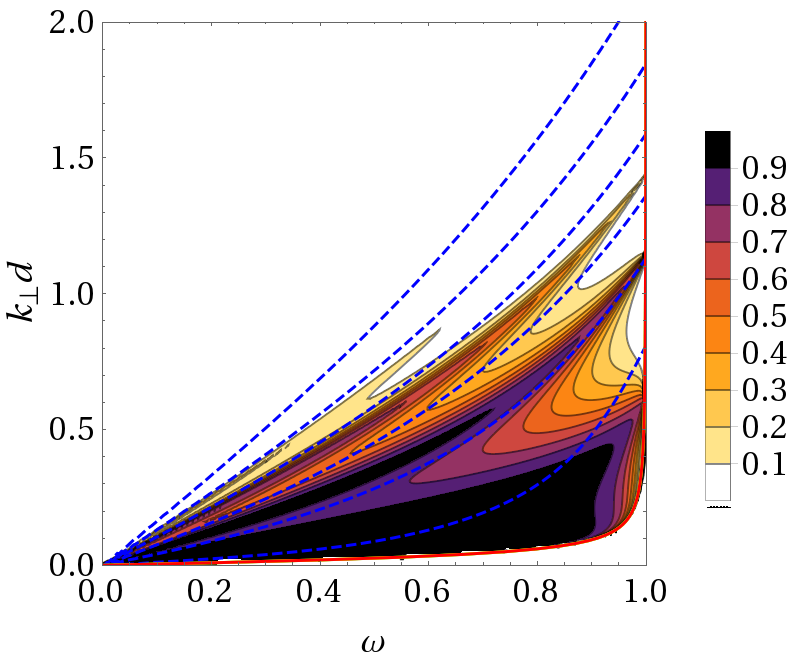}};
		\node at (-0.2,3.1) {\small \textsf{$m=5$}};
		\node at (3.1,2.3) {\small \textsf{$T_{\rm down}$}};
		\node at (0.15,-2.8) {\small \textsf{$/ \bar{N}$}};
	\end{tikzpicture}}\\
	\subfloat{
	\begin{tikzpicture}
		\node at (0,0) {\includegraphics[width=0.41\textwidth]{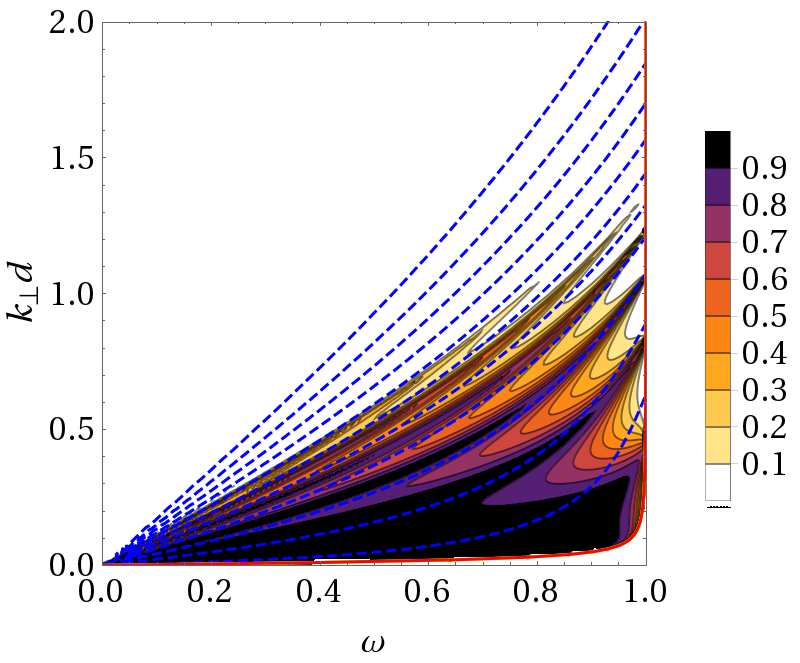}};
		\node at (-0.2,3.1) {\small \textsf{$m=10$}};
		\node at (3.1,2.3) {\small \textsf{$T_{\rm down}$}};
		\node at (0.15,-2.8) {\small \textsf{$/ \bar{N}$}};
	\end{tikzpicture}}
	\subfloat{
	\begin{tikzpicture}
		\node at (0,0) {\includegraphics[width=0.33\textwidth]{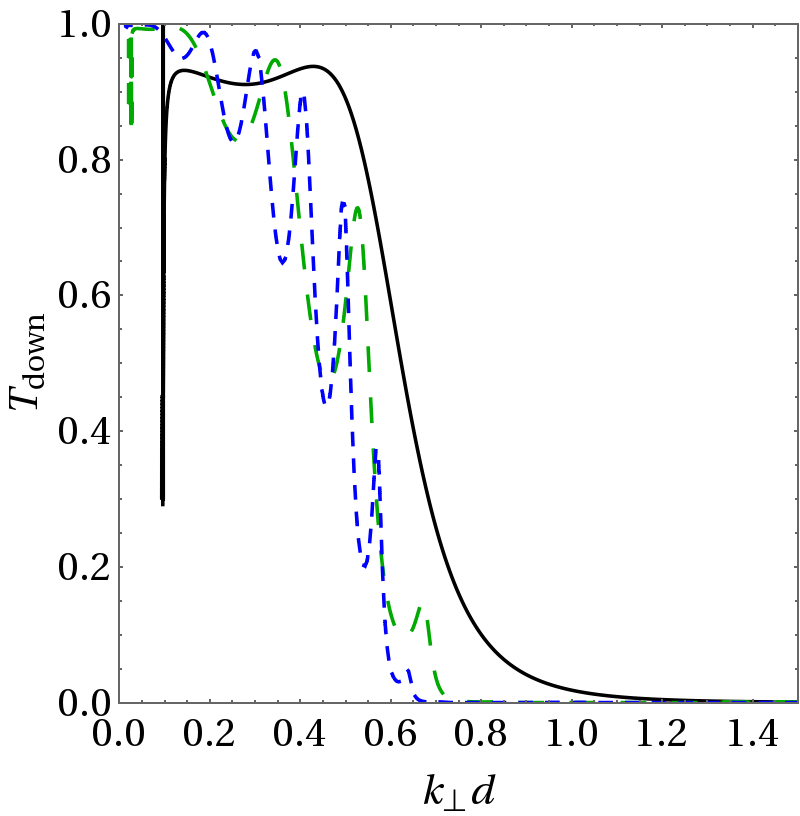}};
		\node at (2.25,2.3) {\includegraphics[width=0.07\textwidth]{"Transmission/m_legend".png}};
	\end{tikzpicture}}
    \caption{Same as Figure~\ref{fig:transmission_fixed_region_small}, except that the staircase is larger relative to the radius of the planet such that $(m+1)\epsilon=1$.}
    \label{fig:transmission_fixed_region_large}
\end{figure*}

\begin{figure*}
    \subfloat{
	\begin{tikzpicture}
		\node at (0,0) {\includegraphics[width=0.41\textwidth]{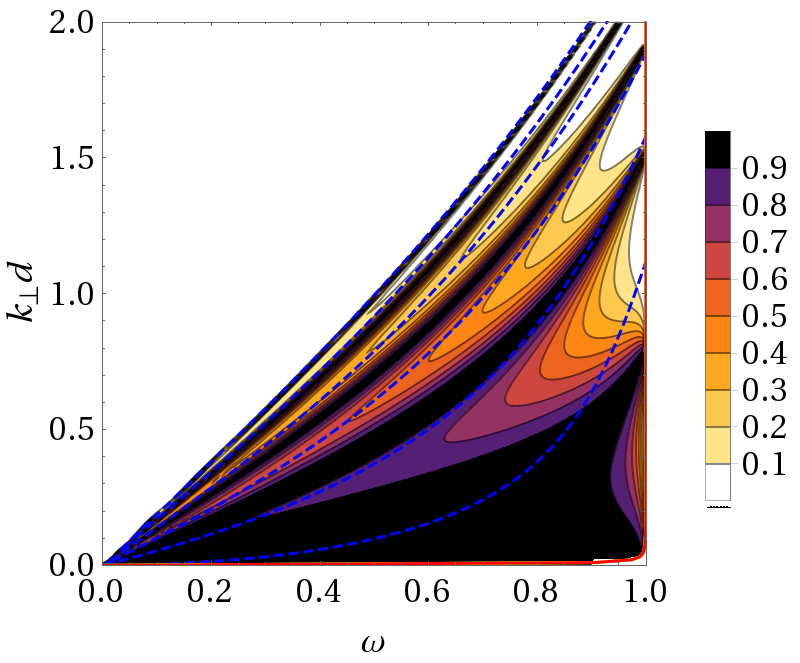}};
		\node at (-0.2,3.1) {\small \textsf{$\epsilon=0.01$}};
		\node at (3.1,2.3) {\small \textsf{$T_{\rm down}$}};
		\node at (0.15,-2.8) {\small \textsf{$/ \bar{N}$}};
	\end{tikzpicture}}
    \subfloat{
	\begin{tikzpicture}
		\node at (0,0) {\includegraphics[width=0.41\textwidth]{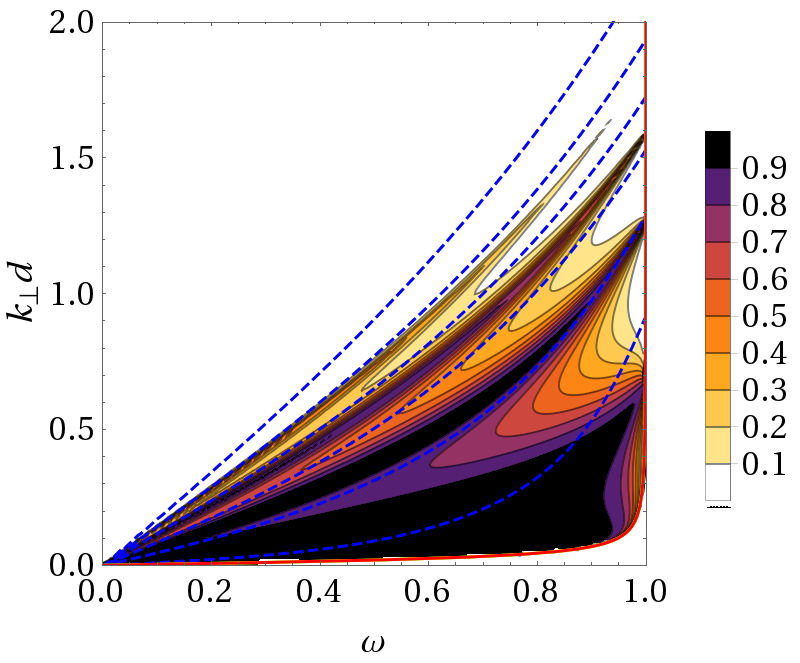}};
		\node at (-0.2,3.1) {\small \textsf{$\epsilon=0.1$}};
		\node at (3.1,2.3) {\small \textsf{$T_{\rm down}$}};
		\node at (0.15,-2.8) {\small \textsf{$/ \bar{N}$}};
	\end{tikzpicture}}\\
	\subfloat{
	\begin{tikzpicture}
		\node at (0,0) {\includegraphics[width=0.44\textwidth]{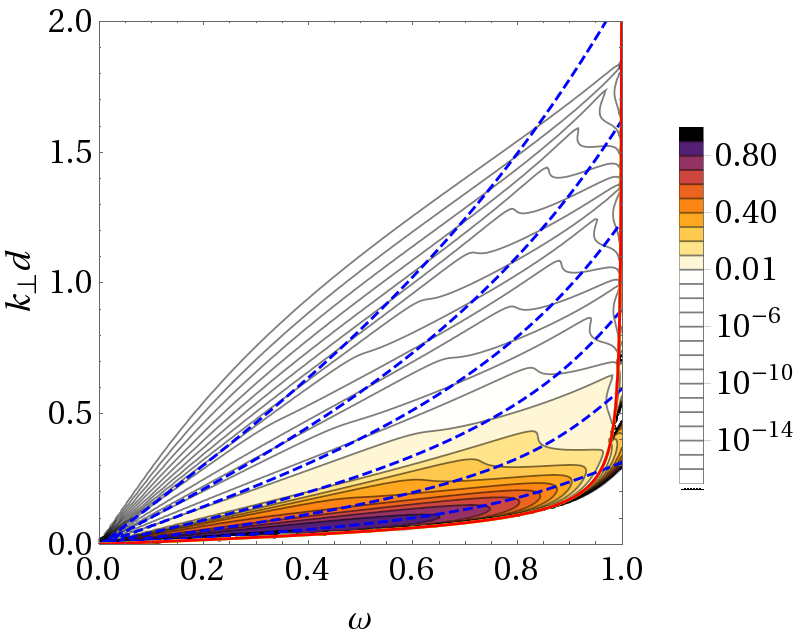}};
		\node at (-0.35,3.2) {\small \textsf{$\epsilon=1$}};
		\node at (3.1,2.3) {\small \textsf{$T_{\rm down}$}};
		\node at (0,-2.9) {\small \textsf{$/ \bar{N}$}};
	\end{tikzpicture}}
	\subfloat{
	\begin{tikzpicture}
		\node at (0,0) {\includegraphics[width=0.33\textwidth]{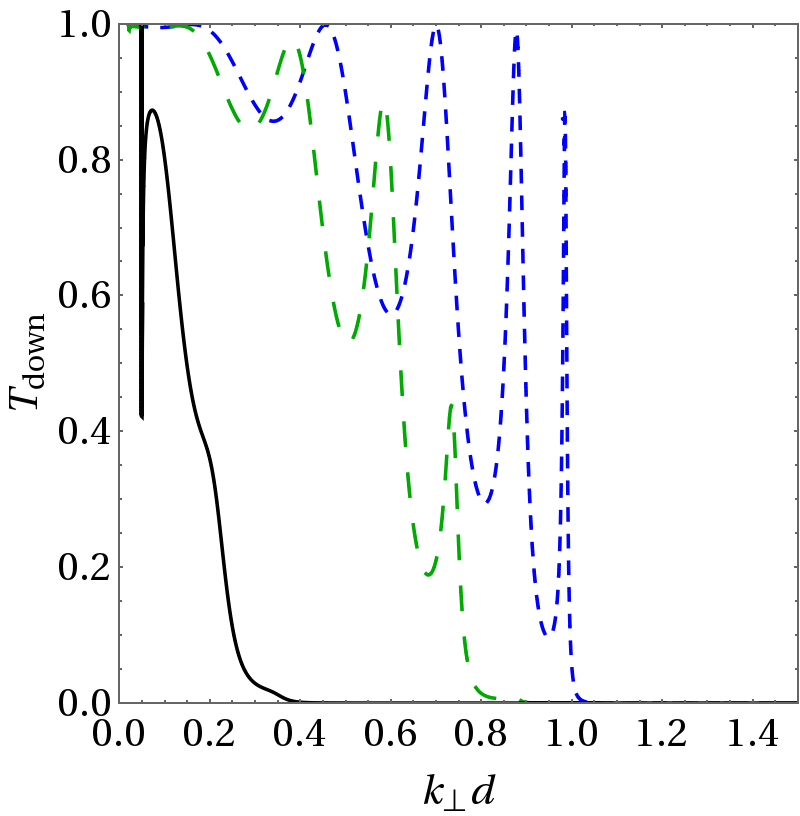}};
		\node at (2.25,2.3) {\includegraphics[width=0.07\textwidth]{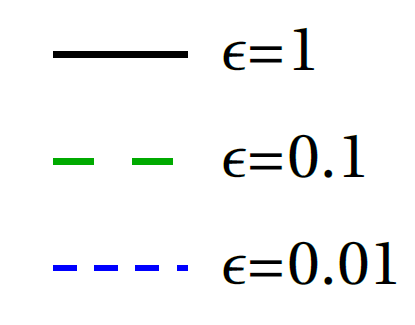}};
	\end{tikzpicture}}
    \caption{Transmission coefficient for a downward propagating wave $T_\mathrm{down}$ as a function of incident wave frequency ($\omega/\bar{N}$) and scaled horizontal wavenumber $k_\perp d=\sqrt{l(l+1)}\epsilon$, for a range of relative step sizes $\epsilon$. Top left panel shows $m=5$ steps, with $\epsilon=0.01$, the top right and bottom left panels show the same case with $\epsilon=0.1$ and $\epsilon=1$, respectively. Each panel has $N_a=N_b=\bar{N}=1$. Over-plotted are the free modes of the same staircase (blue dashed lines), the frequency limits for wave propagation in the end regions and for the staircase if this was instead uniformly-stratified (red). The bottom right panel shows a 1D profile at $\omega=0.5$, for $\epsilon=0.01$ (blue), $\epsilon=0.1$ (green) and $\epsilon=1$ (black).}
    \label{fig:transmission_epsilon}
\end{figure*}

\subsection{Transmission coefficient}

We would like to analyse how efficiently an incident internal gravity wave is transmitted (and how much is reflected) when it propagates through a staircase. To do so, we define the transmission coefficient to be the ratio of the radial energy flux of the incident wave ($F_\mathrm{in}$) with that of the outgoing wave ($F_\mathrm{tr}$), 
\begin{equation}
T=\frac{F_{\mathrm{tr}}}{F_{\mathrm{in}}},
\end{equation}
where the energy flux is defined using the standard definition for a linear wave
\begin{equation}
F=\pi r^2 \int_0^{\pi} \operatorname{Re}[-i \omega \xi_r p^*] \sin \theta d\theta,
\end{equation}
where $p^*$ is the complex conjugate of the pressure perturbation. We are only concerned with the ratio of the energy flux at different radial locations, and therefore it is not necessary to evaluate the energy flux exactly. As a result, we drop unnecessary factors from this analysis, and hence find
\begin{equation}
F \propto \operatorname{Im}[\omega \tilde{\xi}_r \tilde{p}^*]r^2.
\end{equation}
We then use Eqn.~(\ref{eq:dxi}) to eliminate $\tilde{p}^*$, so that
\begin{equation}
F \propto \operatorname{Im}\Bigg[\omega^3 \Bigg(r^2\tilde{\xi}_r\frac{\mathrm{d} \tilde{\xi}_r^*}{\mathrm{d} r}+2r\tilde{\xi}_r \tilde{\xi}_r^* \bigg) \Bigg] r^2,
\end{equation}
and using Eqn.~(\ref{eq:xi}) we find
\begin{equation}
F \propto \operatorname{Im}\bigg[r^2\lambda_{\pm}^{*}|A/B_n|^2r^{2\operatorname{Re}[\lambda_{\pm}]-1}+2r|A/B_n|^2r^{2\operatorname{Re}[\lambda_{\pm}]}\bigg] r^2.
\end{equation}
Therefore, the flux in the downward and upward propagating waves is \begin{equation}
F_{\mathrm{down}} \propto \operatorname{Im}[\lambda_{+}^{*}]|A_n|^2,
\end{equation}
and
\begin{equation}
F_{\mathrm{up}} \propto \operatorname{Im}[\lambda_{-}^{*}]|B_n|^2,
\end{equation}
which allow us define two different transmission coefficients depending on the direction of propagation of the incident wave. For a downward propagating wave,
\begin{equation}
T_{\mathrm{down}}=\frac{|A_0|^2}{|A_{m+1}|^2}\frac{\operatorname{Im}[\lambda^*_{a_+}]}{\operatorname{Im}[\lambda^*_{b_+}]},
\end{equation}
and for an upward propagating wave,
\begin{equation}
T_{\mathrm{up}}=\frac{|B_{m+1}|^2}{|B_0|^2}\frac{\operatorname{Im}[\lambda^*_{b_-}]}{\operatorname{Im}[\lambda^*_{a_-}]}.
\end{equation}
These can be shown to be equivalent to the transmission coefficient obtained in the Cartesian case \citep{Andre2017}. The transmission is observed to depend on both the amplitudes and vertical wave numbers of the solution in the end regions. The wavenumber ratio arises because the group velocity varies in the end regions if $N_a\ne N_b$.

We employ the same interface conditions as in Section~\ref{sec:set_up}, and the matrix $X$ is constructed as before. If the wave propagates downwards, from the top of the staircase towards the centre of the planet, then we have an incident and a reflected component in each layer, except that the final layer is defined to have $B_0=0$. We must have
\begin{equation}
\begin{bmatrix}
A_{m+1} \\ B_{m+1}
\end{bmatrix}
= X
\begin{bmatrix}
A_{0} \\ 0
\end{bmatrix},
\end{equation}
so that the transmission coefficient becomes 
\begin{equation}\label{eq:Tdown}
T_{\mathrm{down}}=\frac{1}{|X_{1,1}|^2}\frac{\operatorname{Im}[\lambda^*_{a_+}]}{\operatorname{Im}[\lambda^*_{b_+}]}.
\end{equation}
For an upward propagating wave, starting near the centre of the planet and propagating outwards, there is similarly no reflected wave in the upper layer ($A_{m+1}=0$), so that 
\begin{equation}
\begin{bmatrix}
0 \\ B_{m+1}
\end{bmatrix}
= X
\begin{bmatrix}
A_{0} \\ B_{0}
\end{bmatrix},
\end{equation}
giving a transmission coefficient, 
\begin{equation}\label{eq:Tup}
T_{\mathrm{up}}=\frac{1}{|X^{-1}_{2,2}|^2}\frac{\operatorname{Im}[\lambda^*_{b_-}]}{\operatorname{Im}[\lambda^*_{a_-}]}.
\end{equation}
Eqns.~(\ref{eq:Tdown}) and (\ref{eq:Tup}) allow us to determine the transmission of an incident down-going or up-going wave through a density staircase. The properties of the staircase enter through the entries of the $X$ matrix, and that of the incident wave and the end regions enter through the wavenumber ratio. As a result of the spherical geometry, it is possible for $T_{\mathrm{up}}$ and $T_{\mathrm{down}}$ to differ for the same incident wave and staircase/end region properties, unlike in the Cartesian case. We expect the transmission to recover the Cartesian results when $\epsilon \ll 1$ (and $r_0 \gg (m+1) d$), at least for waves with $l\gg 1$. On the other hand, spherical effects are expected to become important when $\epsilon \sim 1$ (or $r_0 \sim (m+1) d$).

\subsection{Results for wave transmission}

We present our results for the transmission coefficient as a function of incident wave frequency $\omega$, and horizontal wavenumber $k_\perp=\sqrt{l(l+1)}/r$, where $r$ will take the value of the location of the first interface for the incident wave, in a series of plots for various parameter values (varying $m$, $\epsilon$, $N_a$ and $N_b$). We have treated $k_\perp d$ as a continuous parameter to aid plotting and interpretation, although $l$ strictly only takes integer values and therefore gives discrete values for $k_\perp d$. Unless specified otherwise, we show the downward transmission coefficient in these figures, according to Eqn.~(\ref{eq:Tdown}), though we explore the difference between this and the upward propagation result in one case below.

In each figure, we also over-plot the frequencies of the free modes of the staircase computed from Eqn.~(\ref{eq:T11}) using blue dashed lines, in the case where the staircase is sandwiched between two convective layers (decaying boundary conditions), following  section~\ref{sec:free_modes_decay}. The frequency cut-off for wave propagation in the end regions, according to Eqn.~(\ref{eq:limit}) is shown by the solid coloured lines. The yellow and green lines show the criterion for solutions in the top and bottom layer to be propagative, respectively, while the red line shows the region in which the wave is propagative in the staircase if this were instead a uniformly-stratified layer with the same mean stratification. For cases in which $N_a=N_b=\bar{N}$ only the red line is shown and for cases where $N_a=N_b \neq \bar{N}$ only the red and yellow lines are shown. 

We begin by verifying our method by reproducing results from Cartesian geometry. To do so, we take the double limit $l \gg 1$, and $(m+1)\epsilon \ll 1$,  where for the latter we simply choose $\epsilon\ll 1$. Figure~\ref{fig:transmission_fixed_region_small} shows the transmission through a one ($m=1$), five $(m=5)$, and ten ($m=10$) step staircase, assuming $\epsilon$ is small. These agree quantitatively with Figures 9a and 18b in \citet{Andre2017}.
We observe that long wavelength (low wavenumber) waves are near-perfectly transmitted. This limit is when the waves ``sees" the staircase as a continuously-stratified medium, and is little affected by the discreteness of the steps. On the other hand, shorter wavelength waves, such that  $k_{\perp}d \sim 1$ are only transmitted when they are resonant with a free mode of the staircase. As a result, we observe bands of enhanced transmission that align well with the free modes of the staircase as calculated in \S~\ref{sec:free_modes_decay}. 

The number of peaks of enhanced transmission is always one smaller than the number of free modes of the staircase. The transmission peaks do not lie directly on top of the free modes of the staircase, with the agreement depending on the parameters adopted. This is presumably because the stratified end regions modify the wave frequencies. 

\begin{figure}
    \subfloat{
	\begin{tikzpicture}
		\node at (0,0) {\includegraphics[width=0.41\textwidth]{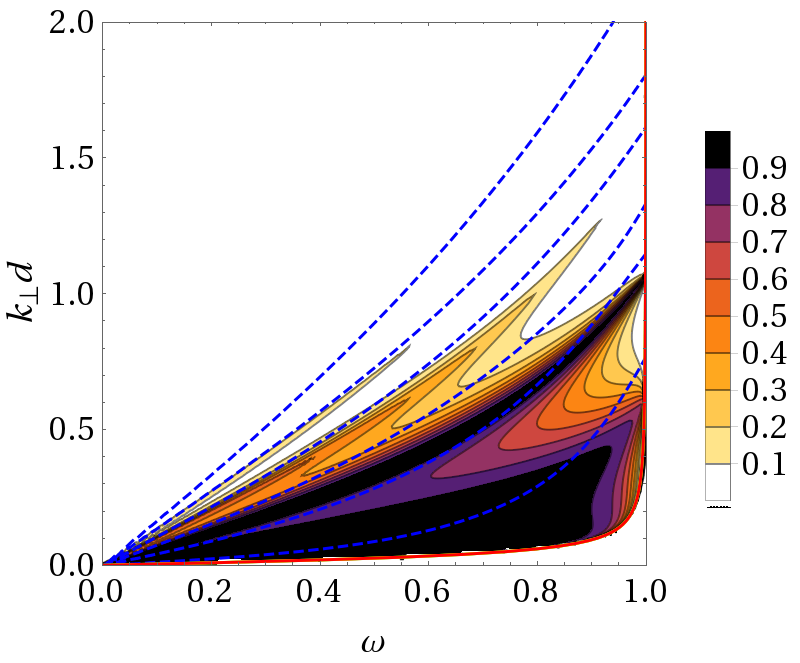}};
		\node at (-0.2,3.1) {\small \textsf{$\gamma=0.2$}};
		\node at (3.1,2.3) {\small \textsf{$T_{\rm down}$}};
		\node at (0.15,-2.8) {\small \textsf{$/ \bar{N}$}};
	\end{tikzpicture}}\\
    \subfloat{
	\begin{tikzpicture}
		\node at (0,0) {\includegraphics[width=0.41\textwidth]{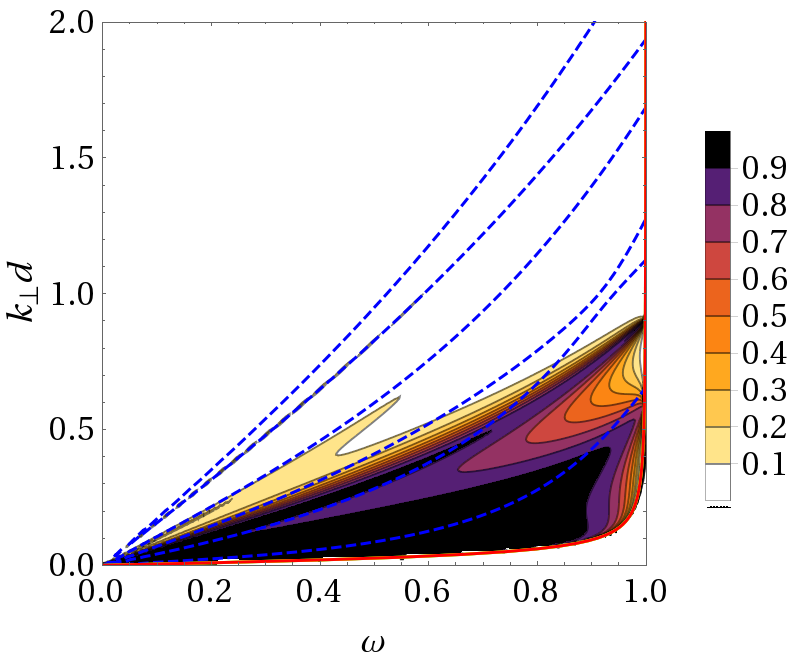}};
		\node at (-0.2,3.1) {\small \textsf{$\gamma=0.6$}};
		\node at (3.1,2.3) {\small \textsf{$T_{\rm down}$}};
		\node at (0.15,-2.8) {\small \textsf{$/ \bar{N}$}};
	\end{tikzpicture}}
    \caption{Transmission coefficient for a downward propagating wave $T_\mathrm{down}$ as a function of incident wave frequency ($\omega/\bar{N}$) and scaled horizontal wavenumber $k_\perp d=\sqrt{l(l+1)}\epsilon$, for non-uniform step size and $x=(m+1)\epsilon=1$. Over-plotted are the free modes of the same staircase (blue dashed lines), the frequency limits for wave propagation in the end regions and for the staircase if this was instead uniformly-stratified (red). Top: $\gamma=0.2$. Bottom: $\gamma=0.6$.}\label{fig:transmission_uneven_steps}
\end{figure}

\subsubsection{Dependence on $\epsilon$ (relative step size)} \label{stepsize}

In spherical geometry, the transmission depends on the relative step size, $\epsilon$, in addition to how this modifies $k_\perp$. This differs from the Cartesian case \citep{Andre2017}. 
First, we explore the dependence on step size by fixing the total size of the staircase $x=(m+1)\epsilon$ and increasing the number of steps, $m$. Figure~\ref{fig:transmission_fixed_region_small} and Figure~\ref{fig:transmission_fixed_region_large} show the overall transmission for $x=0.01$ and $x=1$ respectively. In the case of a small staircase ($x=0.01$) $\epsilon$ remains small for all panels leading to little change in the region of transmission. The only observable effects are the additional and narrower bands of enhanced transmission, reducing the overall transmission. In the case of the large staircase the variation in $\epsilon$ has a greater effect as the size of the staircase is comparable to the staircase radius $r_0$. We can see the reduced size of the transmission region as $\epsilon$ increases, as well as the additional bands observed before. The frequency range in which a wave-like solution can exist (described by Eqn.~(\ref{eq:limit})) also becomes smaller.

By analogy with Eqn.~(\ref{krcalc}), we expect that as $\omega$ increases $k_r$ will decrease, therefore the staircase should have the largest effect on transmission at high $k_\perp$ and low $\omega$ values. This is shown in Figure~\ref{fig:transmission_fixed_region_large} by observing that the peaks at the largest $k_\perp d$ for a given $\omega$ are affected the most strongly as $\epsilon$ is increased. 

Additionally we explore how the transmission depends on $\epsilon$ as the step number remains constant. Figure~\ref{fig:transmission_epsilon} shows transmission decreasing as $\epsilon$ is increased. As $\epsilon$ is increased the peaks of transmission at high $k_\perp$ values become sufficiently small that these are only visible with extra contours for smaller $T$ values. This behaviour is due to the fact that, as $\epsilon$ is increased (for fixed $m$, $\Delta \rho$ and $\bar{N}$), the total size of the staircase increases, thus the total size of the evanescent layers increases, leading to reduced transmission. An additional effect of $\epsilon$ observed here is that as $\epsilon$ is increased, the transmission peaks shift from lying below to above the free mode predictions. 

\subsubsection{Non-uniform step size}
\label{nonuniform}
In reality, we might expect the sizes and density jumps of the convective layers to vary. To explore this effect, we consider non-uniformly sized convective layers by building upon the Cartesian analysis \citep{Sutherland2016,Andre2017}.
The location of each interface is now taken to be
\begin{eqnarray}
&& r_n=1+n \epsilon_n, \\
&& \epsilon_n=\epsilon \bigg(1+\frac{\gamma}{n}\sigma_n\bigg),
\end{eqnarray}
where $\gamma$ is a free parameter taken to be less than 1, and $\sigma_n$ is a random number between $-1$ and $1$ for $n=1,\dots,(m-1)$, and $\sigma_0=0$ and $\sigma_m=0$.

Figure~\ref{fig:transmission_uneven_steps} shows the transmission for two cases with the same set of $\sigma_n$ values with $\gamma=0.2$ and $\gamma=0.6$. We observe the location of the bands of enhanced transmission have shifted to align with the now irregular spacing of the free modes. Overall, the transmission of waves is reduced by the non-uniform step size and continues to decrease as $\gamma$ is increased. The bands of enhanced transmission become narrower. We note that this remains true for a small shift in the interface locations ($\gamma = 0.2$), where the effect on the free modes is small but the effect on transmission is still significant.

\begin{figure}
    \subfloat{
	\begin{tikzpicture}
		\node at (0,0) {\includegraphics[width=0.41\textwidth]{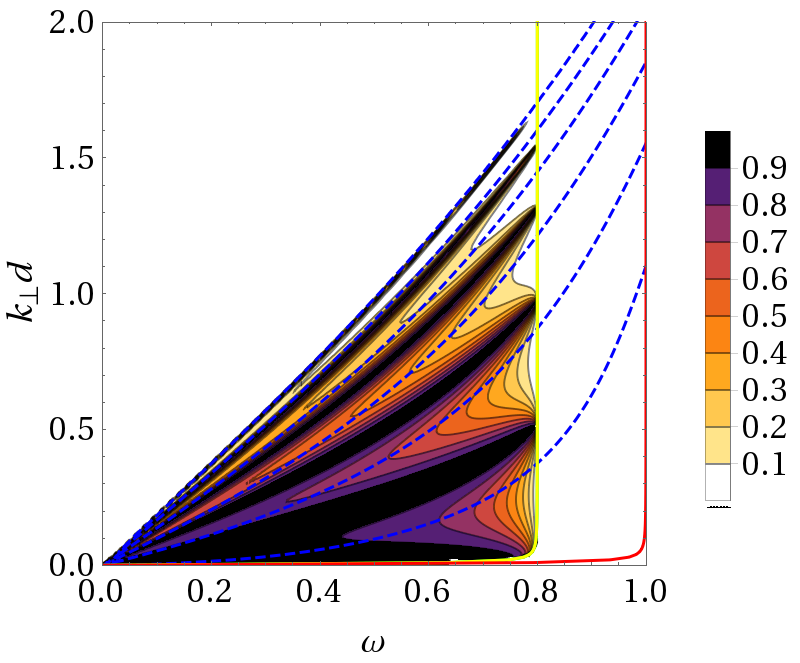}};
		\node at (-0.2,3.1) {\small \textsf{$N_{a}=N_{b}=0.8$}};
		\node at (3.2,2.5) {\small \textsf{$T_{\rm down}$}};
		\node at (0.15,-2.8) {\small \textsf{$/ \bar{N}$}};
	\end{tikzpicture}}\\
    \subfloat{
	\begin{tikzpicture}
		\node at (0,0) {\includegraphics[width=0.41\textwidth]{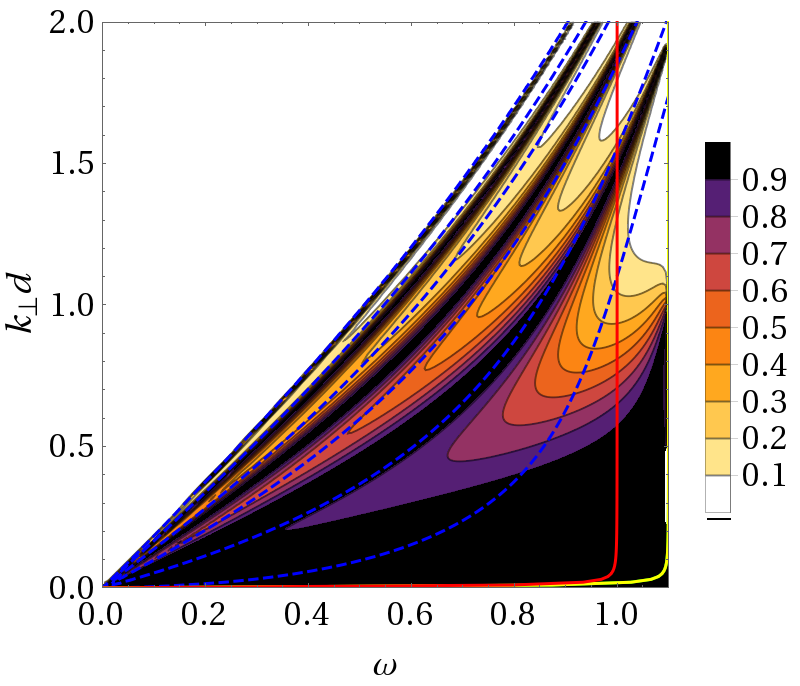}};
		\node at (-0.2,3.1) {\small \textsf{$N_{a}=N_{b}=1.1$}};
		\node at (3.2,2.5) {\small \textsf{$T_{\rm down}$}};
		\node at (0.15,-2.95) {\small \textsf{$/ \bar{N}$}};
	\end{tikzpicture}}\\
	\subfloat{
	\begin{tikzpicture}
		\node at (0,0) {\includegraphics[width=0.41\textwidth]{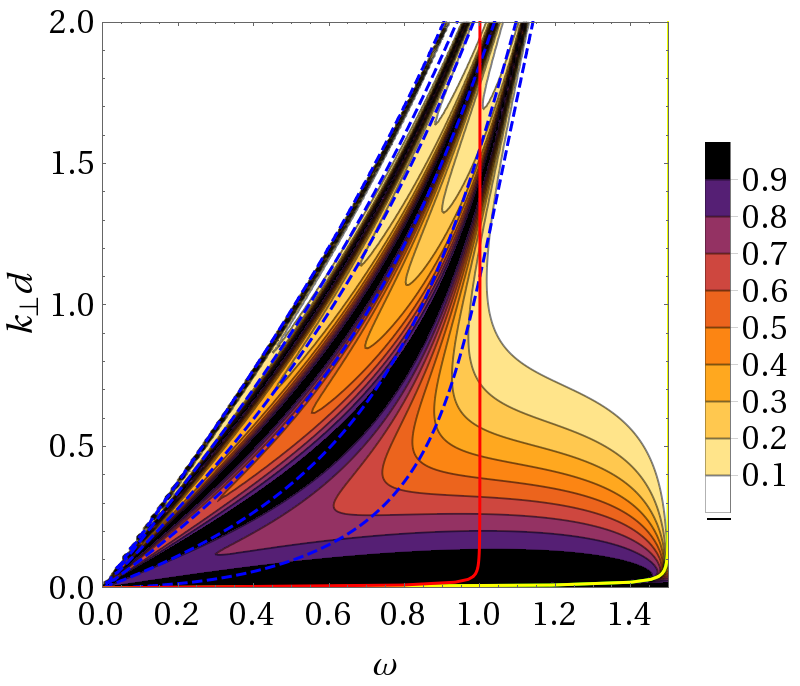}};
		\node at (-0.2,3.1) {\small \textsf{$N_{a}=N_{b}=1.5$}};
		\node at (3.2,2.5) {\small \textsf{$T_{\rm down}$}};
		\node at (0.15,-2.95) {\small \textsf{$/ \bar{N}$}};
	\end{tikzpicture}}
    \caption{Transmission coefficient for a downward propagating wave $T_\mathrm{down}$ as a function of incident wave frequency ($\omega/\bar{N}$) and scaled horizontal wavenumber $k_\perp d=\sqrt{l(l+1)}\epsilon$, for a range of stratification values in the adjacent regions, $N_a$, $N_b$ and a fixed step number, $m=5$ and a fixed staircase size $(m+1)\epsilon$=0.1. Top, middle and bottom panels show $N_a=N_b=0.8,1.1,1.5$, respectively. Over-plotted are the free modes of the same staircase (blue dashed lines), the frequency limits for wave propagation in the end regions (yellow) and for the staircase if this was instead uniformly-stratified (red).}\label{fig:transmission_stratification}
\end{figure}
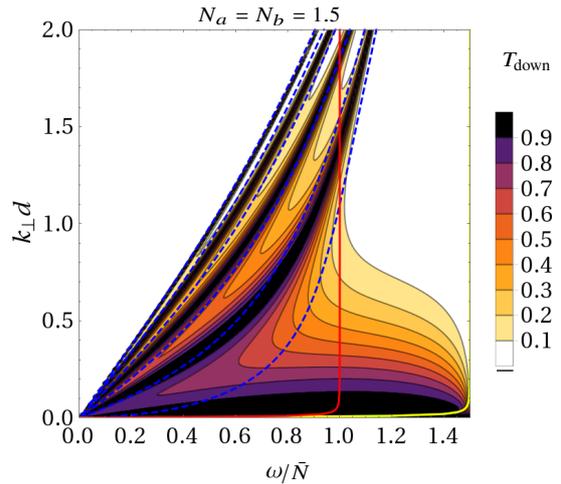

\subsubsection{Changing the properties of the end regions ($N_a, N_b$)}
\label{endregions}

The stratification at the bottom and top of the staircase ($N_a$ and $N_b$) can be varied independently of other staircase properties. Figure~\ref{fig:transmission_stratification} shows that as the stratification is altered such that the stratification is different from the mean stratification of the staircase, the bands of enhanced transmission become narrower with reduced transmission for adjacent non-resonant modes.

As we require wavelike solutions at both the bottom and top of the staircase, the range of frequencies transmitted are constrained by the smallest buoyancy frequency in these regions ($N_a$ and $N_b$), as defined by Eqn.~(\ref{eq:limit}). As the wave is always evanescent inside the staircase the value of $\bar{N}$ does not restrict the range of frequencies transmitted. This allows the staircase to increase the range of transmitted waves to frequencies larger than that of the mean stratification, which would not be transmitted by a uniformly-stratified medium -- see the bottom panel of Figure.~\ref{fig:transmission_stratification}, for example. 

\begin{figure}
    \subfloat{
	\begin{tikzpicture}
		\node at (0,0) {\includegraphics[width=0.41\textwidth]{"Transmission/Tdown_m=5_eps=0-1_N=1_Na=1_Nb=1".png}};
		\node at (3.1,2.3) {\small \textsf{$T_{\rm down}$}};
		\node at (0.15,-2.8) {\small \textsf{$/ \bar{N}$}};
	\end{tikzpicture}}\\
	\subfloat{
	\begin{tikzpicture}
		\node at (0,0) {\includegraphics[width=0.41\textwidth]{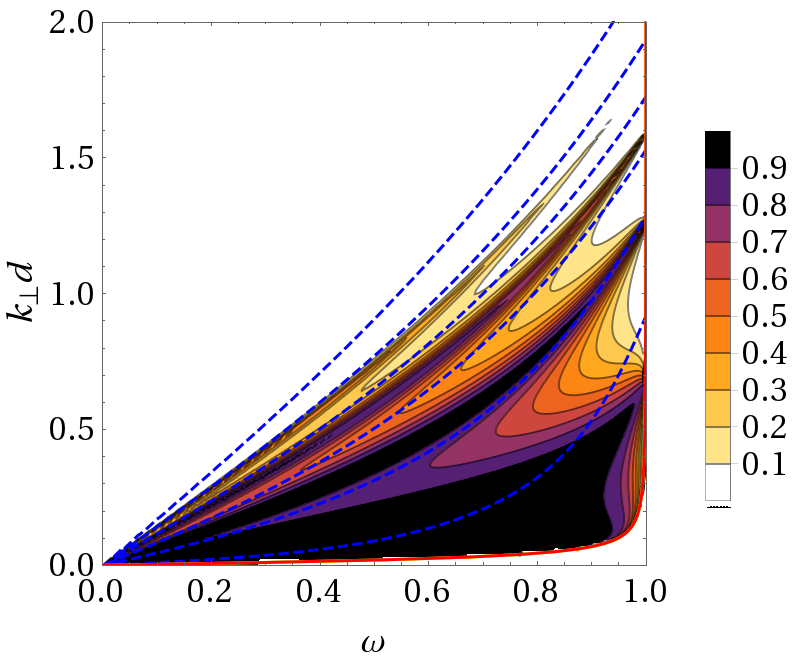}};
		\node at (3.1,2.3) {\small \textsf{$T_{\rm up}$}};
		\node at (0.15,-2.8) {\small \textsf{$/ \bar{N}$}};
	\end{tikzpicture}} 
    \caption{Comparison of the transmission coefficient for a downward ($T_\mathrm{down}$; top panel) and upward ($T_\mathrm{up}$; bottom panel) propagating incident wave as a function of the scaled wave frequency ($\omega/\bar{N}$) and horizontal wavenumber $k_\perp$ (specified in the text). Both panels have $m=5$ steps, $\epsilon=0.1$, and $N_a=N_b=\bar{N}=1$. Over-plotted are the free modes of the same staircase (blue dashed lines) and the frequency limits for wave propagation in the end regions and for the staircase if this was instead uniformly-stratified (red). This shows the symmetry between upward and downward propagating waves, even when $\epsilon$ is no longer small.} \label{fig:transmission_direction}
\end{figure}

\begin{figure*}
    \subfloat{
	\begin{tikzpicture}
		\node at (0,0) {\includegraphics[width=0.41\textwidth]{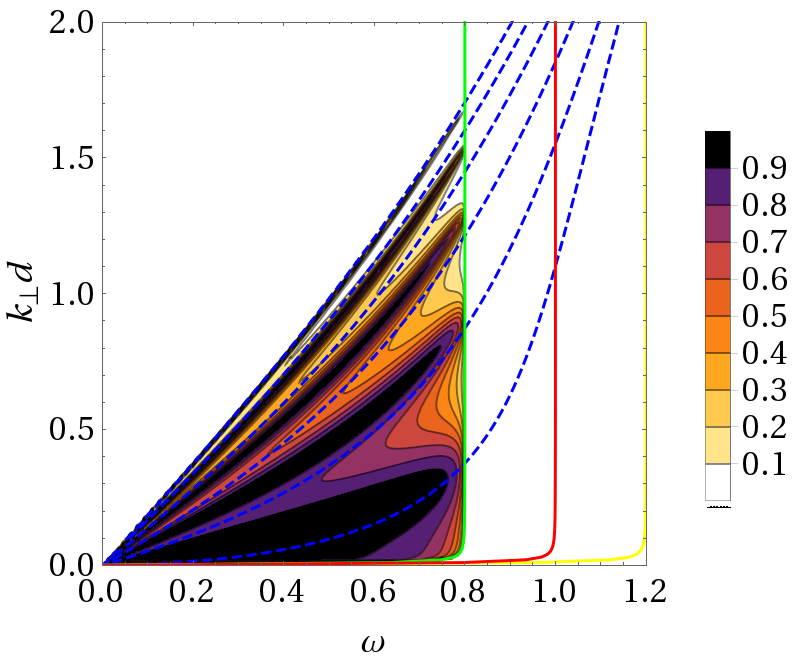}};
		\node at (-0.2,3.1) {\small \textsf{$N_a=0.8$, $N_b=1.2$, $x=0.1$}};
		\node at (3.1,2.3) {\small \textsf{$T_{\rm down}$}};
		\node at (0.15,-2.8) {\small \textsf{$/ \bar{N}$}};
	\end{tikzpicture}}
    \subfloat{
	\begin{tikzpicture}
		\node at (0,0) {\includegraphics[width=0.41\textwidth]{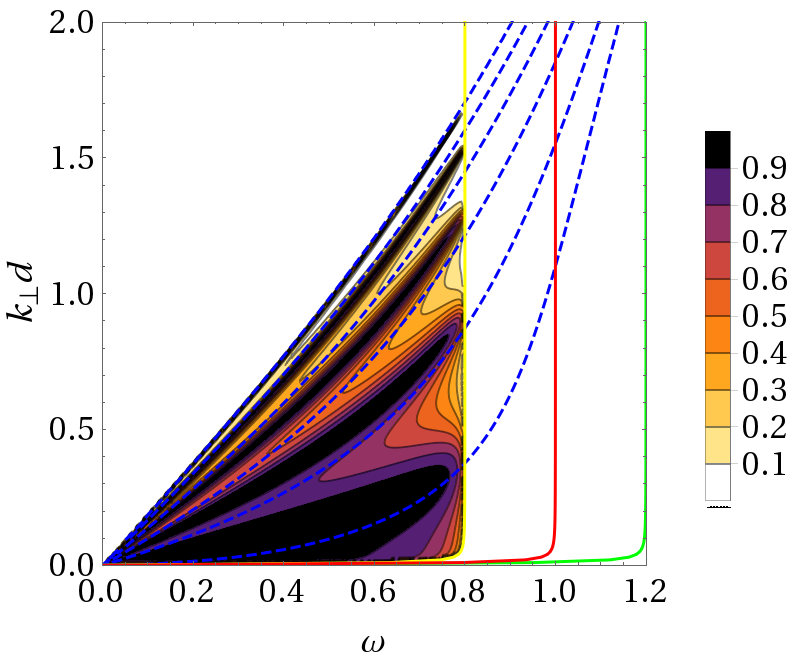}};
		\node at (-0.2,3.1) {\small \textsf{$N_a=1.2$, $N_b=0.8$, $x=0.1$}};
		\node at (3.1,2.3) {\small \textsf{$T_{\rm down}$}};
		\node at (0.15,-2.8) {\small \textsf{$/ \bar{N}$}};
	\end{tikzpicture}}\\
    \subfloat{
	\begin{tikzpicture}
		\node at (0,0) {\includegraphics[width=0.41\textwidth]{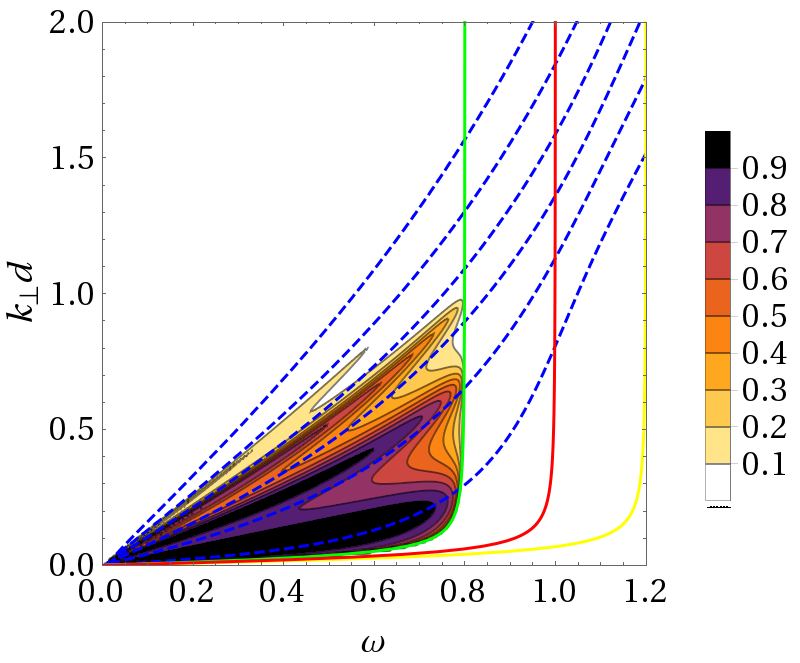}};
		\node at (-0.2,3.2) {\small \textsf{$N_a=0.8$, $N_b=1.2$, $x=1$}};
		\node at (3.1,2.3) {\small \textsf{$T_{\rm down}$}};
		\node at (0.15,-2.8) {\small \textsf{$/ \bar{N}$}};
	\end{tikzpicture}}
	\subfloat{
	\begin{tikzpicture}
		\node at (0,0) {\includegraphics[width=0.41\textwidth]{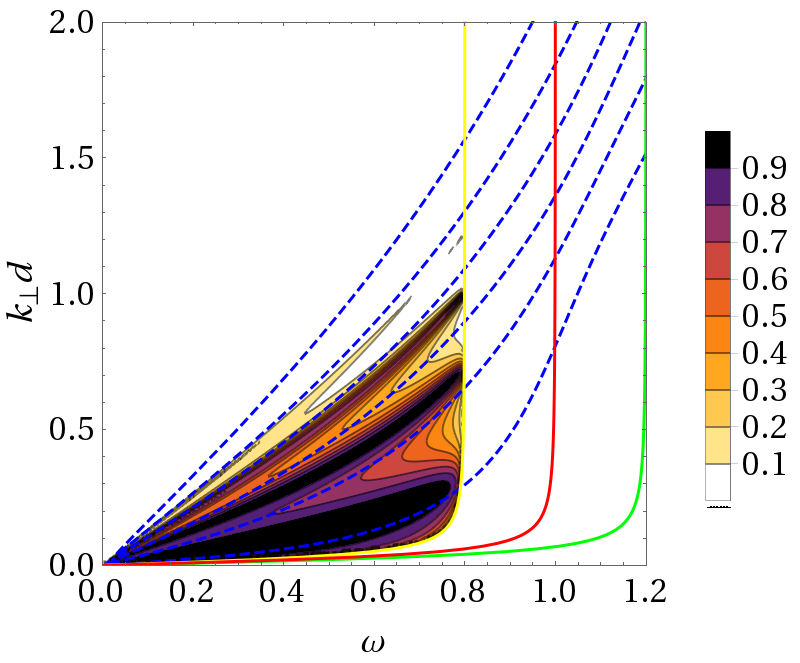}};
		\node at (-0.2,3.4) {\small \textsf{$N_a=1.2$, $N_b=0.8$, $x=1$}};
		\node at (3.1,2.3) {\small \textsf{$T_{\rm down}$}};
		\node at (0.15,-2.8) {\small \textsf{$/ \bar{N}$}};
	\end{tikzpicture}}
    \caption{Transmission coefficient for a downward propagating wave $T_\mathrm{down}$ as a function of incident wave frequency ($\omega/\bar{N}$) and scaled horizontal wavenumber $k_\perp d=\sqrt{l(l+1)}\epsilon$, for a range of stratification in the adjacent regions, $N_a$, $N_b$ and a fixed step number, $m=5$.  Four cases $x=1$ and $x=0.1$ and $N_a=0.8$, $N_b=1.2$ and $N_a=1.2$, $N_b=0.8$. Over-plotted are the free modes of the same staircase (blue dashed lines), the frequency limits for wave propagation in the end regions (yellow for the top region and green for the bottom, respectively) and for the staircase if this was instead uniformly-stratified (red).} \label{fig:transmission_stratification_diff}
\end{figure*}
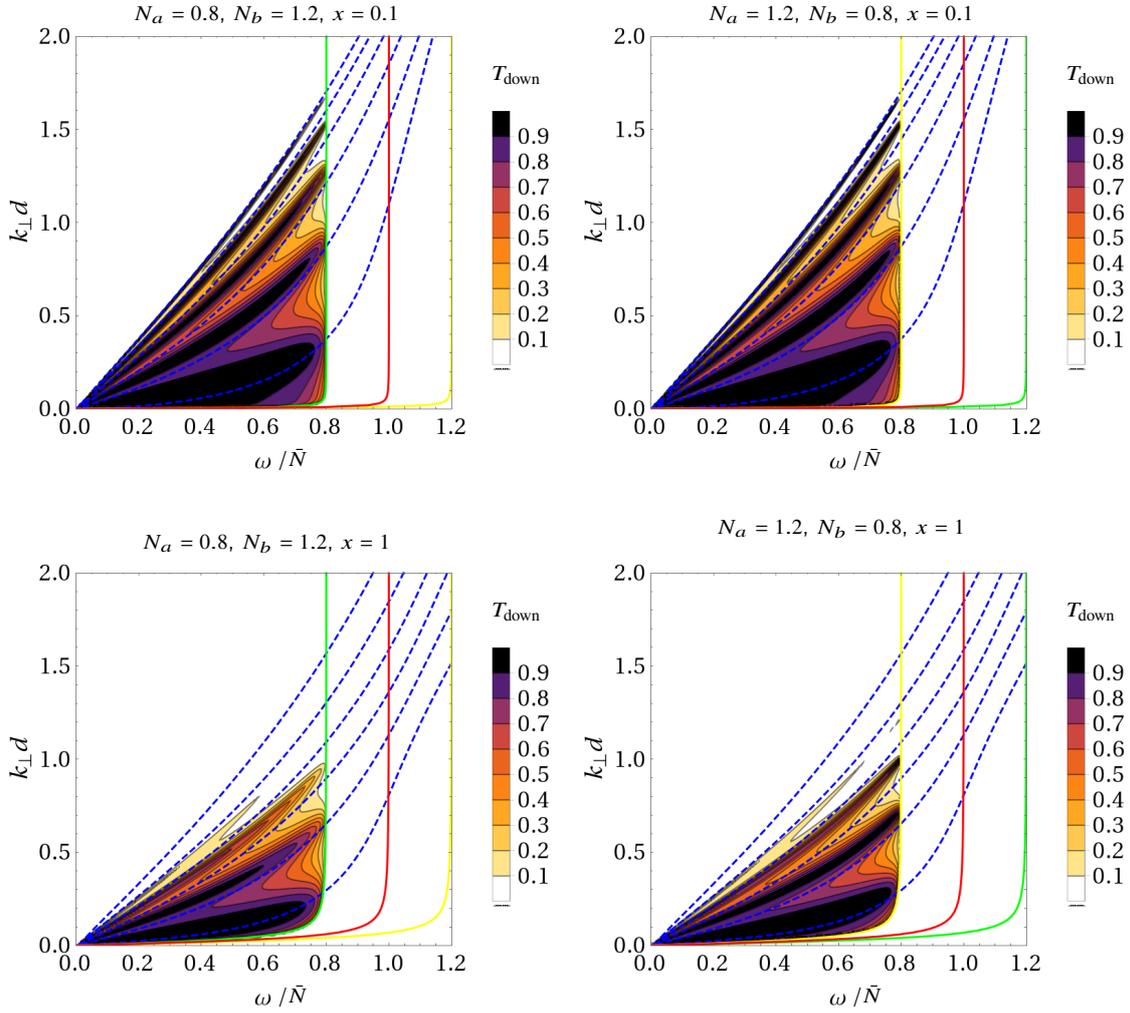

\subsubsection{Testing up/down symmetry} \label{symmetry}

We always observe that the upward and downward transmission differs only by the definition of incident $k_\perp d$. This symmetry is expected in the Cartesian limit due to the up/down symmetry of the Boussinesq system \citep[e.g.][]{sutherland_2010}. However, this symmetry no longer holds in spherical geometry. Figure~\ref{fig:transmission_direction} shows the upward and downward transmission, where in both cases $k_{\perp}d$ value is taken at the top of the staircase, $k_\perp = \frac{\sqrt{l(l+1)}}{1+m \epsilon}$. The transmission is identical in both cases when we scale the $y$-axis in this way. If we were instead to plot the same data as a function of the incident wavenumber, this would only re-scale the $y$-axis values in the right panel. This is consistent with the transmission peaks aligning with the free modes of the staircase, which do not depend on the direction of propagation of the incident wave.

In Cartesian geometry the transmission is also symmetric with respect to exchanging $N_a$ and $N_b$, which ultimately results from the up/down symmetry of the Boussinesq system in that case. This can be observed when looking at transmission in the Cartesian limit (with $\epsilon=0.01$) in  Figure~\ref{fig:transmission_stratification_diff}. On the other hand, when we increase $\epsilon$, spherical effects become important and the symmetry between upward and downward propagating waves does not hold when $N_a$ and $N_b$ are swapped. This shows that the Boussinesq symmetry previously observed no longer holds in the global case. In all cases the effect of reducing the stratification on the transmission is seen in agreement with discussion in \S~\ref{endregions}. 

\section{Conclusions}
\label{Conclusions}

Recent observations of Jupiter and Saturn with Juno and Cassini (e.g.~\citealt{Fuller2014,Wahl17,Guillot2018,Iess2019,Debras2019}) indicate that the heavy elements in these planets are probably distributed throughout the gaseous envelope rather than being solely confined to a central core.
The resulting compositional gradients can inhibit ordinary convection but enable double-diffusive convection (also referred to as semi-convection). This is thought to readily produce a layered structure in the density profile \citep{Garaud2018}, consisting of convective regions separated by thin diffusive stably-stratified interfaces. We refer to such a layered structure as a density staircase. These have been observed on Earth in the Artic ocean, in an analogous situation in which there are competing gradients of both heat and salt \citep{Arctic2016,Shibley2017}.

A layered density structure could play an important role in affecting the propagation of waves in planetary interiors. Previous work has analysed the free modes of a density staircase \citep{Belyaev2015,Andre2017} and quantified the transmission of waves through such a structure \citep{Sutherland2016,Andre2017}. These previous calculations adopted a local Cartesian model to study a small patch of a density staircase. Such a local model is a sensible starting point to study this problem because the individual steps are believed to be very small relative to the planetary radius. But such models neglect any global effects that could arise in spherical geometry. We have built upon these works by adopting a simplified global (spherical) Boussinesq model. Our model allows us to analyse the propagation of waves with wavelengths comparable with the radius of the stratified layer, which may be important for the inner regions of these planets, and also those with small harmonic degrees that may be the easiest to observe. Global effects may also be important for the modes of an extended staircase region, and are likely to be required to study tidal forcing self-consistently (this is work in progress).

We have presented idealised calculations to study the properties of waves in stably-stratified planetary layers containing a layered density structure. As a first step to tackling this problem in a global model, we have omitted planetary rotation and adopted a simplified Boussinesq model in spherical geometry. We have analysed the properties of the free modes as well as the transmission of internal waves through a density staircase. Our main result is that wave propagation is strongly affected by the presence of a density staircase. This extends and confirms prior work in Cartesian geometry \citep{Belyaev2015,Sutherland2016,Andre2017}.

We have determined the free modes in a region containing a density staircase. These consist of both internal and interfacial gravity waves, with the presence of the latter depending on the properties of the surrounding fluid. Solid wall boundary conditions do not exhibit modes with interfacial-like behaviour, whereas a staircase embedded in a convective medium (decaying boundary conditions) has a clear interfacial wave solution.

We have compared the free modes in a density staircase with those of a continuously-stratified layer. In the limit of infinitely many steps, the frequencies of the free modes converge towards those of a continuously-stratified medium. However, for a finite number of steps, the modes of a staircase typically have larger frequencies than those of a continuously-stratified medium. We have quantified this frequency shift due to the presence of a staircase as a function of its properties, as well as the shift in the period spacing between adjacent modes. In both cases we find they scale as $(m+1)^{-2}$, where $m$ is the number of steps in the staircase. This is consistent with the Cartesian results of \cite{Belyaev2015}. For the largest wavelength modes with low harmonic degrees, the shift is found to be very small if there are as many as $10^6$ steps, so this may be difficult to detect observationally. But if such a signal is detected by analysing the properties of the mixed f-g modes that are resonant with density waves in the rings (e.g.~\citealt{MarleyPorco1993,Fuller2014,Hedman2013,Hedman2019}), for example, then this could constrain the properties of any stable layer that is present in the planetary interior. We note that semi-convection in massive stars ($M_{*} \gtrsim 15 M_{\odot}$) could also produce stable layers that could be constrained in a similar way using asteroseismology \citep{Schwarzschild1958, Sakashita1959}. 

The transmission of internal waves through a density staircase was shown to be a strong function of the properties of the incident wave and of the staircase. Waves with large wavelengths are efficiently transmitted, but shorter wavelength waves (comparable with a step-size) are strongly affected by the staircase. Efficient transmission for short-wavelength waves only occurs when the incident wave is resonant with a free mode of the staircase. This agrees with prior results in Cartesian geometry \citep{Andre2017}. Spherical geometry introduces an additional frequency cut-off to the propagation of waves, and affects the transmission when the staircase size is comparable with the distance from the centre of the planet.

Future work should study the effects of rotation to determine how inertial waves are affected by a density staircase in spherical geometry. This will involve two-dimensional numerical computations (e.g.~\citealt{OgilvieLin2004,Ogilvie2007,Rieutord2010}). The importance of a density staircase on tidal dissipation in global models should also be explored, building upon the prior Cartesian numerical calculations of \cite{Andre2019}. The effects of differential rotation are also worth exploring (e.g.~\citealt{Baruteau2013,Favier2014,Guenel2016a}), as are the impact of magnetic fields (e.g.~\citealt{BL2014,OgilvieLin2018,Wei2018}), particularly since recent Juno observations indicate the important role of magnetic fields in controlling the interior differential rotation \citep{Guillot2018}.  
Finally, nonlinear effects could be analysed, since higher harmonics are generated when a wave passes through a staircase \citep{Wunsch2018}.

\section*{Acknowledgements}

We would like to thank the reviewer for their careful reading of the manuscript and for constructive comments that allowed us to improve the paper. CMP was supported by an STFC PhD studentship 2024753.  
AJB was supported by STFC grants ST/R00059X/1 and ST/S000275/1. RH was supported by STFC grant ST/S000275/1. QA and SM acknowledge support from ERC through the SPIRE grant 647383 and from the PLATO grant at the Department of Astrophysics at CEA-Saclay.



\label{lastpage}
\bibliographystyle{mnras}
\bibliography{bib} 

\begin{thebibliography}{}
\makeatletter
\relax
\def\mn@urlcharsother{\let\do\@makeother \do\$\do\&\do\#\do\^\do\_\do\%\do\~}
\def\mn@doi{\begingroup\mn@urlcharsother \@ifnextchar [ {\mn@doi@}
  {\mn@doi@[]}}
\def\mn@doi@[#1]#2{\def\@tempa{#1}\ifx\@tempa\@empty \href
  {http://dx.doi.org/#2} {doi:#2}\else \href {http://dx.doi.org/#2} {#1}\fi
  \endgroup}
\def\mn@eprint#1#2{\mn@eprint@#1:#2::\@nil}
\def\mn@eprint@arXiv#1{\href {http://arxiv.org/abs/#1} {{\tt arXiv:#1}}}
\def\mn@eprint@dblp#1{\href {http://dblp.uni-trier.de/rec/bibtex/#1.xml}
  {dblp:#1}}
\def\mn@eprint@#1:#2:#3:#4\@nil{\def\@tempa {#1}\def\@tempb {#2}\def\@tempc
  {#3}\ifx \@tempc \@empty \let \@tempc \@tempb \let \@tempb \@tempa \fi \ifx
  \@tempb \@empty \def\@tempb {arXiv}\fi \@ifundefined
  {mn@eprint@\@tempb}{\@tempb:\@tempc}{\expandafter \expandafter \csname
  mn@eprint@\@tempb\endcsname \expandafter{\@tempc}}}

\bibitem[\protect\citeauthoryear{{Andr{\'e}}, {Barker}  \&
  {Mathis}}{{Andr{\'e}} et~al.}{2017}]{Andre2017}
{Andr{\'e}} Q.,  {Barker} A.~J.,   {Mathis} S.,  2017, \mn@doi [\aap]
  {10.1051/0004-6361/201730765}, \href
  {https://ui.adsabs.harvard.edu/\#abs/2017A&A...605A.117A} {605, A117}

\bibitem[\protect\citeauthoryear{{Andr{\'e}}, {Mathis}  \&
  {Barker}}{{Andr{\'e}} et~al.}{2019}]{Andre2019}
{Andr{\'e}} Q.,  {Mathis} S.,   {Barker} A.~J.,  2019, \mn@doi [\aap]
  {10.1051/0004-6361/201833674}, \href
  {https://ui.adsabs.harvard.edu/abs/2019A&A...626A..82A} {626, A82}

\bibitem[\protect\citeauthoryear{{Baglin}, {Auvergne}, {Barge}, {Buey},
  {Catala}, {Michel}, {Weiss}  \& {COROT Team}}{{Baglin}
  et~al.}{2002}]{Baglin2002}
{Baglin} A.,  {Auvergne} M.,  {Barge} P.,  {Buey} J.~T.,  {Catala} C.,
  {Michel} E.,  {Weiss} W.,   {COROT Team} 2002, in {Battrick} B.,  {Favata}
  F.,  {Roxburgh} I.~W.,   {Galadi} D.,  eds,  ESA Special Publication Vol.
  485, Stellar Structure and Habitable Planet Finding. pp 17--24

\bibitem[\protect\citeauthoryear{{Barker} \& {Lithwick}}{{Barker} \&
  {Lithwick}}{2014}]{BL2014}
{Barker} A.~J.,  {Lithwick} Y.,  2014, \mn@doi [\mnras]
  {10.1093/mnras/stt1884}, \href
  {https://ui.adsabs.harvard.edu/abs/2014MNRAS.437..305B} {437, 305}

\bibitem[\protect\citeauthoryear{{Baruteau} \& {Rieutord}}{{Baruteau} \&
  {Rieutord}}{2013}]{Baruteau2013}
{Baruteau} C.,  {Rieutord} M.,  2013, \mn@doi [Journal of Fluid Mechanics]
  {10.1017/jfm.2012.605}, \href
  {https://ui.adsabs.harvard.edu/abs/2013JFM...719...47B} {719, 47}

\bibitem[\protect\citeauthoryear{{Belyaev}, {Quataert}  \& {Fuller}}{{Belyaev}
  et~al.}{2015}]{Belyaev2015}
{Belyaev} M.~A.,  {Quataert} E.,   {Fuller} J.,  2015, \mn@doi [\mnras]
  {10.1093/mnras/stv1446}, \href
  {https://ui.adsabs.harvard.edu/abs/2015MNRAS.452.2700B} {452, 2700}

\bibitem[\protect\citeauthoryear{{Berardo} \& {Cumming}}{{Berardo} \&
  {Cumming}}{2017}]{Berardo2017}
{Berardo} D.,  {Cumming} A.,  2017, \mn@doi [\apj] {10.3847/2041-8213/aa81c0},
  \href {https://ui.adsabs.harvard.edu/abs/2017ApJ...846L..17B} {846, L17}

\bibitem[\protect\citeauthoryear{{Bolton} et~al.,}{{Bolton}
  et~al.}{2017}]{BoltonJUNO2017}
{Bolton} S.~J.,  et~al., 2017, \mn@doi [Science] {10.1126/science.aal2108},
  \href {https://ui.adsabs.harvard.edu/abs/2017Sci...356..821B} {356, 821}

\bibitem[\protect\citeauthoryear{{Chabrier} \& {Baraffe}}{{Chabrier} \&
  {Baraffe}}{2007}]{ChabrierBaraffe2007}
{Chabrier} G.,  {Baraffe} I.,  2007, \mn@doi [\apj] {10.1086/518473}, \href
  {https://ui.adsabs.harvard.edu/abs/2007ApJ...661L..81C} {661, L81}

\bibitem[\protect\citeauthoryear{{Chaplin} \& {Miglio}}{{Chaplin} \&
  {Miglio}}{2013}]{Chaplin2013}
{Chaplin} W.~J.,  {Miglio} A.,  2013, \mn@doi [\araa]
  {10.1146/annurev-astro-082812-140938}, \href
  {https://ui.adsabs.harvard.edu/abs/2013ARA&A..51..353C} {51, 353}

\bibitem[\protect\citeauthoryear{Christensen-Dalsgaard}{Christensen-Dalsgaard}{1997}]{JCDLectures}
Christensen-Dalsgaard J.,  1997, Lecture Notes on Stellar Oscillations

\bibitem[\protect\citeauthoryear{{Christensen-Dalsgaard}}{{Christensen-Dalsgaard}}{2002}]{JCD2002}
{Christensen-Dalsgaard} J.,  2002, \mn@doi [Reviews of Modern Physics]
  {10.1103/RevModPhys.74.1073}, \href
  {https://ui.adsabs.harvard.edu/abs/2002RvMP...74.1073C} {74, 1073}

\bibitem[\protect\citeauthoryear{{Cowling}}{{Cowling}}{1941}]{Cowling1941}
{Cowling} T.~G.,  1941, \mn@doi [\mnras] {10.1093/mnras/101.8.367}, \href
  {https://ui.adsabs.harvard.edu/abs/1941MNRAS.101..367C} {101, 367}

\bibitem[\protect\citeauthoryear{{Debras} \& {Chabrier}}{{Debras} \&
  {Chabrier}}{2019}]{Debras2019}
{Debras} F.,  {Chabrier} G.,  2019, \mn@doi [The Astrophysical Journal]
  {10.3847/1538-4357/aaff65}, \href
  {https://ui.adsabs.harvard.edu/abs/2019ApJ...872..100D} {872, 100}

\bibitem[\protect\citeauthoryear{Dintrans, Rieutord  \& Valdettaro}{Dintrans
  et~al.}{1999}]{Dintrans1999}
Dintrans B.,  Rieutord M.,   Valdettaro L.,  1999, \mn@doi [Journal of Fluid
  Mechanics] {10.1017/S0022112099006308}, 398, 271

\bibitem[\protect\citeauthoryear{{Duguid}, {Barker}  \& {Jones}}{{Duguid}
  et~al.}{2020}]{Duguid2019}
{Duguid} C.~D.,  {Barker} A.~J.,   {Jones} C.~A.,  2020, \mn@doi [\mnras]
  {10.1093/mnras/stz2899}, \href
  {https://ui.adsabs.harvard.edu/abs/2020MNRAS.491..923D} {491, 923}

\bibitem[\protect\citeauthoryear{{Favier}, {Barker}, {Baruteau}  \&
  {Ogilvie}}{{Favier} et~al.}{2014}]{Favier2014}
{Favier} B.,  {Barker} A.~J.,  {Baruteau} C.,   {Ogilvie} G.~I.,  2014, \mn@doi
  [\mnras] {10.1093/mnras/stu003}, \href
  {https://ui.adsabs.harvard.edu/abs/2014MNRAS.439..845F} {439, 845}

\bibitem[\protect\citeauthoryear{{Fortney} \& {Nettelmann}}{{Fortney} \&
  {Nettelmann}}{2010}]{Fortney2010}
{Fortney} J.~J.,  {Nettelmann} N.,  2010, \mn@doi [\ssr]
  {10.1007/s11214-009-9582-x}, \href
  {https://ui.adsabs.harvard.edu/abs/2010SSRv..152..423F} {152, 423}

\bibitem[\protect\citeauthoryear{{Fuller}}{{Fuller}}{2014}]{Fuller2014}
{Fuller} J.,  2014, \mn@doi [\icarus] {10.1016/j.icarus.2014.08.006}, \href
  {https://ui.adsabs.harvard.edu/abs/2014Icar..242..283F} {242, 283}

\bibitem[\protect\citeauthoryear{{Fuller}, {Luan}  \& {Quataert}}{{Fuller}
  et~al.}{2016}]{Fuller2016}
{Fuller} J.,  {Luan} J.,   {Quataert} E.,  2016, \mn@doi [\mnras]
  {10.1093/mnras/stw609}, \href
  {https://ui.adsabs.harvard.edu/abs/2016MNRAS.458.3867F} {458, 3867}

\bibitem[\protect\citeauthoryear{{Garaud}}{{Garaud}}{2018}]{Garaud2018}
{Garaud} P.,  2018, \mn@doi [Annual Review of Fluid Mechanics]
  {10.1146/annurev-fluid-122316-045234}, \href
  {https://ui.adsabs.harvard.edu/abs/2018AnRFM..50..275G} {50, 275}

\bibitem[\protect\citeauthoryear{{Gaulme}, {Schmider}, {Gay}, {Guillot}  \&
  {Jacob}}{{Gaulme} et~al.}{2011}]{Gaulme2011}
{Gaulme} P.,  {Schmider} F.~X.,  {Gay} J.,  {Guillot} T.,   {Jacob} C.,  2011,
  \mn@doi [\aap] {10.1051/0004-6361/201116903}, \href
  {https://ui.adsabs.harvard.edu/abs/2011A&A...531A.104G} {531, A104}

\bibitem[\protect\citeauthoryear{{Ghaemsaidi}, {Dosser}, {Rainville}  \&
  {Peacock}}{{Ghaemsaidi} et~al.}{2016}]{Arctic2016}
{Ghaemsaidi} S.~J.,  {Dosser} H.~V.,  {Rainville} L.,   {Peacock} T.,  2016,
  \mn@doi [Journal of Fluid Mechanics] {10.1017/jfm.2015.682}, \href
  {https://ui.adsabs.harvard.edu/abs/2016JFM...789..617G} {789, 617}

\bibitem[\protect\citeauthoryear{{Gilliland} et~al.,}{{Gilliland}
  et~al.}{2010}]{Kepler2010}
{Gilliland} R.~L.,  et~al., 2010, \mn@doi [\pasp] {10.1086/650399}, \href
  {https://ui.adsabs.harvard.edu/abs/2010PASP..122..131G} {122, 131}

\bibitem[\protect\citeauthoryear{{Goldreich} \& {Nicholson}}{{Goldreich} \&
  {Nicholson}}{1977}]{GN1977}
{Goldreich} P.,  {Nicholson} P.~D.,  1977, \mn@doi [\icarus]
  {10.1016/0019-1035(77)90163-4}, \href
  {https://ui.adsabs.harvard.edu/abs/1977Icar...30..301G} {30, 301}

\bibitem[\protect\citeauthoryear{{Gough}}{{Gough}}{1993}]{Gough1993}
{Gough} D.~O.,  1993, in Astrophysical Fluid Dynamics - Les Houches 1987.
  Elsevier Science Ltd, pp 399--560

\bibitem[\protect\citeauthoryear{Guenel, Baruteau, Mathis  \& Rieutord}{Guenel
  et~al.}{2016}]{Guenel2016a}
Guenel M.,  Baruteau C.,  Mathis S.,   Rieutord M.,  2016, \mn@doi [Astronomy &
  Astrophysics] {10.1051/0004-6361/201527621}, 589

\bibitem[\protect\citeauthoryear{{Guillot}}{{Guillot}}{2005}]{Guillot2005}
{Guillot} T.,  2005, \mn@doi [Annual Review of Earth and Planetary Sciences]
  {10.1146/annurev.earth.32.101802.120325}, \href
  {https://ui.adsabs.harvard.edu/abs/2005AREPS..33..493G} {33, 493}

\bibitem[\protect\citeauthoryear{{Guillot}, {Stevenson}, {Hubbard}  \&
  {Saumon}}{{Guillot} et~al.}{2004}]{Guillot2004}
{Guillot} T.,  {Stevenson} D.~J.,  {Hubbard} W.~B.,   {Saumon} D.,  2004, {The
  interior of Jupiter}.
Cambridge University Press, pp 35--57

\bibitem[\protect\citeauthoryear{{Guillot} et~al.,}{{Guillot}
  et~al.}{2018}]{Guillot2018}
{Guillot} T.,  et~al., 2018, \mn@doi [\nat] {10.1038/nature25775}, \href
  {https://ui.adsabs.harvard.edu/abs/2018Natur.555..227G} {555, 227}

\bibitem[\protect\citeauthoryear{{Hedman} \& {Nicholson}}{{Hedman} \&
  {Nicholson}}{2013}]{Hedman2013}
{Hedman} M.~M.,  {Nicholson} P.~D.,  2013, \mn@doi [\aj]
  {10.1088/0004-6256/146/1/12}, \href
  {https://ui.adsabs.harvard.edu/abs/2013AJ....146...12H} {146, 12}

\bibitem[\protect\citeauthoryear{{Hedman}, {Nicholson}  \& {French}}{{Hedman}
  et~al.}{2019}]{Hedman2019}
{Hedman} M.~M.,  {Nicholson} P.~D.,   {French} R.~G.,  2019, \mn@doi [\aj]
  {10.3847/1538-3881/aaf0a6}, \href
  {https://ui.adsabs.harvard.edu/abs/2019AJ....157...18H} {157, 18}

\bibitem[\protect\citeauthoryear{{Helled} \& {Stevenson}}{{Helled} \&
  {Stevenson}}{2017}]{Helled2017}
{Helled} R.,  {Stevenson} D.,  2017, \mn@doi [\apj] {10.3847/2041-8213/aa6d08},
  \href {https://ui.adsabs.harvard.edu/abs/2017ApJ...840L...4H} {840, L4}

\bibitem[\protect\citeauthoryear{{Helled}, {Nettelmann}  \& {Guillot}}{{Helled}
  et~al.}{2019}]{HelledUrNep2019}
{Helled} R.,  {Nettelmann} N.,   {Guillot} T.,  2019, arXiv e-prints, \href
  {https://ui.adsabs.harvard.edu/abs/2019arXiv190904891H} {p. arXiv:1909.04891}

\bibitem[\protect\citeauthoryear{{Iess} et~al.,}{{Iess}
  et~al.}{2019}]{Iess2019}
{Iess} L.,  et~al., 2019, \mn@doi [Science] {10.1126/science.aat2965}, \href
  {https://ui.adsabs.harvard.edu/abs/2019Sci...364.2965I} {364, aat2965}

\bibitem[\protect\citeauthoryear{{Ioannou} \& {Lindzen}}{{Ioannou} \&
  {Lindzen}}{1993a}]{IoannouLindzen1993p1}
{Ioannou} P.~J.,  {Lindzen} R.~S.,  1993a, \mn@doi [\apj] {10.1086/172437},
  \href {https://ui.adsabs.harvard.edu/abs/1993ApJ...406..252I} {406, 252}

\bibitem[\protect\citeauthoryear{{Ioannou} \& {Lindzen}}{{Ioannou} \&
  {Lindzen}}{1993b}]{IoannouLindzen1993p2}
{Ioannou} P.~J.,  {Lindzen} R.~S.,  1993b, \mn@doi [\apj] {10.1086/172438},
  \href {https://ui.adsabs.harvard.edu/abs/1993ApJ...406..266I} {406, 266}

\bibitem[\protect\citeauthoryear{{Kippenhahn}, {Weigert}  \&
  {Weiss}}{{Kippenhahn} et~al.}{2012}]{Kippenhahn2012}
{Kippenhahn} R.,  {Weigert} A.,   {Weiss} A.,  2012, {Stellar Structure and
  Evolution}, \mn@doi{10.1007/978-3-642-30304-3.
}

\bibitem[\protect\citeauthoryear{Lainey, Arlot, Karatekin  \&
  Van~Hoolst}{Lainey et~al.}{2009}]{Lainey2009}
Lainey V.,  Arlot J.-E.,  Karatekin O.,   Van~Hoolst T.,  2009, \mn@doi
  [Nature] {10.1038/nature08108}, 459, 957

\bibitem[\protect\citeauthoryear{Lainey et~al.,}{Lainey
  et~al.}{2012}]{Lainey2012}
Lainey V.,  et~al., 2012, \mn@doi [The Astrophysical Journal]
  {10.1088/0004-637X/752/1/14}, 752

\bibitem[\protect\citeauthoryear{Lainey et~al.,}{Lainey
  et~al.}{2017}]{Lainey2017}
Lainey V.,  et~al., 2017, \mn@doi [Icarus]
  {https://doi.org/10.1016/j.icarus.2016.07.014}, 281, 286

\bibitem[\protect\citeauthoryear{{Leconte} \& {Chabrier}}{{Leconte} \&
  {Chabrier}}{2012}]{Leconte2012}
{Leconte} J.,  {Chabrier} G.,  2012, \mn@doi [\aap]
  {10.1051/0004-6361/201117595}, \href
  {https://ui.adsabs.harvard.edu/abs/2012A&A...540A..20L} {540, A20}

\bibitem[\protect\citeauthoryear{{Leconte} \& {Chabrier}}{{Leconte} \&
  {Chabrier}}{2013}]{Leconte2013}
{Leconte} J.,  {Chabrier} G.,  2013, \mn@doi [Nature Geoscience]
  {10.1038/ngeo1791}, \href
  {https://ui.adsabs.harvard.edu/abs/2013NatGe...6..347L} {6, 347}

\bibitem[\protect\citeauthoryear{{Lin} \& {Ogilvie}}{{Lin} \&
  {Ogilvie}}{2018}]{OgilvieLin2018}
{Lin} Y.,  {Ogilvie} G.~I.,  2018, \mn@doi [\mnras] {10.1093/mnras/stx2764},
  \href {https://ui.adsabs.harvard.edu/abs/2018MNRAS.474.1644L} {474, 1644}

\bibitem[\protect\citeauthoryear{{Liu}, {M\"{u}ller}, {Zheng}, {Helled}, {Lin}
  \& {Isella}}{{Liu} et~al.}{2019}]{Liu2019}
{Liu} H.,  {M\"{u}ller} {Zheng} {Helled} {Lin}  {Isella} 2019, \nat, 572

\bibitem[\protect\citeauthoryear{{Lozovsky}, {Helled}, {Rosenberg}  \&
  {Bodenheimer}}{{Lozovsky} et~al.}{2017}]{Lozovsky2017}
{Lozovsky} M.,  {Helled} R.,  {Rosenberg} E.~D.,   {Bodenheimer} P.,  2017,
  \mn@doi [\apj] {10.3847/1538-4357/836/2/227}, \href
  {https://ui.adsabs.harvard.edu/abs/2017ApJ...836..227L} {836, 227}

\bibitem[\protect\citeauthoryear{{Maeder}}{{Maeder}}{2009}]{Maeder2009}
{Maeder} A.,  2009, {Physics, Formation and Evolution of Rotating Stars},
  \mn@doi{10.1007/978-3-540-76949-1.
}

\bibitem[\protect\citeauthoryear{{Marley} \& {Porco}}{{Marley} \&
  {Porco}}{1993}]{MarleyPorco1993}
{Marley} M.~S.,  {Porco} C.~C.,  1993, \mn@doi [\icarus]
  {10.1006/icar.1993.1189}, \href
  {https://ui.adsabs.harvard.edu/abs/1993Icar..106..508M} {106, 508}

\bibitem[\protect\citeauthoryear{{Mathis} \& {Remus}}{{Mathis} \&
  {Remus}}{2013}]{Mathis2013}
{Mathis} S.,  {Remus} F.,  2013, {Tides in Planetary Systems and in Multiple
  Stars: a Physical Picture}.
pp 111--147, \mn@doi{10.1007/978-3-642-30648-8_4}

\bibitem[\protect\citeauthoryear{{Miguel}, {Guillot}  \& {Fayon}}{{Miguel}
  et~al.}{2016}]{Miguel2016}
{Miguel} Y.,  {Guillot} T.,   {Fayon} L.,  2016, \mn@doi [\aap]
  {10.1051/0004-6361/201629732}, \href
  {https://ui.adsabs.harvard.edu/abs/2016A&A...596A.114M} {596, A114}

\bibitem[\protect\citeauthoryear{{Moll}, {Garaud}, {Mankovich}  \&
  {Fortney}}{{Moll} et~al.}{2017}]{Moll2017}
{Moll} R.,  {Garaud} P.,  {Mankovich} C.,   {Fortney} J.~J.,  2017, \mn@doi
  [\apj] {10.3847/1538-4357/aa8d74}, \href
  {https://ui.adsabs.harvard.edu/abs/2017ApJ...849...24M} {849, 24}

\bibitem[\protect\citeauthoryear{{Nettelmann}, {Fortney}, {Moore}  \&
  {Mankovich}}{{Nettelmann} et~al.}{2015}]{Nettelmann2015}
{Nettelmann} N.,  {Fortney} J.~J.,  {Moore} K.,   {Mankovich} C.,  2015,
  \mn@doi [\mnras] {10.1093/mnras/stu2634}, \href
  {https://ui.adsabs.harvard.edu/abs/2015MNRAS.447.3422N} {447, 3422}

\bibitem[\protect\citeauthoryear{Ogilvie}{Ogilvie}{2014}]{Ogilvie2014}
Ogilvie G.~I.,  2014, Annual Review of Astronomy and Astrophysics, 52, 171

\bibitem[\protect\citeauthoryear{{Ogilvie} \& {Lin}}{{Ogilvie} \&
  {Lin}}{2004}]{OgilvieLin2004}
{Ogilvie} G.~I.,  {Lin} D.~N.~C.,  2004, \mn@doi [\apj] {10.1086/421454}, \href
  {https://ui.adsabs.harvard.edu/abs/2004ApJ...610..477O} {610, 477}

\bibitem[\protect\citeauthoryear{{Ogilvie} \& {Lin}}{{Ogilvie} \&
  {Lin}}{2007}]{Ogilvie2007}
{Ogilvie} G.~I.,  {Lin} D.~N.~C.,  2007, \mn@doi [\apj] {10.1086/515435}, \href
  {https://ui.adsabs.harvard.edu/abs/2007ApJ...661.1180O} {661, 1180}

\bibitem[\protect\citeauthoryear{{Podolak}, {Helled}  \& {Schubert}}{{Podolak}
  et~al.}{2019}]{PodolakUrNep2019}
{Podolak} M.,  {Helled} R.,   {Schubert} G.,  2019, \mn@doi [\mnras]
  {10.1093/mnras/stz1467}, \href
  {https://ui.adsabs.harvard.edu/abs/2019MNRAS.487.2653P} {487, 2653}

\bibitem[\protect\citeauthoryear{{Remus}, {Mathis}, {Zahn}  \&
  {Lainey}}{{Remus} et~al.}{2012}]{Remus2012}
{Remus} F.,  {Mathis} S.,  {Zahn} J.~P.,   {Lainey} V.,  2012, \mn@doi [\aap]
  {10.1051/0004-6361/201118595}, \href
  {https://ui.adsabs.harvard.edu/abs/2012A&A...541A.165R} {541, A165}

\bibitem[\protect\citeauthoryear{{Rieutord} \& {Valdettaro}}{{Rieutord} \&
  {Valdettaro}}{2010}]{Rieutord2010}
{Rieutord} M.,  {Valdettaro} L.,  2010, \mn@doi [Journal of Fluid Mechanics]
  {10.1017/S002211200999214X}, \href
  {https://ui.adsabs.harvard.edu/abs/2010JFM...643..363R} {643, 363}

\bibitem[\protect\citeauthoryear{Sakashita \& Hayashi}{Sakashita \&
  Hayashi}{1959}]{Sakashita1959}
Sakashita S.,  Hayashi C.,  1959, \mn@doi [Progress of Theoretical Physics]
  {10.1143/PTP.22.830}, 22, 830

\bibitem[\protect\citeauthoryear{{Schwarzschild} \& {H{\"a}rm}}{{Schwarzschild}
  \& {H{\"a}rm}}{1958}]{Schwarzschild1958}
{Schwarzschild} M.,  {H{\"a}rm} R.,  1958, \mn@doi [\apj] {10.1086/146548},
  \href {https://ui.adsabs.harvard.edu/abs/1958ApJ...128..348S} {128, 348}

\bibitem[\protect\citeauthoryear{{Shibley}, {Timmermans}, {Carpenter}  \&
  {Toole}}{{Shibley} et~al.}{2017}]{Shibley2017}
{Shibley} N.~C.,  {Timmermans} M.~L.,  {Carpenter} J.~R.,   {Toole} J.~M.,
  2017, \mn@doi [Journal of Geophysical Research (Oceans)]
  {10.1002/2016JC012419}, \href
  {https://ui.adsabs.harvard.edu/abs/2017JGRC..122..980S} {122, 980}

\bibitem[\protect\citeauthoryear{{Spiegel} \& {Veronis}}{{Spiegel} \&
  {Veronis}}{1960}]{Spiegel1960}
{Spiegel} E.~A.,  {Veronis} G.,  1960, \mn@doi [\apj] {10.1086/146849}, \href
  {https://ui.adsabs.harvard.edu/abs/1960ApJ...131..442S} {131, 442}

\bibitem[\protect\citeauthoryear{{Stevenson}}{{Stevenson}}{1982}]{Stevenson1982}
{Stevenson} D.~J.,  1982, \mn@doi [Annual Review of Earth and Planetary
  Sciences] {10.1146/annurev.ea.10.050182.001353}, \href
  {https://ui.adsabs.harvard.edu/abs/1982AREPS..10..257S} {10, 257}

\bibitem[\protect\citeauthoryear{{Stevenson} \& {Salpeter}}{{Stevenson} \&
  {Salpeter}}{1977}]{Stevenson1977}
{Stevenson} D.~J.,  {Salpeter} E.~E.,  1977, \mn@doi [\apjs] {10.1086/190479},
  \href {https://ui.adsabs.harvard.edu/abs/1977ApJS...35..239S} {35, 239}

\bibitem[\protect\citeauthoryear{Sutherland}{Sutherland}{2010}]{sutherland_2010}
Sutherland B.~R.,  2010, Internal Gravity Waves.
Cambridge University Press, \mn@doi{10.1017/CBO9780511780318}

\bibitem[\protect\citeauthoryear{{Sutherland}}{{Sutherland}}{2016}]{Sutherland2016}
{Sutherland} B.~R.,  2016, \mn@doi [Physical Review Fluids]
  {10.1103/PhysRevFluids.1.013701}, \href
  {https://ui.adsabs.harvard.edu/\#abs/2016PhRvF...1a3701S} {1, 013701}

\bibitem[\protect\citeauthoryear{{Thompson}}{{Thompson}}{2006}]{Thompson2006}
{Thompson} M.~J.,  2006, {An introduction to astrophysical fluid dynamics}

\bibitem[\protect\citeauthoryear{{Vazan}, {Helled}, {Podolak}  \&
  {Kovetz}}{{Vazan} et~al.}{2016}]{Vazan2016}
{Vazan} A.,  {Helled} R.,  {Podolak} M.,   {Kovetz} A.,  2016, \mn@doi [\apj]
  {10.3847/0004-637X/829/2/118}, \href
  {https://ui.adsabs.harvard.edu/abs/2016ApJ...829..118V} {829, 118}

\bibitem[\protect\citeauthoryear{{Vazan}, {Helled}  \& {Guillot}}{{Vazan}
  et~al.}{2018}]{Vazan2018}
{Vazan} A.,  {Helled} R.,   {Guillot} T.,  2018, \mn@doi [\aap]
  {10.1051/0004-6361/201732522}, \href
  {https://ui.adsabs.harvard.edu/abs/2018A&A...610L..14V} {610, L14}

\bibitem[\protect\citeauthoryear{Wahl et~al.,}{Wahl et~al.}{2017}]{Wahl17}
Wahl S.,  et~al., 2017, \mn@doi [Geophysical Research Letters]
  {10.1002/2017GL073160}, 44

\bibitem[\protect\citeauthoryear{{Wei}}{{Wei}}{2018}]{Wei2018}
{Wei} X.,  2018, \mn@doi [\apj] {10.3847/1538-4357/aaa54d}, \href
  {https://ui.adsabs.harvard.edu/abs/2018ApJ...854...34W} {854, 34}

\bibitem[\protect\citeauthoryear{{Wilson} \& {Militzer}}{{Wilson} \&
  {Militzer}}{2012}]{Wilson2012}
{Wilson} H.~F.,  {Militzer} B.,  2012, \mn@doi [\prl]
  {10.1103/PhysRevLett.108.111101}, \href
  {https://ui.adsabs.harvard.edu/abs/2012PhRvL.108k1101W} {108, 111101}

\bibitem[\protect\citeauthoryear{{Wood}, {Garaud}  \& {Stellmach}}{{Wood}
  et~al.}{2013}]{Wood2013}
{Wood} T.~S.,  {Garaud} P.,   {Stellmach} S.,  2013, \mn@doi [\apj]
  {10.1088/0004-637X/768/2/157}, \href
  {https://ui.adsabs.harvard.edu/abs/2013ApJ...768..157W} {768, 157}

\bibitem[\protect\citeauthoryear{{Wunsch}}{{Wunsch}}{2018}]{Wunsch2018}
{Wunsch} S.,  2018, \mn@doi [Physical Review Fluids]
  {10.1103/PhysRevFluids.3.114803}, \href
  {https://ui.adsabs.harvard.edu/abs/2018PhRvF...3k4803W} {3, 114803}

\makeatother
\end{thebibliography}



\appendix
\section{Equivalence with the Boussinesq approximation}\label{app:B_approx}

Our model in \S~\ref{sec:set_up} is equivalent to taking the Boussinesq approximation from the outset. Here we outline the derivation of Eqn.~(\ref{EQ1}) starting from the linearised Boussinesq system (neglecting viscosity and thermal diffusion)
\begin{align}
  &  \dpartial{\vb{u}}{t} = -\frac{1}{\rho_0} \nablab p + b\vb{r}, \\
    & \dpartial{b}{t}+u_r\frac{N^2}{r}  =0,
    \label{eqA:b}
\end{align}
where $b=-\frac{g \rho}{r \rho_0}$ is a buoyancy variable, $N^2$ is defined in \S~\ref{sec:set_up}, and $\vb{u}$ is incompressible. 
The radial and horizontal components of the Eulerian displacement satisfy
\begin{align}\label{eqA:ur}
    &\ddpartial{\xi_r}{t} = -\frac{1}{\rho_0} \dpartial{p}{r} + r b, \\
\label{eqA:uh}
    &\ddpartial{\xib_h}{t} = -\frac{1}{\rho_0} \nablab_h p.
\end{align}
Using incompressibility,
together with Eqn.~(\ref{eqA:uh}), we can eliminate $\xib_h$, to obtain
\begin{equation}\label{eqA:p}
    \frac{1}{r^2} \dpartial{}{r}\bigg(r^2 \ddpartial{\xi_r}{t}\bigg) - \frac{1}{\rho_0} \nablab_h^2 p=0.
\end{equation}
When perturbations are expanded  
using spherical harmonics with harmonic time-dependence (as in \S~\ref{sec:set_up}), and 
with some algebra, Eqns.~(\ref{eqA:b}), (\ref{eqA:ur}), and (\ref{eqA:p}), can be combined to eliminate $\tilde{b}$ and $\tilde{p}$, resulting in Eqn.~(\ref{EQ1}).


\bsp	
\end{document}